\documentclass[acmsmall,screen]{acmart}

\usepackage{tabularx}
\AtBeginDocument{%
  }

\usepackage{etoolbox}
\AtBeginEnvironment{minted}{\footnotesize}

\setcopyright{acmlicensed}
\copyrightyear{2018}
\acmYear{2018}
\acmDOI{XXXXXXX.XXXXXXX}

\acmJournal{JACM}
\acmVolume{37}
\acmNumber{4}
\acmArticle{111}
\acmMonth{8}

\usepackage{CJKutf8}
\usepackage{subfigure}
\usepackage{amsmath}
\usepackage{amsthm}

\usepackage{ragged2e}
\usepackage{wasysym}
\usepackage{listings}

\usepackage{graphicx}  
\usepackage{adjustbox}  
\usepackage{varwidth}   

\usepackage{array}

\usepackage{float}

\usepackage{amssymb}
\usepackage{multirow}
\usepackage{mathrsfs}
\usepackage{algorithmic}
\usepackage[ruled,linesnumbered]{algorithm2e}
\SetKwRepeat{Do}{do}{while}
\usepackage{url}
\usepackage{enumitem}

\usepackage{soul}
\usepackage{balance}
\usepackage{color}
\usepackage{makecell}
\usepackage{pdflscape}

\usepackage{stfloats}
\usepackage{verbatim}
\usepackage{setspace}
\usepackage{threeparttable}
\usepackage{supertabular,booktabs}
\usepackage{rotating}

\usepackage{pifont}

\usepackage{lettrine}
\usepackage{xurl}
\usepackage{hyperref}
\hypersetup{
	colorlinks=true,
	linkcolor=blue,
	citecolor=blue,
	urlcolor=black
}
\makeatletter
\renewcommand\@cite[2]{\textcolor{blue}{[#1]}}
\makeatother
\usepackage[hyphenbreaks]{breakurl}

\usepackage{fancyhdr}

\allowdisplaybreaks[4]

\usepackage{tcolorbox}
\tcbuselibrary{breakable}
\usepackage{soul}  

\usepackage[table]{pict2e}

\usepackage{colortbl}
\usepackage{longtable}
\usepackage{array}
\usepackage{booktabs}

\usepackage{pdflscape} 
\definecolor{lightpink}{rgb}{1.0, 0.87, 0.87}
\definecolor{lightpurple}{rgb}{0.94, 0.85, 0.94}
\definecolor{lightgreen}{rgb}{0.88, 1.0, 0.88}
\definecolor{lightyellow}{rgb}{1.0, 1.0, 0.88}
\definecolor{lightblue}{rgb}{0.88, 0.94, 1.0}
\definecolor{lightgrey}{rgb}{0.8, 0.8, 0.8}
\definecolor{mintfrost}{rgb}{0.98, 0.98, 0.98}   

\definecolor{deepgreen}{rgb}{0.0, 0.75, 0.0}

\newcommand{\hlpink}[1]{{\sethlcolor{lightpink}\hl{#1}}}
\newcommand{\hlpurple}[1]{{\sethlcolor{lightpurple}\hl{#1}}}
\newcommand{\hlgreen}[1]{{\sethlcolor{lightgreen}\hl{#1}}}
\newcommand{\hlyellow}[1]{{\sethlcolor{lightyellow}\hl{#1}}}
\newcommand{\hlblue}[1]{{\sethlcolor{lightblue}\hl{#1}}}
\newcommand{\hlgrey}[1]{{\sethlcolor{lightgrey}\hl{#1}}}
\newcommand{\hlgray}[1]{{\sethlcolor{lightgray}\hl{#1}}}
\newcommand{\hlmintfrost}[1]{{\sethlcolor{mintfrost}\hl{#1}}}

\definecolor{codegreen}{rgb}{0,0.6,0}
\definecolor{codegray}{rgb}{0.5,0.5,0.5}
\definecolor{codepurple}{rgb}{0.58,0,0.82}
\definecolor{backcolour}{rgb}{0.95,0.95,0.92}

\definecolor{green-up}{RGB}{0, 200, 0} 
\definecolor{gray-neutral}{RGB}{120, 120, 120}
\newcommand{\up}{\textcolor{green-up}{$\blacktriangle$}}
\newcommand{\down}{\textcolor{red}{$\blacktriangledown$}}

\lstdefinestyle{mystyle}{
	backgroundcolor=\color{backcolour},
	commentstyle=\color{codegreen},
	keywordstyle=\color{magenta},
	numberstyle=\tiny\color{codegray},
	stringstyle=\color{codepurple},
	basicstyle=\footnotesize\ttfamily,
	breakatwhitespace=false,
	breaklines=true,
	captionpos=b,
	keepspaces=true,
	numbers=left,
	numbersep=5pt,
	showspaces=false,
	showstringspaces=false,
	showtabs=false,
	tabsize=2
}

\newcommand{\modeliconsize}{2.1ex}

\newcommand{\modelicon}[1]{\raisebox{-0.2ex}{\includegraphics[height=\modeliconsize]{#1}}}

\newcommand{\ClaudeIcon}{\modelicon{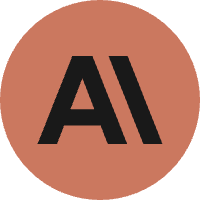}}
\newcommand{\GeminiIcon}{\modelicon{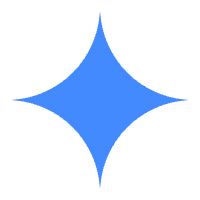}}
\newcommand{\GPTIcon}{\modelicon{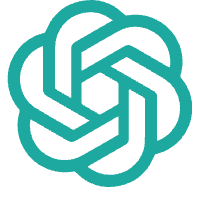}}

\newcommand{\GrokIcon}{\modelicon{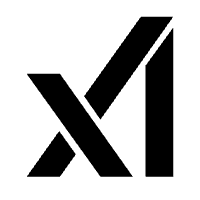}}
\newcommand{\DoubaoIcon}{\modelicon{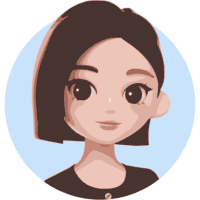}}
\newcommand{\KimiIcon}{\modelicon{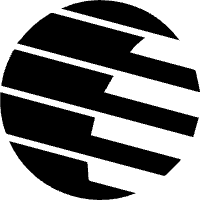}}
\newcommand{\ERNIEIcon}{\modelicon{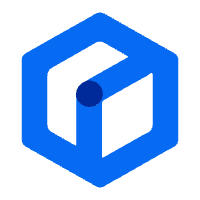}}

\newcommand{\QwenIcon}{\modelicon{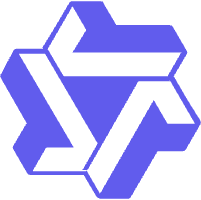}}

\newcommand{\MistralIcon}{\modelicon{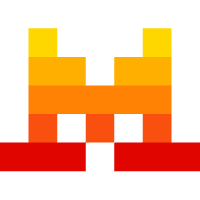}}

\newcommand{\Claude}{\textsc{Claude}\xspace}
\newcommand{\Gemini}{\textsc{Gemini}\xspace}
\newcommand{\GPT}{GPT\xspace}

\newcommand{\Grok}{\textsc{Grok}\xspace}
\newcommand{\Doubao}{\textsc{Doubao}\xspace}
\newcommand{\Kimi}{\textsc{Kimi}\xspace}
\newcommand{\ERNIE}{ERNIE\xspace}

\newcommand{\Qwen}{\textsc{Qwen}\xspace}

\newcommand{\Mistral}{\textsc{Mistral}\xspace}

\newcommand{\ClaudeM}{\ClaudeIcon~\Claude}
\newcommand{\GeminiM}{\GeminiIcon~\Gemini}
\newcommand{\GPTM}{\GPTIcon~\GPT}

\newcommand{\GrokM}{\GrokIcon~\Grok}
\newcommand{\DoubaoM}{\DoubaoIcon~\Doubao}
\newcommand{\KimiM}{\KimiIcon~\Kimi}
\newcommand{\ERNIEM}{\ERNIEIcon~\ERNIE}

\newcommand{\QwenM}{\QwenIcon~\Qwen}

\newcommand{\MistralM}{\MistralIcon~\Mistral}

\newcounter{DaveCommentCounter}
\newcommand{\ddu}[1]{
    \stepcounter{DaveCommentCounter}
    \textcolor{purple}{\textit{/**Dave's comment [\arabic{DaveCommentCounter}]: I don't understand the intended meaning in the next sentence. Please revise/delete/explain. **/}}
}

\newcommand{\dns}[1]{
    \stepcounter{DaveCommentCounter}
    \textcolor{purple}{\textit{/**Dave's comment [\arabic{DaveCommentCounter}]: I'm not sure that I have captured the intended meaning in the next sentence. Please check/confirm. **/}}
}

\newcounter{ChenhuiCommentCounter}
\begin{document}
\title{Large Language Models for Mobile GUI Text Input Generation: An Empirical Study}

\author{Chenhui Cui}
\email{3230002105@student.must.edu.mo}
\orcid{0009-0004-8746-316X}
\affiliation{
  \institution{School of Computer Science and Engineering, Macau University of Science and Technology}
  \city{Macao SAR}
  \country{China}
  \postcode{999078}
}

\author{Tao Li}
\email{3220007015@student.must.edu.mo}
\orcid{0009-0001-7413-9692}
\affiliation{
  \institution{School of Computer Science and Engineering, Macau University of Science and Technology}
  \city{Macao SAR}
  \country{China}
  \postcode{999078}
}

\author{Junjie Wang}
\email{junjie@iscas.ac.cn}
\orcid{0000-0002-9941-6713}
\affiliation{
  \institution{Institute of Software Chinese Academy of Sciences}
  \city{Beijing}
  \country{China}
  \postcode{100190}
}

\author{Chunyang Chen}
\email{chun-yang.chen@tum.de}
\orcid{0000-0003-2011-9618}
\affiliation{
  \institution{Department of Computer Science, Technical University of Munich}
  \city{Heilbronn}
  \country{Germany}
  \postcode{100190}
}

\author{Dave Towey}
\email{dave.towey@nottingham.edu.cn}
\orcid{0000-0003-0877-4353}
\affiliation{
  \institution{School of Computer Science, University of Nottingham Ningbo China}
  \city{Ningbo}
  \state{Zhejiang}
  \country{China}
  \postcode{315100}
}

\author{Rubing Huang}
\email{rbhuang@must.edu.mo}
\orcid{0000-0002-1769-6126}
\affiliation{
  \institution{School of Computer Science and Engineering, Macau University of Science and Technology}
  \city{Macao SAR}
  \country{China}
  \postcode{999078}
}
\affiliation{
  \institution{Macau University of Science and Technology Zhuhai MUST Science and Technology Research Institute}
  \city{Zhuhai}
  \state{Guangdong Province}
  \country{China}
  \postcode{519099}
}

\begin{abstract}
    Mobile applications (apps) have become an essential part of daily life, making their quality assurance ever more important.
    Graphical User Interface (GUI) testing is a widely-used quality-assurance method for mobile apps, but text-input components remain a practical obstacle for automated GUI exploration because many User Interface (UI) pages require semantically appropriate text before they can move to subsequent UI pages.
    Large Language Models (LLMs) have demonstrated excellent generation of such context-aware text inputs, but it remains unclear how different UI-context representations, execution feedback, and human intervention affect LLM-based Android text-input generation.
    This paper reports on a large-scale empirical study that extensively investigates the effectiveness of nine state-of-the-art (SOTA)
    LLMs for Android text-input generation on UI pages.
    We collected 115 Android apps from Google Play and compared three UI-context prompting settings:
    extracted context; 
    UI-hierarchy XML; and 
    screenshot-vision input.
    The extracted-context and UI-hierarchy-XML prompting reach similar page-pass-through rates (PPTRs)
    ---
    measured as the proportion of the LLM-generated text inputs that can result in moving the GUI to the next page
    ---
    of 71.4\% and 71.0\%, respectively;     while 
    the screenshot-vision prompting reaches 65.1\%.
    However, UI-hierarchy-XML and screenshot-vision incur much higher token costs than the extracted-context setting.
    We conducted an experiment to assess the bug-detection capabilities of LLMs by directly generating invalid text inputs.
    The study involved 37 real-world bugs related to text inputs.
    The experimental results show that the bug-detection rates (using LLMs to generate invalid text inputs) were around 51\%, across all nine evaluated LLMs.
    We further evaluated a feedback-enhanced generation protocol that feeds failed parsing information, or execution outcomes, back to the LLM for subsequent generation attempts.
    The feedback protocol achieved the average PPTRs of between 69.18\% and 73.78\%, and an average bug-detection rate of between 50.95\% and 64.46\%.
    We also invited professional testers to manually evaluate, modify, and re-create the LLM-generated text inputs.
    We integrated the text-input generation process into DroidBot, a well-known automated Android testing tool, to augment its UI-exploration capabilities.
    Finally, we present several valuable insights regarding the application of LLMs to Android testing:
    These insights are expected to advance the knowledge and practices within the Android testing community.
\end{abstract}



\received{10 September 2025}
\received[revised]{19 June 2026}

\maketitle

\section{Introduction
\label{sec:introduction}}

The significance of mobile phones is unparalleled in today's era:
Their portability and versatility make them indispensable in our daily lives.
As of June 2025, Google Play\footnote{\url{https://play.google.com/store/apps.}\label{footnote:googleplay}}, the official Android application (app) store, boasts a distribution of over two million apps~\cite{Google2025}.
These apps cover a wide range, from social and streaming media to medical and financial services.
For users, employing these apps has become a fundamental method to understand and interact with the world around them.
However, for testers, ensuring the quality of these apps presents a complex challenge.
Because mobile apps rely on Graphical User Interfaces (GUIs) as a critical bridge for user interaction, GUI testing~\cite{LinYLLY18, GuSMC0YZLS19, CoppolaMT19} has emerged as a popular method for verifying the functional correctness of these apps.
Although manual GUI testing has been widely used in practice, automated GUI testing is more effective at reducing costs and testing more complex GUI scenarios~\cite{HaasEJPA21,LiuCWCHHW23}.
Various approaches have been proposed to support automated mobile GUI testing, including model-based~\cite{GuSMC0YZLS19, LvPZSLY22} and learning-based~\cite{LiY0C19, GaoTGG22}.
These methods are essential for dynamically interacting with apps, enabling scrolling and clicking on the User Interface (UI) pages.
However, these methods lack sufficient attention to intricate interactions within apps, particularly in text-input generation~\cite{LiuCWCHHW23}.

\begin{figure*}
    \centering
    \includegraphics[width=\textwidth]{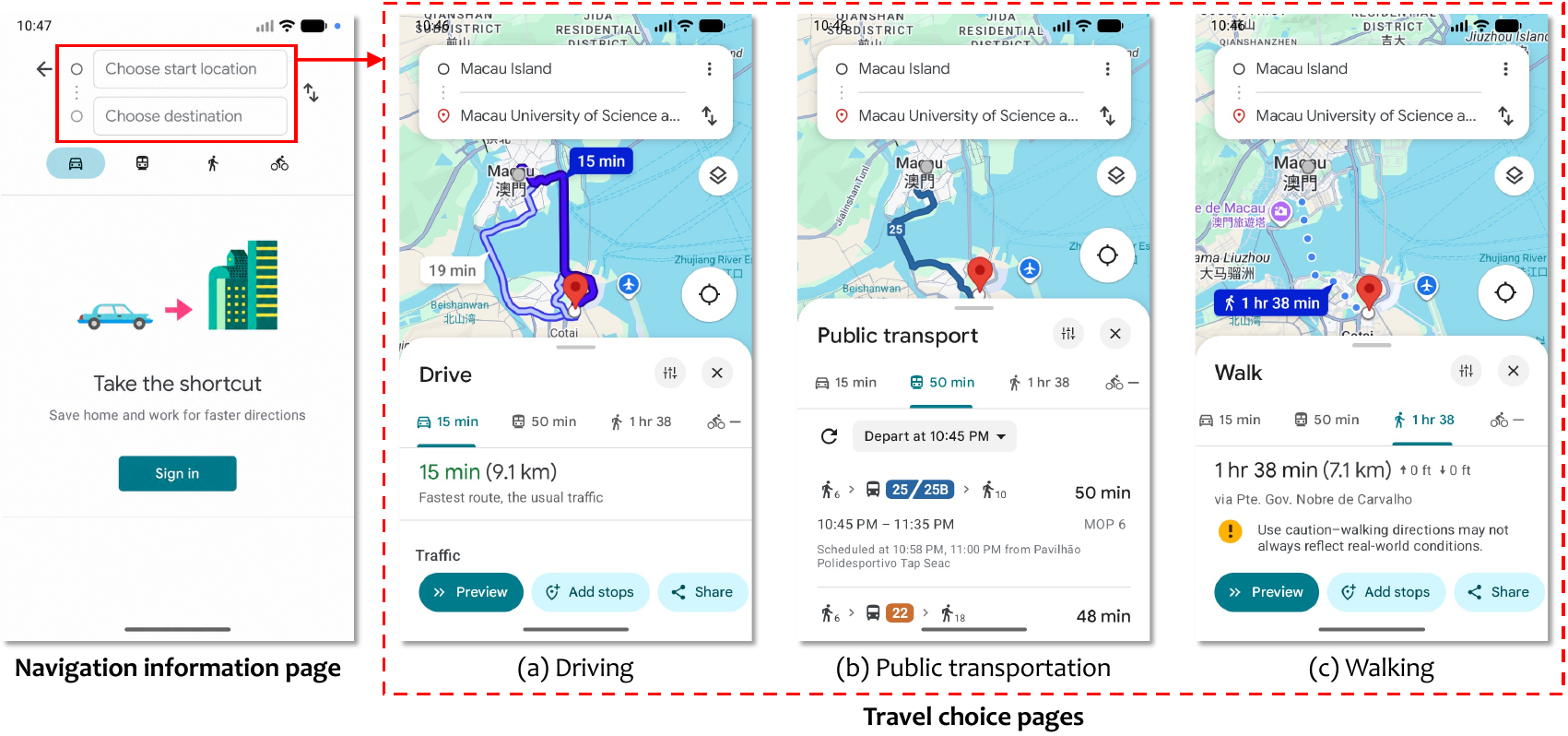}\captionsetup{skip=2pt}
    \caption{Example of text inputs in Google Maps.}
    \label{FIG: google-map}
\end{figure*}

Text inputs are essential elements of mobile GUIs, providing more complex interactions than actions based only on swiping and clicking.
A study by Liu et al.~\cite{LiuCWCHHW23} found that 73\% of the 6950 surveyed apps had UI pages requiring text inputs to move to the next page:
Text inputs thus constrain UI exploration.
In addition, previous studies~\cite{LiuCWCHHW23,ChoudharyGO15,SuMCWYYPLS17} indicate that incomplete UI exploration yields relatively low coverage and limits the effectiveness of testing.
For example, Monkey~\cite{monkey}, a tool that can randomly generate Android test inputs, was found to only achieve approximately half of the target coverage when testing~\cite{ChoudharyGO15}.
Figure~\ref{FIG: google-map} depicts the importance of the text inputs in Google Maps\footnote{\url{https://play.google.com/store/apps/details?id=com.google.android.apps.maps.}}, a popular navigation app.
There are two text-input components on its navigation information page: ``Choose start location'' and ``Choose destination''.
First, the components provide an interface for interacting with users.
Second, when two locations are confirmed, Google Maps presents the travel-choice pages tailored to the specified journey.
This example shows that, in some cases, apps require users to enter specific information in text input components, and it is not possible to proceed to the next UI page without these inputs.

In automated testing, however, it is challenging to generate effective text inputs that are consistent with the context of the relevant UI page~\cite{LiuCWCHHW23, HeZYCLLYZYZD20}:
Different contexts may result in different definitions of text input types for UI pages (such as account passwords, queries, comments, etc.).
The Google Maps content in Figure~\ref{FIG: google-map} is relatively easy for humans to understand: It is a navigation information page requiring start and destination locations.
However, in automated testing, accurately recognizing the need for specific (contextual) text inputs is a complex task.
For example, as a recent study~\cite{WangLYCZDX18} shows, most current automated methods can only generate random (or specific) strings such as ``hello world'',
which may not be consistent with the context.
The ``hello world'' is not a specific location name, and typically does not trigger a transition to the next UI page.
This will be problematic for testing.
Therefore, it is important to generate effective text inputs that meet the UI context for GUI testing~\cite{LiuCWCHHW23, HeZYCLLYZYZD20}.
Large Language Models (LLMs)~\cite{abs-2303-18223}, trained on ultra-large-scale corpora, have provided the potential to understand complex Natural Language Processing (NLP) tasks, and can provide appropriate inputs for many domains, such as finance~\cite{RaoSKMMY25} and medical treatment~\cite{XuCHL25}.
Recently, the Generative Pre-trained Transformer (GPT) series of LLMs has been widely discussed~\cite{LiuCWCHHW23,abs-2310-15657,HuangQZ25}.
They can generate texts, answer questions, and handle other tasks that require simple inputs.
Liu et al.~\cite{LiuCWCHHW23}, for example, built a model based on GPT-3~\cite{gpt3}, \texttt{QTypist}, to automatically generate text inputs for mobile apps.
However, a recent attempt to set up \texttt{QTypist}~\cite{LiuCWCHHW23} revealed that OpenAI had discontinued API access to GPT-3 (which \texttt{QTypist} relies upon).
It is therefore essential to explore the potential and substitutability of different LLMs to guide text-input generation in mobile app GUI testing.
Modern LLMs increasingly support both textual and visual inputs, while many commercial models remain closed-source and difficult to interpret mechanistically.
This leads to a broader question than simply which individual LLM happens to rank highest at one point in time.
For Android GUI text-input generation, among other things, testers need to know:
which UI-context representation should be exposed to LLMs; 
whether execution outcomes should be fed back into subsequent text-input generation attempts; and 
how these choices affect effectiveness, cost, and practical testing workflow.
To the best of our knowledge, no empirical study has systematically compared different LLM prompt settings and feedback protocols for the generation of text inputs for mobile GUI testing.

Motivated by the above facts, we conducted a large-scale empirical study of LLM-based Android text-input generation under different UI-context prompting settings and feedback strategies.
We evaluated nine recent LLMs on 115 UI pages containing 183 text-input components collected from Google Play Store\textsuperscript{\ref{footnote:googleplay}}, and on 37 real-world text-input-related bugs collected from F-Droid\footnote{\url{https://f-droid.org/.}\label{footnote:fdroid}}.
The source code for this study is available at \url{https://github.com/chenhuicui/text-input-generation}.
Based on the experimental results, our main findings can be summarized as follows:

\begin{itemize}[leftmargin=2em, topsep=0pt, itemsep=0pt, parsep=0pt, partopsep=0pt]
    \item
    Extracted-context and UI-hierarchy-XML prompting achieve similar average PPTR values (71.4\% and 71.0\%, respectively); while screenshot-vision prompting reaches 65.1\%.
    Statistical comparisons further indicate that most LLM pairs exhibit no significant differences, suggesting that the prompt-setting strategy has a greater impact on the performance of navigating the GUI to the next page than the choice of LLM.

    \item
    Different UI-context representations involve a clear effectiveness-cost tradeoff.
    XML-based prompting exposes complete structural information, but costs substantially more tokens than extracted-context prompting.
    Screenshot-vision prompting provides visual information, but its effectiveness was nearly the same as, but slightly worse than, the other two settings.

    \item
    The feedback-enhanced generation protocol directly reduces the limitation of one-shot prompting.
    Across the nine LLMs, the feedback protocol improves the average PPTR from 69.18\% to 73.78\% and the average bug detection rate from 50.95\% to 64.46\%.

    \item
    For the randomly-selected evaluation cases, the tester-modified inputs raise bug-detection rate from 36.10\% for LLM-generated inputs to 100.00\%, while tester-generated inputs reach an average bug detection rate of 93.33\%. 

    \item 
    The LLM-generated text inputs can improve the UI-exploration capability of traditional automated Android GUI testing tools.
    When integrated into DroidBot, they increase UI-page exploration by 36.06\% and 26.41\% under the Depth-First Search (DFS) and Breadth-First Search (BFS) strategies\footnote{DFS explores the newest, deepest UI page first, drilling down into one path as far as possible before backtracking to try alternatives. BFS explores all UI components on the current page before moving deeper, systematically covering the app, layer by layer.\label{footnote:dfs_bfs}}, respectively.
\end{itemize}

The main contributions of this paper can be summarized as follows:
\begin{itemize}[leftmargin=2em, topsep=0pt, itemsep=0pt, parsep=0pt, partopsep=0pt]
    \item[(1)]
    This is, we believe, the first comprehensive empirical study of LLM-based Android text-input generation under three UI-context prompting settings: 
    extracted textual context;
    UI-hierarchy XML; and 
    screenshot-based vision input.
    The study evaluates nine recent LLMs on the same page and bug samples, allowing the comparison to focus on UI-context representations, rather than transient LLM rankings alone.

    \item[(2)]
    We evaluate a feedback-enhanced generation protocol for both page-pass-through and bug-detection tasks.
    The protocol uses execution outcomes, parsing failure information, and recent failed-generation history to guide subsequent attempts.

    \item[(3)]
    We quantify human involvement in LLM-generated text inputs by comparing LLM-generated, tester-generated, and tester-modified inputs for both page pass-through and revealing bugs.

    \item[(4)]
    We integrate LLM-generated text inputs into DroidBot and evaluate whether this independent text-input generation step improves complete Android GUI exploration workflows.

    \item[(5)]
    We offer six insights into the use of LLMs for Android testing.
    These will be of interest and benefit to the Android testing community.
\end{itemize}

The rest of this paper is organized as follows:
Section~\ref{SEC: Background} provides necessary background information.
Section~\ref{SEC: Task Formulation} outlines the task formulation for this empirical study.
Section~\ref{SEC: Experimental Study} describes the experimental design.
Section~\ref{SEC: Experimental Results and Analysis} presents and analyzes the experimental results.
Section~\ref{SEC: Related Work} discusses some related work.
Finally, Section~\ref{SEC: Conclusions and Future Work} concludes the paper and discusses potential future work.

\section{Background
\label{SEC: Background}}

This section provides background information on automated Android testing and LLMs.

\subsection{Automated Android Testing
\label{SEC: Android Automated Testing}}

\begin{figure}
	\centering
	\includegraphics[width=0.58\textwidth]{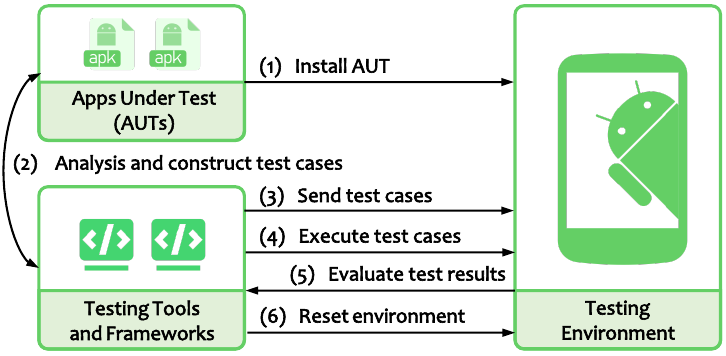}\captionsetup{skip=2pt}
	\caption{Typical workflow of automated Android testing.}
	\label{FIG: Typical workflow of Android automated testing}
\end{figure}

Automated Android testing~\cite{kong2018automated} aims to test the functionality, usability, and compatibility of AUTs running on Android devices.
Figure~\ref{FIG: Typical workflow of Android automated testing} presents a typical workflow of automated Android testing processes, including GUI testing.
In Step (1), the AUTs are installed on devices or emulators within the testing environment, such as a specified version of the Android OS.
In Step (2), an in-depth analysis of the AUTs is conducted to understand the expected behavior, which is then used to construct test cases based on the testing tools and frameworks.
Step (3) sends the test cases into the testing environment.
In Step (4), the test cases are executed in the testing environment with the support of testing tools and frameworks.
After the execution, the test results determine whether the testing tasks pass or fail in Step (5).
In Step (6), the testing environment is reset, which may involve uninstalling the AUTs, removing temporary files, and resetting the system state.

Android is an event-driven platform that relies heavily on user inputs, such as gesture inputs (e.g., click and swipe) and text inputs (e.g., account names and passwords)~\cite{wu2019analyses, jabbarvand2019search}.
It is difficult to automatically generate effective text inputs, because they need to conform to the contextual information on UI pages~\cite{LiuCWCHHW23}, such as hint texts displayed in text-input components (which often describe what should be entered).

\subsection{Large Language Models
\label{SEC: Large Language Models}}

Large Language Models (LLMs)~\cite{abs-2303-18223}  have been trained on ultra-large-scale corpora.
They have demonstrated impressive performance in Natural Language Processing (NLP).
LLMs (like GPT-5.4~\cite{openai_gpt54_2026} and Claude-4.6~\cite{claude_46_2026}) can show understanding of natural language sentences and visual inputs, and generate reasonable texts, which can be used in different downstream tasks related to software testing,
such as unit-test generation~\cite{TufanoDSS22, abs-2307-00588, LemieuxILS23} and
system test-input generation~\cite{LiuCWCHHW23,0010ZPZMZ17,DBLP:conf/pldi/YeTTHFSBW021}.
Previous research studied Vision Language Models (VLMs), which align visual inputs with natural-language representations, and enable models to reason over image-text inputs~\cite{alayrac2022flamingo,liu2023visual}.
For mobile GUI testing, VLMs are relevant because screenshots can carry visual layout, widget appearance, and rendered-text cues that are not always fully preserved in textual UI metadata~\cite{liu2024vision}.
LLMs may have different parameter scales, network structures, and training datasets.
However, users can interact with and query LLMs using natural language;
and use visual inputs for VLMs and multimodal LLMs.

\section{Text-Input Generation and Evaluation Workflow
\label{SEC: Task Formulation}}

\begin{figure*}
	\centering
	\includegraphics[width=\textwidth]{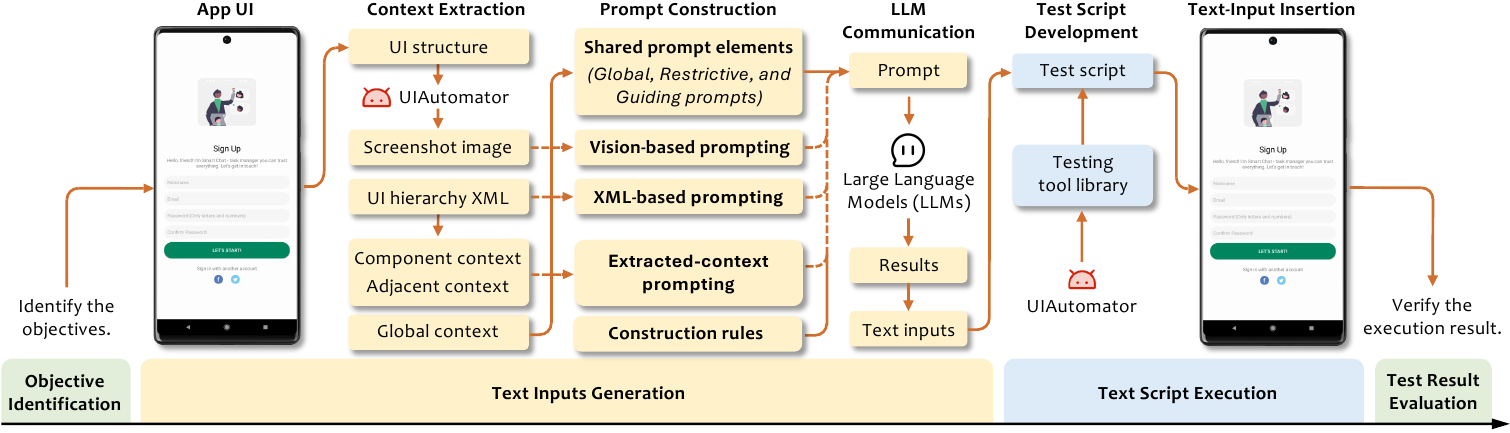}\captionsetup{skip=2pt}
    \caption{Overview of the text-input generation and evaluation workflow.}
	\label{FIG: framework of study design}
\end{figure*}

Figure~\ref{FIG: framework of study design} illustrates the workflow for generating and evaluating text inputs, which includes the following four steps:
(1) \textit{Objective Identification};
(2) \textit{Text-Input Generation};
(3) \textit{Test-Script Execution}; and
(4) \textit{Test-Result Evaluation}.
This section next introduces these steps.

\subsection{Objective Identification
\label{sec:objective-identification}}

The study evaluates how LLM-based text-input generation behaves under different UI-context representations and feedback strategies, using the selected LLMs as the generation backends. 
Accordingly, the study focuses on black-box GUI-level testing\footnote{Black-box GUI-level testing generally refers to testing an app through its displayed GUI and observed execution outcomes, without relying on source-code information~\cite{kong2018automated}.}:
The prompts are constructed from the currently-displayed UI page and execution outcomes, rather than from source code, formal app specifications, or requirement documents.
The design, therefore, supports practical Android GUI testing on real apps, while also making explicit that deeper business-logic constraints require complementary specification-aware techniques.

\subsection{Text-Input Generation
\label{SEC: Test Case Generation}}

This section describes the steps involved in generating text inputs by LLMs, including \textit{context extraction}, \textit{prompt construction}, and \textit{LLM communication}.

\subsubsection{Context Extraction}
\label{SEC: Contextual Integration}

As shown in Figure~\ref{FIG: framework of study design}, we extract the contextual information from the currently-displayed UI page for prompt-construction.
Using UIAutomator~\cite{UIAutomator}, we collect two types of UI artifacts from the UI structure:  
\textit{screenshot images} and \textit{UI-hierarchy XML}.
In addition, based on the basic app information and the UI-hierarchy XML, we further extract three types of textual context:
\textit{component context};
\textit{adjacent context}; and
\textit{global context}.

\begin{itemize}[leftmargin=2em, topsep=0pt, itemsep=0pt, parsep=0pt, partopsep=0pt]
    \item
    The \textit{component context} refers to the essential information about text-input components, including the component type, the hint text displayed within the component (which often provides guidance or a description of what should be entered), and the resource ID (which is a unique identifier to match and position the text-input component on the UI page).

    \item
    The \textit{adjacent context} refers to the texts adjacent to the text-input components.
    We define four adjacent positions around the text-input component: top, bottom, left, and right.
    The text labels at these four positions are referred to as the adjacent context.
    The adjacent context can help interpret the function and purpose of the text-input components.
    Figure~\ref{FIG: com_adj_example} (Appendix~\ref{ASEC: Contextual Integration}) presents an example of extracting the component context and adjacent context from a real-world app.

    \item
    The \textit{global context} refers to higher-level contextual information, such as the app name, the name of the currently running activity, and the number of text-input components on the UI page.
    The global context can provide a broader understanding of the role and intention of these text-input components.
\end{itemize}

Overall, for each UI page, the screenshot image, the UI-hierarchy XML, and the extracted textual context provide different views of the same UI page and are used to construct prompts for the screenshot-vision, UI-hierarchy-XML, and extracted-context prompting settings, respectively.

\begin{table*}
	\centering
	\scriptsize
    \captionsetup{skip=2pt}
	\caption{Definitions of Different Sub-Prompts and Construction Rules.}
	\label{TAB: linguistic patterns of prompts}
	\setlength\tabcolsep{1.6mm}

	\begin{tabular}{c|p{2.1cm}|p{5.0cm}|p{5.0cm}}
		\hline
		\textbf{No.} & \textbf{Sub-Prompts} & \textbf{Patterns / Inputs} & \textbf{Examples} \\
		\hline

		\rowcolor{lightgreen}
		1 & Global ($GloP$)
		& This is an app named \texttt{[MASK]}, and there are \texttt{[MASK]} text-input components on its \texttt{[MASK]} page, including \texttt{[MASK]}.
		& This is an app named \textit{Smart chat}, and there are \textit{2} text-input components on its \textit{sign up} page, including \textit{2} \textit{EditText}. \\
		\hline

		\rowcolor{lightpink}
		2 & Component ($ComP$)
		& For the \texttt{[MASK]} (resource ID: \texttt{[MASK]}), the hint text is ``\texttt{[MASK]}'', and currently displays ``\texttt{[MASK]}''.
		& For the \textit{1st EditText} (resource ID: \textit{et\_nickname}), the hint text is ``\textit{Nickname}'', and currently displays ``\textit{Nickname}''. \\
		\hline

		\rowcolor{lightpurple}
		3 & Adjacent ($AdjP$)
		& The \texttt{[MASK]} adjacent label of \texttt{[MASK]} is ``\texttt{[MASK]}''.
		& The \textit{right} adjacent label of \textit{et\_nickname} is ``\textit{Please input nickname}''. \\
		\hline

		\rowcolor{mintfrost}
		4 & Hierarchy ($XML$)
		&
		\makecell[l]{
		Here is the UI-hierarchy XML that contains\\
		the text-input components:\\
		\texttt{[MASK]}
		}
		&
		\makecell[l]{
		Here is the UI-hierarchy XML that contains\\
		the text-input components:\\
		\textit{\textless hierarchy rotation=``0''\textgreater}\\
		~~~~\textit{\textless node class=``android.widget.FrameLayout''...\textgreater}\\
		~~~~~~~~\textit{\textless node text=``Nickname'' resource ID=``et\_nickname''}\\
		~~~~~~~~~~~~~~~~\textit{class=``android.widget.EditText''...\textgreater}\\
		~~~~~~~~\textit{\textless/node\textgreater}...\\
		~~~~\textit{\textless/node\textgreater}\\
		\textit{\textless/hierarchy\textgreater}
		} \\
		\hline

		\rowcolor{lightgray}
		5 & Image ($ImgP$)
		&
		\makecell[l]{
		Here is the screenshot image of the current\\
		UI page that contains the text-input components:\\
		\texttt{[IMAGE]}
		}
		&
		\makecell[l]{
		Here is the screenshot image of the current\\
		UI page that contains the text-input components:\\
		\textit{A screenshot of the sign-up page with}\\
		\textit{two text-input components.}
		} \\
		\hline

		\rowcolor{lightyellow}
		6 & Restrictive ($ResP$)
		&
		\makecell[l]{
		Please generate \texttt{[MASK]} text input for these\\
		components.\\
		Response format MUST BE:\\
		\textasciigrave\textasciigrave\textasciigrave json \{\\
		~~~~\texttt{[MASK]}\\
		\} \textasciigrave\textasciigrave\textasciigrave
		}
		&
		\makecell[l]{
		Please generate \textit{valid} text input for these\\
		components.\\
		Response format MUST BE:\\
		\textasciigrave\textasciigrave\textasciigrave json \{\\
		~~~~\textit{``et\_nickname'': ``generated\_value''}\\
		\} \textasciigrave\textasciigrave\textasciigrave
		} \\
		\hline

		\rowcolor{lightblue}
		7 & Guiding ($GuiP$)
		& Explain your value selection strategy step-by-step.
		& Explain your value selection strategy step-by-step. \\
		\hline
		\hline

		\multicolumn{2}{c|}{\textbf{Construction Rules}}
		& \multicolumn{2}{p{10.2cm}}{
		(1) \textit{Vision-based prompting}:
		\hlgreen{$GloP$} $+$ \hlgray{$ImgP$} $+$ \hlyellow{$ResP$} $+$ \hlblue{$GuiP$}; \newline
		(2) \textit{XML-based prompting}:
		\hlgreen{$GloP$} $+$ \hlmintfrost{$XML$} $+$ \hlyellow{$ResP$} $+$ \hlblue{$GuiP$}; \newline
		(3) \textit{Extracted-context prompting}:
		\hlgreen{$GloP$} $+$ $n$\hlpink{$ComP$} $+$ $n$\hlpurple{$AdjP$} $+$ \hlyellow{$ResP$} $+$ \hlblue{$GuiP$} ($n \geq 0$).
        } \\
		\hline
	\end{tabular}
\end{table*}

\subsubsection{Prompt Construction}
\label{SEC: Prompt Construction}

After obtaining the contextual information, we construct prompts for the LLMs under three prompting settings:
\textit{extracted-context prompting};
\textit{XML-based prompting}; and 
\textit{vision-based prompting}.
These settings use different representations of the same UI page, enabling the examination of how textual, structural, and visual UI information affect text-input generation.
\begin{itemize}[leftmargin=2em, topsep=0pt, itemsep=0pt, parsep=0pt, partopsep=0pt]
    \item
    The \textit{screenshot-vision prompting} uses the screenshot image of the current UI page as the main UI context.
    The screenshot preserves the rendered layout, visual appearance, and spatial arrangement of UI components.
    It allows multimodal LLMs to observe where the text-input components are placed on the page, what labels or buttons appear nearby, and how these components are visually grouped with other UI elements:
    This may help them to determine the expected content of each input field.

    \item
    The \textit{UI-hierarchy-XML prompting} uses the UI-hierarchy XML obtained by UIAutomator as the main UI context.
    The XML document represents the UI page as structured text, recording the hierarchical relations among UI components and their attributes, such as component classes, resource IDs, texts, hints, and bounds.
    Compared with vision-based prompting, XML-based prompting exposes explicit structural and attribute-level information for the LLMs.

    \item
    Although UI-hierarchy-XML prompting exposes the complete UI hierarchy to the LLMs, the raw XML often contains many irrelevant UI elements and attributes.
    Therefore, directly using the full XML may consume many tokens and introduce unnecessary noise into the prompt.
    To address this issue, inspired by the context-extraction strategy used in \texttt{QTypist}~\cite{LiuCWCHHW23}, we extract only the information that is directly relevant to text-input generation, and organize it into concise natural-language prompts (i.e., the \textit{extracted-context prompting}).
    Specifically, for each text-input component, we construct a \textit{component sub-prompt} ($ComP$) and an \textit{adjacent sub-prompt} ($AdjP$).
    $ComP$ describes the essential attributes of the target component, such as its component type, hint text, and resource ID.
    $AdjP$ describes the surrounding textual labels extracted from the top, bottom, left, and right positions of the component.
    Compared with XML-based prompting, extracted-context prompting reduces the prompt length by filtering out irrelevant UI elements and attributes, while retaining the textual information most useful for inferring the expected content of each input field.
\end{itemize}

To ensure consistency across the three prompting settings, we introduce the following shared prompt elements:
the \textit{global sub-prompt} ($GloP$);
the \textit{restrictive sub-prompt} ($ResP$); and 
the \textit{guiding sub-prompt} ($GuiP$).
$GloP$ describes page-level information, such as the app name, the currently running activity, and the number of text-input components.
$ResP$ specifies the restrictions for text-input generation, ensuring that LLMs produce text inputs in a controlled and predefined format.
$GuiP$ asks the LLMs to generate both text inputs and explanations of why these inputs are generated:
This may help to improve the reasoning quality of the LLM outputs~\cite{abs-2302-10198}.
The final prompt is constructed by combining the shared sub-prompts with one of the three UI context representations.
\begin{itemize}[leftmargin=2em, topsep=0pt, itemsep=0pt, parsep=0pt, partopsep=0pt]
    \item
    For \textit{vision-based prompting}, the final prompt consists of one global sub-prompt, the screenshot image, one restrictive sub-prompt, and one guiding sub-prompt:
    ``$GloP + Image + ResP + GuiP$''.

    \item
    For \textit{XML-based prompting}, the final prompt consists of one global sub-prompt, the UI-hierarchy XML, one restrictive sub-prompt, and one guiding sub-prompt:
    ``$GloP + XML + ResP + GuiP$''.

    \item
    For \textit{extracted-context prompting}, the final prompt consists of one global sub-prompt, $n$ component sub-prompts, $n$ adjacent sub-prompts, one restrictive sub-prompt, and one guiding sub-prompt:
    ``$GloP + nComP + nAdjP + ResP + GuiP~(n \geq 0)$''.
\end{itemize}

The \texttt{[MASK]} fields in the prompt patterns are replaced with the corresponding contextual information during AUT execution.
Table~\ref{TAB: linguistic patterns of prompts} summarizes the sub-prompts, linguistic patterns, and examples.

\begin{figure}
	\centering
	\includegraphics[width=\textwidth]{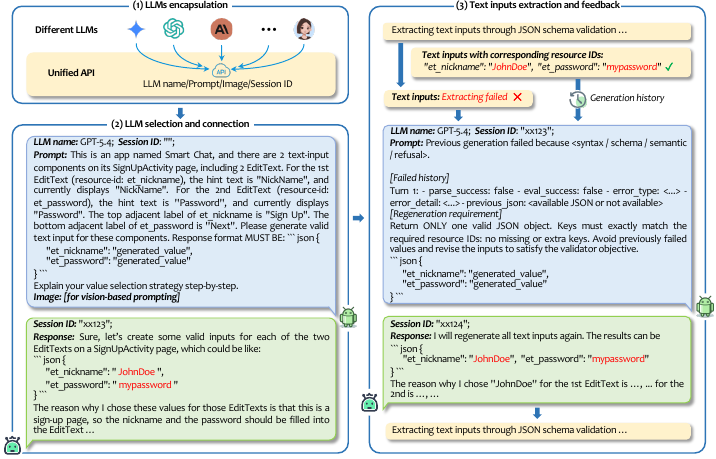}\captionsetup{skip=2pt}
	\caption{The LLM communication process.}
	\label{FIG: The process of LLM communication}
\end{figure}

\subsubsection{LLM Communication}\label{SEC:LLM Communication}

The LLMs generate text inputs by feeding them the final prompt constructed in the previous step.
As shown in Figure~\ref{FIG: The process of LLM communication}, this process comprises three phases:

\begin{itemize}[leftmargin=2em, topsep=0pt, itemsep=0pt, parsep=0pt, partopsep=0pt]
    \item
    \textbf{LLM encapsulation:}
    This phase encapsulates multiple LLMs into a unified API that provides text-input generation capabilities.
    Three parameters are used:
    the \textit{LLM name} (the name of the LLM to be evaluated);
    the final \textit{prompt} text constructed in Section~\ref{SEC: Prompt Construction}; and
    the \textit{session ID} (which is used to retrieve the messages
    ---
    the session ID is blank during the initial API call).

    \item
    \textbf{Model selection and connection:}
    This phase establishes the connection between the testing task and a particular LLM.
    In the example in Figure~\ref{FIG: The process of LLM communication}, GPT-5.4 is selected by calling the encapsulated API with the LLM name ``GPT-5.4''.
    The constructed prompt is also passed to the API.
    Once the LLM generates the result, the API returns the response and the corresponding session ID.

    \item
    \textbf{Text-input extraction and feedback:}
    This phase parses the LLM response to obtain the text inputs for the UI page.
    We use JSON Schema validation to match and extract specific strings from the LLM response, following the format defined by the restrictive sub-prompt ($ResP$).
    Once extracted, the strings are parsed into a sequence of text inputs, each with its corresponding resource ID.
    The extracted text inputs are then used for test-script execution and result evaluation.
    If the response cannot be parsed, or uses incorrect resource IDs, or fails to satisfy the test objective (e.g., does not reveal the target bug), then we construct a feedback prompt: 
    This prompt contains the failure type, the latest error details, the required resource IDs, the required JSON schema, and the recent failed-generation history.
    The LLM is then queried again with the session ID and the feedback prompt to guide the next text-input generation.
\end{itemize}

\subsection{Test-Script Execution
\label{SEC: Test Case Execution and Test Result Evaluation}}

This section illustrates the two main steps for test-script execution, the \textit{test-script development} and \textit{text-input insertion}.

\subsubsection{Test-Script Development}

We develop test scripts using UIAutomator~\cite{UIAutomator} to control the testing process and to help enter text into text-input components.
The test scripts can automatically install and test the AUT on the test device, enabling testers to create specific interaction-based test logic.
A basic test script involves three main components: 
\textit{setup};
\textit{test}; and 
\textit{teardown}.
The setup component initializes the testing environment by installing and launching the AUT and connecting to the LLMs.
The test component defines the primary logic for testing UI actions and LLM communication.
The teardown component resets the testing environment to the original state.
The test scripts provide three types of functions to support the testing process:
\begin{itemize}[leftmargin=2em, topsep=0pt, itemsep=0pt, parsep=0pt, partopsep=0pt]
    \item
    The \textit{UI actions} comprise two types of functions:
    \textit{click} and \textit{swipe}.
    The click functions simulate a click operation on a UI component identified by its displayed text, resource ID, or specific screen coordinates.
    The swipe functions simulate swipe operations from one point to another.

    \item
    There are three basic \textit{logic} functions:
    \textit{findViewByID}; 
    \textit{findViewByText}; and 
    \textit{findViewByXPath}.
    These functions can be used to locate a UI component by its resource ID, text, and XPath attributes.

    \item
    There are two basic \textit{assert} functions:
    \textit{assertAppear}; and 
    \textit{assertDisappear}:
    These functions verify whether or not a UI component appears on the UI page.    
\end{itemize}

\subsubsection{Text-Input Insertion}

We can invoke the logic functions to automatically insert the generated text inputs into the corresponding text-input components.
Specifically, the LLMs are provided with the context of the text inputs with their corresponding resource ID attributes in a JSON format
---
as defined in our restrictive prompt (Section~\ref{SEC: Prompt Construction}).
Thus, a mapping between the text inputs and the text-input components is established.
After obtaining the text inputs from the LLM, UIAutomator locates all the text-input components, and fills them with the corresponding text (based on their resource ID attributes).

\subsection{Test-Result Evaluation
\label{sec: Test Result Evaluation}}

Once the generated text inputs have been inserted into the text-input components, some UI actions are needed to launch a new UI page.
For example, clicking the submit button triggers the login process.

After clicking, we can check whether or not a specific UI component appears, and the test can be evaluated by invoking assertion functions.
Our method currently relies on manual identification of required UI actions and the development of corresponding test scripts.
This follows a user-centric approach, simulating interactions exclusively through GUI-level operations for test-script generation, deliberately avoiding source-code inspection techniques (such as listener exploration) to derive UI components or their associated actions.
Further details about the UI actions used in our experiments are provided in the experimental setting section (Section~\ref{sec:Settings of Effectiveness Evaluation}).
During our experiments, we observed delays in page launches after UI actions were executed.
These delays hindered the immediate determination of test results.
To address this challenge, we developed a waiting mechanism that repeatedly checks if specific conditions are met within a predefined time window.
This mechanism ensures that evaluations occur only after the required conditions are satisfied.
For instance, it can be used to wait for a specific text component to disappear from the current UI page.
A test is considered successful if the assertion passes; 
otherwise, it is considered to have failed.
In the feedback-based generation process, this assertion result is further used as the semantic validation signal for the generated text inputs.
If the generated inputs fail to satisfy the test objective, such as failing to pass through a page or failing to reveal the target bug, the failure information is fed back to the LLM (Section~\ref{SEC:LLM Communication}).
Specifically, as shown in Figure~\ref{FIG: The process of LLM communication}, we construct a feedback prompt that summarizes the failure type, the latest error detail, the required input schema, and recent generation history:
We then query the LLM again under the same session to guide the next generation.
To ensure reliable and consistent test results, we introduced a repetition parameter, $t$, that specifies the expected number of times that the test script is to be executed.
This parameter mitigates randomness and uncertainty during testing, including for:
(1) network-connectivity fluctuations between test scripts and LLM APIs; and
(2) connection instabilities with testing environments (e.g., Android Debug Bridge (ADB) interface reliability during device operations).
Incorporation of these mechanisms enhances the robustness and validity of our testing framework.

\section{Experimental Study
\label{SEC: Experimental Study}}

This section presents the research questions and then introduces the experimental design and settings used to evaluate the text-input generation capabilities of different LLMs.

\subsection{Research Questions}
\label{SEC: Research Questions}

As explained in Section~\ref{sec:objective-identification}, this study aims to evaluate how LLMs generate text inputs for Android GUI testing under different prompting settings and enhancement strategies.
The following four research questions guided our investigation:

\begin{description}[leftmargin=0pt]
    \item[\textbf{RQ1:}]
    \textbf{[UI-Context Prompting Effectiveness]}
    How effective are different UI-context prompting settings for Android text-input generation?
    \begin{itemize}[leftmargin=20pt, labelindent=0pt, itemindent=0pt]
        \item
        \textbf{RQ1.1:} (Page Pass-Through)
        How effective are extracted-context, XML-based, and vision-based prompting settings at generating valid text inputs that enable page pass-through?

        \item
        \textbf{RQ1.2:} (Bug Detection)
        How effective are extracted-context, XML-based, and vision-based prompting settings at generating invalid text inputs that reveal text-input-related bugs?
    \end{itemize}
\end{description}

\begin{description}[leftmargin=0pt]
    \item[\textbf{RQ2:}]
    \textbf{[Feedback Protocol]}
    How does an LLM-based text-input feedback-generation protocol compare with no feedback generation, across different UI-context prompting settings?
    \begin{itemize}[leftmargin=20pt, labelindent=0pt, itemindent=0pt]
        \item
        \textbf{RQ2.1:} (Feedback for Page Pass-Through)
        For page pass-through, how does feedback valid-input generation compare with no-feedback generation?

        \item
        \textbf{RQ2.2:} (Feedback for Bug Detection)
        For bug detection, how does feedback invalid-input generation compare with no-feedback generation?
    \end{itemize}
\end{description}

\begin{description}[leftmargin=0pt]
    \item[\textbf{RQ3:}]
    \textbf{[Human Assessment and Modification]}
    How do human testers evaluate and improve LLM-generated text inputs?
    \begin{itemize}[leftmargin=20pt, labelindent=0pt, itemindent=0pt]
        \item
        \textbf{RQ3.1:} (Human Assessment)
        How do human testers rate LLM-generated text inputs for page pass-through and bug-detection tasks?

        \item
        \textbf{RQ3.2:} (Human Modification)
        Can human modifications further improve generated inputs for page pass-through and bug detection?
    \end{itemize}
\end{description}

\begin{description}[leftmargin=0pt]
    \item[\textbf{RQ4:}]
    \textbf{[Practical Significance]}
    How should LLMs be used to support Android quality assurance?
    \begin{itemize}[leftmargin=20pt, labelindent=0pt, itemindent=0pt]
        \item
        \textbf{RQ4.1:} (Tool Improvement)
        Can LLM-generated text inputs improve traditional automated Android GUI testing tools?

        \item
        \textbf{RQ4.2:} (Testing Insights)
        What insights can be provided for testers using LLMs for Android testing?
    \end{itemize}
\end{description}

\subsection{LLM Selection
\label{subsec:llms-selection}}

This study aimed to evaluate the effectiveness of text inputs generated by different LLMs for mobile GUI testing.
We first consulted the Chatbot Arena+ leaderboard~\cite{chatbot_arena} as the primary screening source\footnote{We used the Chatbot Arena+ snapshot dated April 10, 2026, available at \url{https://openlm.ai/chatbot-arena}.}.
This leaderboard and its underlying evaluation framework have been widely adopted in prior studies to assess or analyze LLM capabilities~\cite{ChaJSS26,XianF0RG25,Krechetova25}.
To keep the experimental cost manageable, we retained the top-50 ranked models from Chatbot Arena+ as the initial candidate pool.
To avoid overrepresenting multiple similar models from the same provider, we further retained at most one representative model family per provider.
In addition, the selected LLMs had to be released within six months of the Chatbot Arena+ snapshot retrieval date (April 10, 2026):
This was because LLMs evolve rapidly, and a long time gap would reduce comparability among selected models.
Following this process, we obtained 13 initial candidate LLMs.

To further examine the competitiveness of these candidates from the user-preference perspective, we additionally checked them against the LLM Arena leaderboard~\cite{arena_text_2026}, a widely used public leaderboard reflecting model competitiveness based on large-scale user voting\footnote{We used the LLM Arena leaderboard snapshot dated April 14, 2026, available at \url{https://arena.ai/leaderboard/text}.}.
All retained candidates in our final selection appeared in the top 100 of the leaderboard~\cite{arena_text_2026}, providing additional cross-benchmark support for the representativeness of the final model set.

\begin{table*}
	\centering
	\scriptsize
    \captionsetup{skip=2pt}
	\caption{Large Language Models used in the Experiments.}
	\label{TAB:llm-baselines}
	\setlength\tabcolsep{1mm}
	\begin{tabular}{c|l|l|c|c|c|c|c|c}
		\hline
		\textbf{No.} & \multicolumn{1}{c|}{\textbf{Model}} & \multicolumn{1}{c|}{\textbf{Version}} & \textbf{Provider} & \textbf{CA+ Rank} & \textbf{LA Rank} & \textbf{Price} & \textbf{Release Date} & \textbf{Source Type} \\ \hline

		1  & \GeminiM   & \texttt{gemini-3.1-pro-preview}   & Google    & 1/50  & 4/339  & \$2 / \$12      & 2026-02-19 & Proprietary \\
		2  & \ClaudeM   & \texttt{claude-opus-4-6-thinking} & Anthropic & 2/50  & 1/339  & \$5 / \$25      & 2026-02-05 & Proprietary \\
		3  & \GrokM     & \texttt{grok-4.20-0309-reasoning} & xAI       & 3/50  & 8/339  & \$5 / \$6       & 2026-03-09 & Proprietary \\
		4  & \GPTM      & \texttt{gpt-5.4-2026-03-05}       & OpenAI    & 4/50  & 7/339  & \$2.50 / \$15   & 2026-03-05 & Proprietary \\

		5  & \DoubaoM   & \texttt{Doubao-Seed-2.0-pro}      & ByteDance & 15/50 & 20/339 & N/A             & 2026-02-16 & Proprietary \\
		6  & \QwenM     & \texttt{qwen3.5-397b-a17b}        & Alibaba   & 27/50 & 36/339 & \$0.39 / \$2.34 & 2026-02-16 & Open-weight \\
		7  & \ERNIEM    & \texttt{ernie-5.0-thinking-latest}& Baidu     & 24/50 & 30/339 & N/A             & 2026-01-15 & Proprietary \\
		8  & \KimiM     & \texttt{kimi-k2.5-thinking}       & Moonshot  & 26/50 & 28/339 & \$0.60 / \$3    & 2026-01-26 & Open-weight \\
		9  & \MistralM  & \texttt{mistral-large-3}          & Mistral   & 41/50 & 73/339 & \$0.50 / \$1.50 & 2025-12-02 & Open-weight \\
		\hline
	    \multicolumn{9}{p{0.98\linewidth}}{\noindent\justifying
        \textbf{Note:}
        \textit{\textbf{CA+ Rank} denotes the model's position in the Chatbot Arena+ ranking.
        \textbf{LA Rank} denotes the corresponding position of the same model or model family in the LM Arena leaderboard snapshot.
        \textbf{Price} denotes the input and output prices per million tokens.}}
    \end{tabular}
\end{table*}

Among the 13 initial candidates, based on model availability and modality support, we excluded four: 
Muse Spark~\cite{meta_muse_spark_2026};
DeepSeek-V3.2~\cite{deepseek_v32_release_2025};
GLM-5.1~\cite{zai_glm51_2026}; and 
MiMo-V2-Pro~\cite{xiaomi_mimo_v2_pro_2026}.
They were excluded for the following reasons:
(1) Although Muse Spark quickly appeared on public leaderboards, Meta stated that it would be offered only in private preview to select partners, and would not be generally available.
Therefore, we could not obtain stable public access for reproducible evaluation, and thus did not include Muse Spark in the final model set.
(2) This study aimed to compare extracted-context, XML-based, and vision-based prompting settings.
We therefore required that each selected model support both textual and visual inputs, within the same model version.
According to this criterion, we excluded DeepSeek-V3.2, GLM-5.1, and MiMo-V2-Pro.
Finally, nine LLMs were retained for evaluation.
Table~\ref{TAB:llm-baselines} summarizes the key information of the selected LLMs, comprising six proprietary models and three open-weight\footnote{Open-weight indicates that the model weights are publicly released under the provider's model license, but does not imply that the training data, training code, or full development process is open-source.} models.
Because many evaluated LLMs are proprietary and their training data, parameter counts, and derivative relationships are not fully observable, we use the LLM set primarily to control for contemporary LLM availability and modality support.
We therefore avoid attributing performance differences to unverified internal model properties, and focus the analysis on prompting settings, feedback protocol, and observable overheads.

\subsection{Experimental Setup
\label{SEC: Experimental Setup}}

We designed a series of experiments to address each research question.

\subsubsection{Settings for the UI-Context Prompting Effectiveness Evaluation (RQ1)}
\label{sec:Settings of Effectiveness Evaluation}

\begin{figure}
    \centering
    \subfigure[Categories of the text inputs under test.
    \label{fig:-the-categories-of-the-text-inputs-under-test}]{
        \includegraphics[width=0.48\columnwidth]{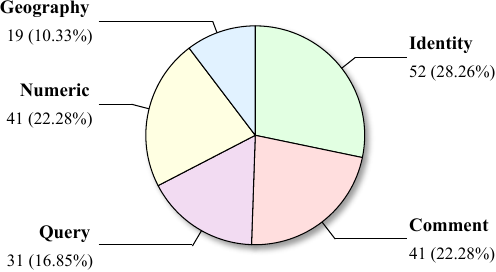}
    }
    \hfill
    \subfigure[Categories of the bugs under test.
    \label{FIG: The categories of the bugs under test}]{
        \includegraphics[width=0.47\columnwidth]{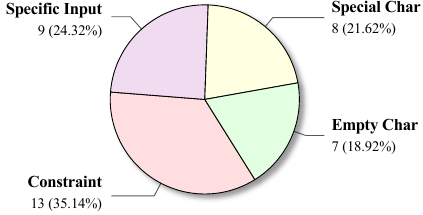}
    }\captionsetup{skip=2pt}
    \caption{Categories of the text inputs and bugs under test.}
    \label{fig:test-categories}
\end{figure}

To answer \textbf{RQ1}, we selected a set of apps for the experiments using the selection process from previous studies~\cite{LiuCWCHHW23}.
For \textbf{RQ1.1}, we developed a collecting script to identify the top 100 apps of each category from Google Play\textsuperscript{\ref{footnote:googleplay}}, the official Android application store, which has also often been used in previous studies~\cite{LiuCWCHHW23}:
Only the most recent apps, with at least one update since April 2025, were kept.
This process yielded 1075 apps across 30 categories.
We then used a filtering script to automatically traverse all the XML layout files in each candidate app.
Because these XML files represent components within a UI page (e.g., ``EditText'' or ``TextView''), the script searches for the keywords ``EditText'', ``AutoCompleteTextView'', and ``MultiAutoCompleteTextView'', and excludes any apps that do not contain them.
These apps were the focus, because they are the primary components for text input.
There are other UI components for text interaction that we did not consider, such as ``TextView'' and ``Spinner'':
This was because ``TextView'' is usually used to display text rather than receive input, and ``Spinners'' only allow users to pick from predefined options.
We then manually launched each candidate app in our experimental environment to verify the text input components.
Finally, 115 apps were selected for the experiments.
We then randomly selected one UI page containing at least one text-input component from each app:
This gave us 115 UI pages, across which we identified 184 text-input components.
Following Liu et al.~\cite{LiuCWCHHW23}, we categorized these 184 text-input components into five types, as shown in Figure~\ref{fig:-the-categories-of-the-text-inputs-under-test}.
The identity-type inputs accounted for the largest proportion (28.26\%), followed by numeric (22.28\%) and comment-type (22.28\%); while the query-type (16.85\%) and geography-type (10.33\%) inputs were the least common.
For each UI page, we evaluated three prompting settings that expose different UI-context representations to the LLMs:
extracted textual context;
XML hierarchy; and 
screenshot-based vision input.
This enabled a direct comparison between textual and visual UI representations, with the same sample set.

We manually identified the UI components responsible for launching a new screen and developed test scripts to simulate user actions that trigger these components.
Table~\ref{ATAB:UIPAGESDETAILS} (in Appendix~\ref{ASEC:Details of UI Pages}) summarizes the different UI actions across the 115 UI pages in the experiments.
Among the 115 UI pages, 60 involved clicking a button with specific text;
22 required triggering the action by clicking the ``Enter'' key; and
the remaining 33 required clicking a button at specific coordinates on the screen.

To answer \textbf{RQ1.1}, each LLM was evaluated on the 115 UI pages under the three prompting settings.
Following the setup used in \texttt{QTypist}~\cite{LiuCWCHHW23}, each LLM was allowed up to three attempts for each model-page-setting combination
---
i.e., the repetition parameter $t$, defined in Section~\ref{sec: Test Result Evaluation}, was set to $3$.
This resulted in 1035 generation tasks per LLM and 9315 tasks across the nine LLMs.

For \textbf{RQ1.2}, we used the 200 most popular applications from F-Droid\textsuperscript{\ref{footnote:fdroid}}, another widely used platform \cite{abs-2310-15657}, which provides direct access to transparent bug reporting and issue tracking for its hosted apps.
We also collected all the corresponding issue reports from GitHub.
Using ``Text'' as the keyword, we filtered the issues related to text inputs.
Finally, 31 buggy apps with 37 issue reports were identified:
All issues were reproduced in the experimental environment.

Figure~\ref{FIG: The categories of the bugs under test} shows the bug-type classifications, and
Table~\ref{TAB:bugs_list} (in Appendix~\ref{ASEC:Details of Bugs}) provides the details (including app names, bug categories, bug descriptions, and links to the issue reports).
The identified bugs were categorized into four types:
\begin{itemize}[leftmargin=2em, topsep=0pt, itemsep=0pt, parsep=0pt, partopsep=0pt]
    \item
    \textit{Constraint} (35.14\%):
    These bugs occur when the input fails to meet specific requirements (such as text length or format).

    \item
    \textit{Specific Input} (24.32\%):
    These bugs are revealed only when specific character sequences are entered, such as ``http''.

    \item
    \textit{Special Char} (21.62\%):
    These bugs are revealed when the input contains certain specific special characters (such as spaces, punctuation, or symbols).

    \item
    \textit{Empty Char} (18.92\%):
    These bugs are revealed by empty inputs.
\end{itemize}

We modified the restrictive prompt (Section~\ref{SEC: Prompt Construction}) by replacing the term ``valid'' with ``invalid'' to directly ask the LLMs to generate invalid text inputs.
To answer \textbf{RQ1.2}, each LLM generated up to 30 invalid text inputs for each of the 37 buggy pages under each prompting setting.
This resulted in 3330 generation tasks per LLM, and 29,970 tasks across the nine LLMs.

\subsubsection{Settings for the Feedback-Protocol Evaluation (RQ2)}
\label{subsubsec:settings-of-rq2}

To answer \textbf{RQ2}, we conducted an independent feedback-enhancement evaluation.
The evaluation used the same 115 UI pages and 37 buggy pages as RQ1, and retained the three prompting settings to support comparisons with no-feedback generation.
For \textbf{RQ2.1}, each model-page-setting combination began with an initial valid-input generation attempt under the corresponding prompting setting.
If extraction failed, or if the generated text inputs failed to pass through the page, then the failure was incorporated into a feedback prompt.
This feedback prompt included the failure type, the latest error details, the expected resource IDs, required JSON schema, and recent generation history.
The process reused the same LLM session across attempts, and stopped once the page-pass-through succeeded, or after a total of three attempts.
For \textbf{RQ2.2}, each buggy-page-setting combination analogously began with an initial invalid-input generation attempt.
After each unsuccessful attempt, the generation continued (with feedback) until a bug was revealed, or until a total of 30 invalid-input attempts were made.

\subsubsection{Settings for Human Assessment and Modification (RQ3)}
\label{subsubsec:settings-of-rq3}
To answer \textbf{RQ3}, we focused on tester assessment, reconstruction, and modification.
For \textbf{RQ3.1}, 20 software testers from well-known Internet enterprises and research institutions were invited to assess the quality of the LLM-generated text inputs, from a practical testing perspective.
We provided each tester with all 115 UI images (with 183 text-input components) and the corresponding text inputs generated by the nine different LLMs.
Following previous studies~\cite{LiuCWCHHW23,Metric-ShaoHWXZ19,oda2015learning}, the testers independently evaluated the quality of text inputs online\footnote{Although the questionnaire was in Chinese, we refer to the English translations in this paper.\label{footnot:web}}.

For \textbf{RQ3.2}, we conducted a controlled experiment with professional software testers to compare LLM-generated, tester-generated, and tester-modified text inputs.
We selected text-input cases from both the page-pass-through task (RQ1.1) and the bug-detection task (RQ1.2):
The selected cases covered both objectives, allowing tester reconstruction and modification to be compared for page-pass-through and bug-detection effectiveness under the same human-in-the-loop protocol.
We invited three professional testers (with over 15, five, and three years of software-testing experience, respectively) to participate in two parallel tasks:
(1) Reconstruct new text inputs without referring to the LLM outputs for the selected cases; and (2) modify the LLM-generated text inputs for the same selected cases.
Each tester received the same cases.
Details of the selected page-pass-through and bug-detection cases are provided in Appendices \ref{ASEC:Details of UI Pages} and \ref{ASEC:Details of Bugs}.

\subsubsection{Settings for Practical Significance Investigation (RQ4)}\label{subsubsec:settings-of-rq4}
For \textbf{RQ4.1}, we integrated LLM into DroidBot~\cite{LiYGC17}, a widely used automated Android testing framework.
This integration evaluated the compatibility of the independent text-input generation approach with complete testing workflows.
We examined two UI exploration strategies implemented in DroidBot:
(1) Depth-First Search (DFS); and
(2) Breadth-First Search (BFS)\textsuperscript{\ref{footnote:dfs_bfs}}.
We randomly selected 30 apps from Google Play\textsuperscript{\ref{footnote:googleplay}} to be the AUTs (Table~\ref{ATAB:tool_improve} in Appendix~\ref{ASEC:Details of Apps}).

For \textbf{RQ4.2}, we combined findings from the other research questions to provide insights for testers who use LLMs for Android GUI testing.

\subsection{Evaluation Metrics
\label{SEC: Evaluation Metrics}}

In this section, we present the evaluation metrics used in our experiments.

\subsubsection{Metrics for the Page-Pass-Through Evaluation}
\label{SEC: Metric for Effectiveness Evaluation}
We used the \textit{page-pass-through rate} (PPTR)\footnote{This metric was originally referred to as the ``\textit{passing rate}''~\cite{LiuCWCHHW23}.}, a metric for GUI testing~\cite{LiuCWCHHW23}, to evaluate the effectiveness of moving the GUI to the next page.
A UI page is considered to be a successful pass-through if the following conditions are met:
(1) a new specific element is detected on the page; and
(2) a previously-displayed element can no longer be detected on the page.
For attempt-level reporting, we use PPTR@$i$ to denote the cumulative proportion of UI pages that are successfully passed through within the first $i$ attempts.
We also use $\Delta@i$ to denote the number of pages newly passed at attempt $i$.
Thus, for page pass-through,
$@1=\Delta@1$,
$@2 = @1 + \Delta@2$, and
$@3 = @2 + \Delta@3$.
These cumulative metrics are reported for $i\in\{1,2,3\}$, matching the three-attempt setting used in the page-pass-through experiments.
The token (including input, output, and total tokens) and latency values are reported as the average cost per generation attempt, so these metrics reflect per-attempt overhead, not the cumulative cost.

\subsubsection{Metrics for Bug Detection}
We used the \textit{bug-detection rate} (BDR)~\cite{abs-2310-15657} to evaluate the bug-detection capabilities of the different LLMs:
The BDR is defined as the proportion of inputs that successfully revealed bugs.
Following previous work \cite{abs-2310-15657}, each LLM was prompted to generate 30 invalid text inputs for a single buggy page:
The number and proportion of bugs successfully revealed were then recorded.
We denote cumulative bug detection within the first $i$ invalid-input attempts as BDR@$i$.
Because bug-revealing may require more trials than page pass-through, we reported BDR@1, BDR@5, BDR@10, and BDR@30.
The corresponding $\Delta@i$ values denote newly revealed bugs between the previous reporting checkpoint and attempt $i$.
BDR@$i$ is task-specific:
it measures whether the generated invalid text inputs reveal the target text-input-related bug, and should not be interpreted as a complete measure of functional correctness or business-logic coverage of the specific app.
We also measure the token and latency values of the average cost per generation attempt.

\subsubsection{Metrics for Testers' Evaluation}
Following previous studies~\cite{LiuCWCHHW23,Metric-ShaoHWXZ19,oda2015learning}, the testers indicated their agreement with the questionnaire statements using a 5-point Likert scale:
strongly disagree (1);
disagree (2);
neutral (3);
agree (4); and
strongly agree (5).
We used Kendall's $W$ value~\cite{LiuCWCHHW23, Schmidt1997Delphi} to measure the agreement among different testers.
\citet{Schmidt1997Delphi} interpreted Kendall's $W$ values as indicating different levels of expert consensus: 
$W<0.3$ suggests weak agreement;
$W\approx0.5$ suggests moderate agreement; and 
$W>0.7$ suggests strong agreement.

\subsubsection{Metrics for Practical Significance Investigation}
We used the number of triggered activities and UI pages~\cite{SuMCWYYPLS17,HeZYCLLYZYZD20} as metrics to evaluate the UI-exploration ability of both the LLM-assisted and the original DroidBot:
These metrics are provided by DroidBot.

\subsection{Statistical Analysis}

We adopted the two-tailed non-parametric \textit{Mann-Whitney U test} (U-test)~\cite{Wilcoxon1945} and the \textit{Vargha and Delaney's effect size} ($\hat{A}_{12}$)~\cite{Vargha2000}, both of which have been widely used to describe statistical differences between compared methods~\cite{Arcuri14,Huang24}.

The U-test uses $p$-values to identify significant differences in the results (at a significance level of $0.05$) of two methods, $\mathcal{M}_{1}$ and $\mathcal{M}_{2}$, with a $p$-value of less than $0.05$ indicating a significant difference~\cite{Huang24,Huang23vpp}.
$\hat{A}_{12}$ compares $\mathcal{M}_{1}$ and $\mathcal{M}_{2}$ as:
\begin{equation}
	\hat{A}_{12}(\mathcal{M}_{1}, \mathcal{M}_{2}) = \frac{R_1/P - (P + 1)/2}{Q},
\end{equation}
where $R_1$ is the rank sum assigned to $\mathcal{M}_{1}$; and
$P$ and $Q$ are the numbers of samples (observations) in $\mathcal{M}_{1}$ and $\mathcal{M}_{2}$, respectively.
In general,
$\hat{A}_{12}(\mathcal{M}_{1}, \mathcal{M}_{2})=0.50$ indicates that $\mathcal{M}_{1}$ and $\mathcal{M}_{2}$ have equal performance;
$\hat{A}_{12}(\mathcal{M}_{1}, \mathcal{M}_{2})>0.50$ indicates that $\mathcal{M}_{1}$ performs better than $\mathcal{M}_{2}$; and
$\hat{A}_{12}(\mathcal{M}_{1}, \mathcal{M}_{2})<0.50$ indicates that $\mathcal{M}_{2}$ performs better than $\mathcal{M}_{1}$.

\subsection{Running Environment}

All experiments were run on emulators under Android 10.
The emulator requires 4GB of RAM.
The test scripts were developed in Python on a MacBook Pro laptop with an M4 Pro chip and 48GB of RAM.

\section{Experimental Results and Discussion
\label{SEC: Experimental Results and Analysis}}

This section presents the experimental results, and answers the research questions from Section~\ref{SEC: Research Questions}.
In the results, for any two methods, $\mathcal{M}_{1}$ and $\mathcal{M}_{2}$, the \ding{109} symbol means that there was no statistical difference between them (the $p$-value was greater than 0.05);
the \ding{52} symbol indicates that $\mathcal{M}_{1}$ was significantly better ($p$-value was less than 0.05, and $\hat{A}_{12}(\mathcal{M}_{1},\mathcal{M}_{2})$ was greater than 0.50); and
the \ding{54} symbol indicates that $\mathcal{M}_{2}$ was significantly better ($p$-value was less than 0.05, and the $\hat{A}_{12}(\mathcal{M}_{1},\mathcal{M}_{2})$ was less than 0.50).

\subsection{Effectiveness Evaluation (RQ1)
\label{SEC:Effectiveness Evaluation (RQ1)}}

This section discusses the effectiveness of the text inputs generated by the different LLMs under different prompt-setting strategies, from the perspective of \textit{page pass-through}, and \textit{bug detection}.

\subsubsection{Page Pass-Through (RQ1.1)}
\label{sec:Page Pass-Through (RQ1.1)}

\begin{table*}
    \centering
    \scriptsize
    \captionsetup{skip=2pt}
    \caption{Average effectiveness and token/time overhead per attempt across prompt settings for PPTR.}
    \label{TAB:result-RQ1pptr-summary}
    \setlength\tabcolsep{3mm}
    
    \begin{tabular}{lrrrrrrr}
        \hline
        \textbf{Setting} & \textbf{@1} & \textbf{@2} & \textbf{@3} & \textbf{Input Token} & \textbf{Output Token} & \textbf{Total Token} & \textbf{Latency (s)} \\
        \hline
        Extracted Context & 67.5\% & 70.5\% & 71.4\% & 249.1 & 1,023.5 & 1,272.7 & 26.9 \\
        UI-hierarchy XML & 67.5\% & 70.5\% & 71.0\% & 7,061.9 & 834.7 & 7,896.6 & 21.6 \\
        Screenshot Vision & 61.4\% & 63.9\% & 65.1\% & 4,829.9 & 834.6 & 5,664.5 & 25.3 \\
        \hline
    \end{tabular}
\end{table*}

\begin{figure}
	\centering
	\includegraphics[width=0.76\textwidth]{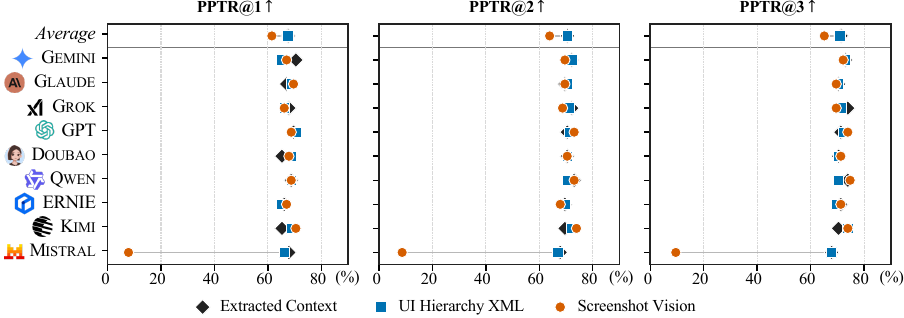}\captionsetup{skip=2pt}
	\caption{Prompt-setting impact on page pass-through effectiveness.}
	\label{fig:page_pass_prompt_setting_effectiveness}
\end{figure}

\begin{figure}
	\centering
	\includegraphics[width=\textwidth]{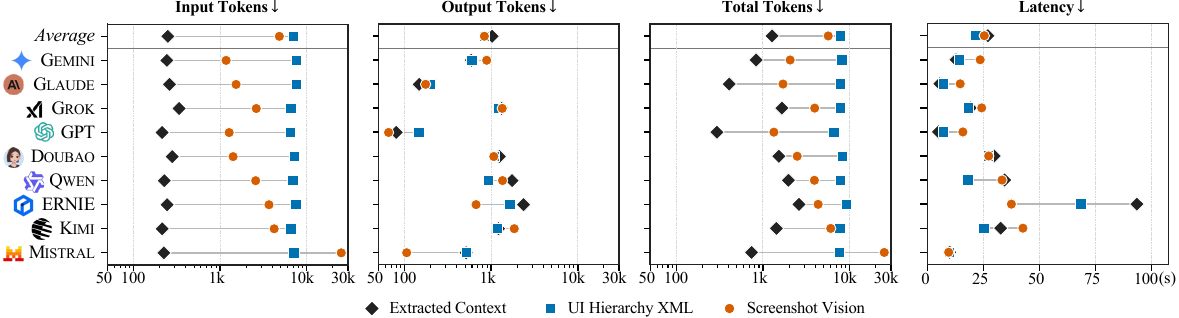}\captionsetup{skip=2pt}
	\caption{Prompt-setting token and time overhead per attempt for page pass-through.}
	\label{fig:page_pass_prompt_setting_overhead}
\end{figure}

Table~\ref{TAB:result-RQ1pptr-summary}, Figure~\ref{fig:page_pass_prompt_setting_effectiveness}, and Figure~\ref{fig:page_pass_prompt_setting_overhead} show the page-pass-through effectiveness and overhead under the three UI-context prompting settings.
Appendix Tables~\ref{TAB:result-RQ1.1} and~\ref{TAB:result-RQ1.1-stat} provide the detailed per-LLM trends and statistical comparisons.
Based on the results, we have the following observations:
\begin{itemize}[leftmargin=2em, topsep=0pt, itemsep=0pt, parsep=0pt, partopsep=0pt]
    \item 
    Extracted-context prompting and UI-hierarchy-XML prompting had very similar cumulative effectiveness, with PPTR@3 values of 71.4\% and 71.0\%, respectively;
    the screenshot-vision prompting achieved 65.1\%.
    The corresponding PPTR@1 values were 67.5\%, 67.5\%, and 61.4\%, indicating that most successful page pass-through cases were already solved in the first attempt, and that the second and third attempts added only limited additional coverage.
    These results do not allow for recommendation of a particular LLM, especially because Appendix Table~\ref{TAB:result-RQ1.1-stat} shows that most LLMs are not statistically distinguishable for a specific prompting setting.

    \item 
    The overhead difference is much larger than the effectiveness difference.
    Extracted-context prompting costs an average of 1272.7 tokens per attempt, whereas UI-hierarchy-XML prompting costs 7896.6 and screenshot-vision prompting costs 5664.5.
    Thus, XML exposes a more complete structure, and screenshot vision adds visual information, but neither provides a proportional PPTR gain in this page-pass-through task.
\end{itemize}

Based on the results among the three settings, extracted-context prompting is identified as the practical default when the token budget is a concern:
It preserves the component, adjacent-label, and global context needed by the task, while avoiding the token cost of full UI serialization or screenshot-based prompting.

To better understand the performance of the different sub-prompts defined in the extracted-context prompting (Section~\ref{SEC: Prompt Construction}), we conducted a prompt-ablation experiment.
Based on the results in Table~\ref{TAB: prompt-ablation}, we had the following key findings (more detailed analysis about the prompt ablation is discussed in Appendix~\ref{asec:prompt-ablation}):
\begin{itemize}[leftmargin=2em, topsep=0pt, itemsep=0pt, parsep=0pt, partopsep=0pt]
    \item 
    Removing the global, component, adjacent, and guiding sub-prompts decreases the average PPTR by 1.4, 2.3, 0.5, and 1.3 percentage points (pp), corresponding to relative decreases of 1.83\%, 3.29\%, 0.61\%, and 1.71\%, respectively.
    
    \item 
    The component sub-prompt has the largest average effect because it contains the target widget type, hint text, displayed text, and resource ID.
    
    \item 
    No single sub-prompt dominates the entire task, so the evidence supports preserving a compact combination of relevant context, rather than expanding the prompt to include all available UI details.
\end{itemize}

We also observed that the effectiveness of page pass-through could be affected by the dependencies and validation constraints among multiple text-input components within the same UI page.
To further investigate this phenomenon, we conducted a multi-component ablation study to examine the impact of page-level constraints. The experimental settings and detailed analysis are provided in Appendix~\ref{ASEC:Multi-Component Ablation Settings}.
Table~\ref{TAB:multi-component-ablation} examines pages containing multiple text-input components using the combinatorial strength $\tau$, where $\tau$ denotes the number of actively filled components. 
Some pages can be successfully passed only when all, or nearly all, of the required fields are completed. 
For example, the PPTR of P045 increases from 0.0\% at $\tau=0$ to 96.3\% at $\tau=5$;
while P059 remains at 0.0\% through $\tau=5$, and only reaches 100.0\% when $\tau=8$. 
These results suggest that many valid-input failures stem from page-level field dependencies, or missing critical fields, rather than from limitations of the LLM itself.

This observation also helps to explain why richer contextual information can be beneficial in principle, yet simply providing additional XML structures or screenshots does not necessarily improve PPTR. 
Unless the generated values satisfy the effective constraints among page components, the page may still reject the inputs. 
A more comprehensive analysis of the multi-component ablation results is provided in Appendix~\ref{ASEC:Multi-Component Ablation Results}.

\begin{tcolorbox}[breakable,colframe=black,arc=1mm,left={1mm},top={0mm},bottom={1mm},right={1mm},boxrule={0.25mm},before upper={\raisebox{-0.25\height}{\includegraphics[scale=0.013]{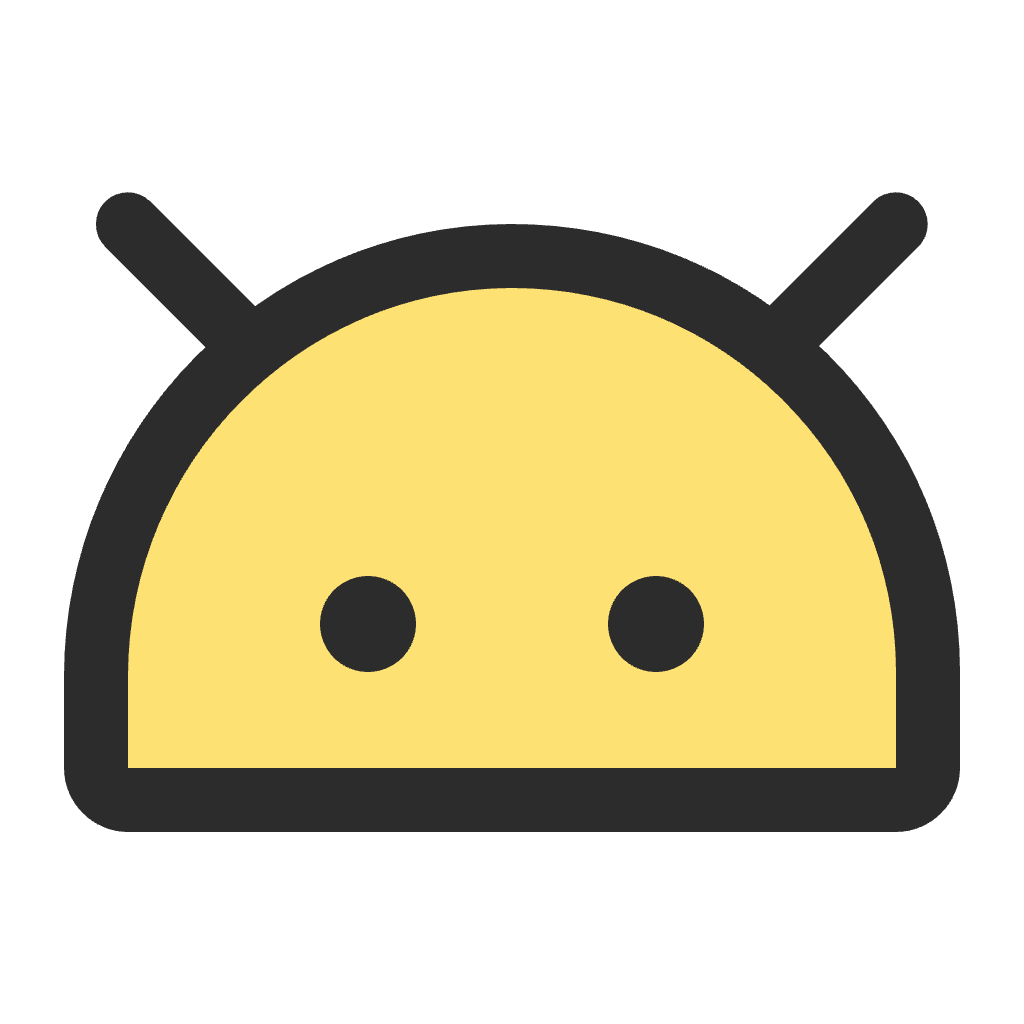}}~}]
    \textit{\textbf{Summary of Answers to RQ1.1:}}
    For page pass-through, the extracted-context and UI-hierarchy-XML prompting achieved similar PPTR values, but the extracted-context prompting was substantially cheaper, in terms of overhead.
    We therefore recommend the extracted-context prompting as the practical default for this setting, not because it makes a particular LLM better, but because it preserves task-relevant context at a much lower cost.
    The prompt-ablation and multi-component ablation analyses showed that component context and multi-field dependencies explain important parts of the observed PPTR behavior.
\end{tcolorbox}

\begin{table*}
    \centering
    \scriptsize
    \captionsetup{skip=2pt}
    \caption{Average effectiveness and token/time overhead per attempt across prompt settings for BDR.}
    \label{TAB:resultbdr-RQ1-summary}
    \setlength\tabcolsep{2.35mm}
   
    \begin{tabular}{lrrrrrrrr}
        \hline
        \textbf{Setting} & \textbf{@1} & \textbf{@5} & \textbf{@10} & \textbf{@30} & \textbf{Input Token} & \textbf{Output Token} & \textbf{Total Token} & \textbf{Latency (s)} \\
        \hline
        Extracted Context & 36.0\% & 42.9\% & 47.1\% & 51.1\% & 320.9 & 1,572.7 & 1,893.7 & 36.7 \\
        UI-hierarchy XML & 35.1\% & 44.7\% & 47.1\% & 51.1\% & 11,505.2 & 1,237.6 & 12,742.7 & 30.4 \\
        Screenshot Vision & 32.7\% & 44.1\% & 46.5\% & 50.8\% & 15,551.9 & 1,504.0 & 17,055.9 & 39.0 \\
        \hline
    \end{tabular}
\end{table*}

\begin{figure}
	\centering
	\includegraphics[width=\textwidth]{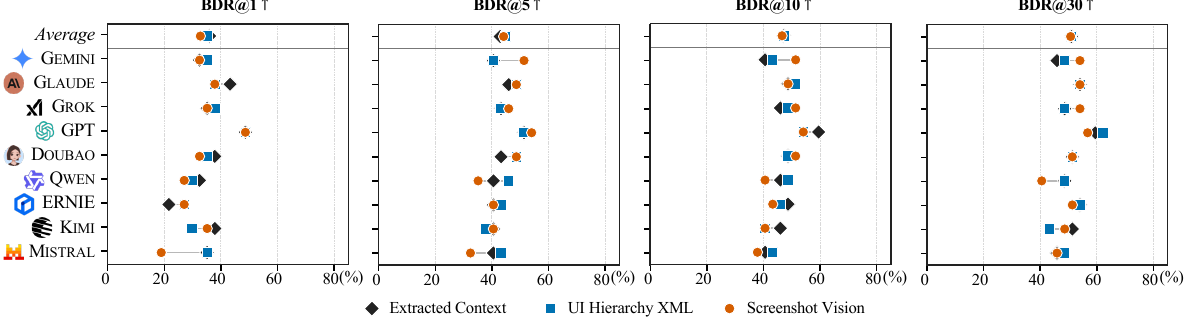}\captionsetup{skip=2pt}
	\caption{Prompt-setting impact on bug-revealing effectiveness.}
	\label{fig:bug_reveal_prompt_setting_effectiveness}
\end{figure}

\begin{figure}
	\centering
	\includegraphics[width=\textwidth]{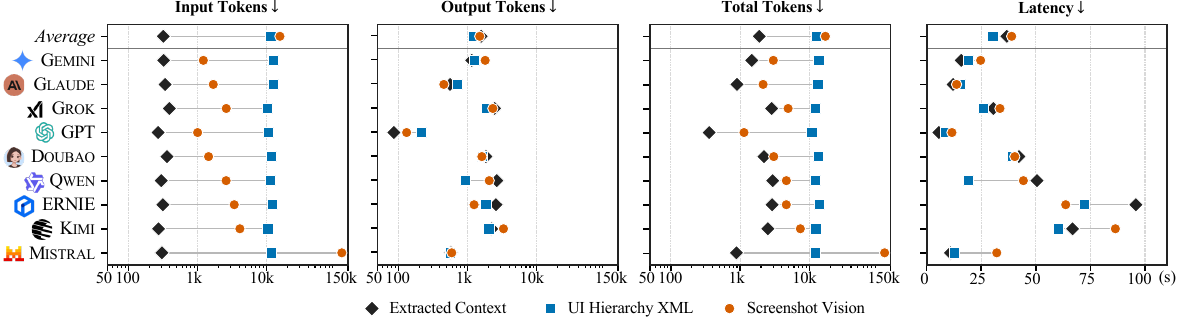}\captionsetup{skip=2pt}
	\caption{Prompt-setting token and time overhead per attempt for bug revealing.}
	\label{fig:bug_reveal_prompt_setting_overhead}
\end{figure}

\subsubsection{Bug Detection (RQ1.2)}
Table~\ref{TAB:resultbdr-RQ1-summary}, Figure~\ref{fig:bug_reveal_prompt_setting_effectiveness}, and Figure~\ref{fig:bug_reveal_prompt_setting_overhead} show the bug-detection effectiveness and overhead.
Appendix Tables~\ref{TAB:result-RQ1.2} and~\ref{TAB:result-RQ1.2-stat} provide the detailed per-LLM trends and statistical comparisons.
Based on the results, we have the following observations:
\begin{itemize}[leftmargin=2em, topsep=0pt, itemsep=0pt, parsep=0pt, partopsep=0pt]
    \item 
    The final bug-detection effectiveness was nearly identical across the three UI-context prompting settings:
    The extracted-context and UI-hierarchy-XML prompting both achieved a BDR@30 of 51.1\%, while the screenshot-vision prompting reached 50.8\%. 
    Similar trends were observed at intermediate cutoffs, indicating a comparable cumulative search process: 
    All three settings increased from detecting roughly one-third of bugs at BDR@1 to approximately one-half at BDR@30.
    
    \item 
    Richer UI context increased the cost more than it improved the final effectiveness.
    The extracted-context prompting cost an average of 1893.7 total tokens per attempt, while the UI-hierarchy-XML and screenshot-vision prompting cost 12,742.7 and 17,055.9, respectively.
    Thus, for the bug detection task, extracted-context prompting was again the cost-effective default among the evaluated settings.
    
    \item 
    Table~\ref{TAB:result-RQ1.2-stat} shows that BDR@30 differences within each setting were (mostly) not statistically distinguishable across LLMs.
    The BDR results, therefore, should not be used to claim that one LLM is generally best for Android bug detection.
    They also show that adding visual input alone is not enough to solve the invalid-input problem:
    The difficult cases usually required a value that reached a specific validation branch, or known bug-triggering pattern, which may not be visible in the UI screenshot.
\end{itemize}

Based on the findings presented above, it is clear that there are challenges to using LLMs to directly generate effective invalid text inputs for bug detection.
We next discuss some possible reasons for this:
\begin{itemize}[leftmargin=2em, topsep=0pt, itemsep=0pt, parsep=0pt, partopsep=0pt]
    \item
    LLMs are typically trained on extensive corpora that focus on common invalid input types, such as empty characters, special symbols, or formatting errors (e.g., incorrect email formats).
    However, these ``common invalid'' types may not be sufficient to reveal the bugs that may be revealed by specific inputs:
    For example, a bug may be revealed by the string ``http:/'', where the app unexpectedly converts part of this valid string into an emoji (e.g., ``http://www.google.com'' is converted into ``\includegraphics[scale=0.028]{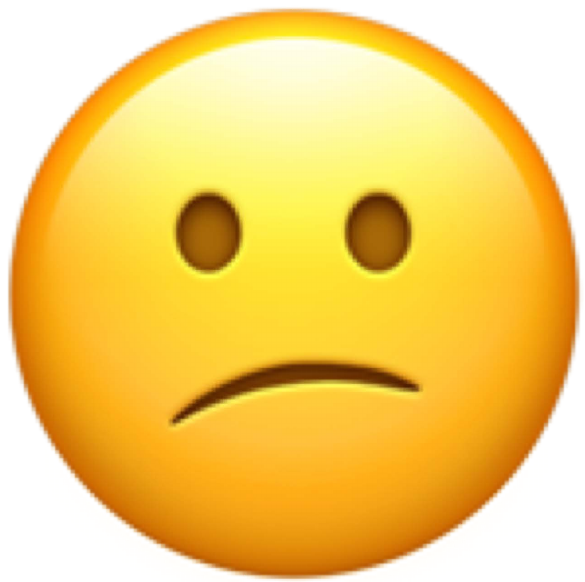}/www.google.com'').
    Therefore, the LLMs cannot depend only on this to effectively identify bugs.
    One potential approach to address this is to fine-tune or retrain LLMs on domain-specific data, which may yield more realistic invalid inputs.
    Furthermore, incorporating text inputs that have previously revealed bugs (in the specific app) may further assist LLMs in generating invalid inputs that more accurately reflect real-world scenarios.

    \item 
    Static generation may not be sufficient for LLMs to effectively learn how to generate bug-revealing inputs.
    In our experiments, the LLMs generated each invalid text input independently, without incorporating any feedback from previously generated texts to inform subsequent generations. 
    However, bug detection is inherently dynamic.
    Therefore, inspiration may be drawn from Adaptive Random Testing (ART)~\cite{HuangARTSurvey}, which dynamically adjusts test-case selection based on feedback to improve bug-detection capability:
    If a specific text input does not reveal a bug, then the next input should be as different as possible, to enhance the likelihood of revealing one.
    Although static prompts (including global, adjacent, and component prompts) can provide helpful UI context, they may not systematically guide LLMs towards the most effective invalid inputs.
    It is necessary to integrate feedback from test-execution results 
    --- 
    such as whether or not a bug was revealed, or the type of errors encountered
    ---
    to improve the bug-revealing ability. 

    \item 
    During the experiments, in some cases, it was observed that the LLMs generated abstract descriptions of test inputs (e.g., ``input with 5000 * symbols'') rather than the actual text. 
    Because these descriptions were meant to represent specific inputs, we did not manually convert them into actual strings of 5000 asterisks. 
    Adjusting the prompts to further guide the LLMs toward producing invalid text inputs would be both feasible and beneficial:
    For instance, a revised prompt such as 
    ``\textit{Generate a concrete string containing the exact characters, not a description}'', could help to ensure that the output aligns with the intended requirements.
\end{itemize}

These results motivated the feedback-enhanced protocol evaluated in RQ2 and the tester-modification analysis in RQ3.2, both of which add signals beyond generic invalid-input generation.

\begin{tcolorbox}[breakable,colframe=black,arc=1mm,left={1mm},top={0mm},bottom={1mm},right={1mm},boxrule={0.25mm},before upper={\raisebox{-0.25\height}{\includegraphics[scale=0.013]{figures/icons/android}}~}]
    \textit{\textbf{Summary of Answers to RQ1.2:}}
    Without making use of execution feedback, the three UI-context settings detected a similar proportion of text-input-related bugs, within 30 attempts.
    The extracted-context prompting had the lowest cost, while the XML and screenshot vision added substantial token overheads, without any clear BDR advantages in this task.
\end{tcolorbox}

\subsection{Feedback Protocol (RQ2)}

This section discusses the effectiveness of the text inputs generated by the different LLMs under the different prompt-setting strategies (and with feedback protocol), in terms of the \textit{page pass-through}, and the \textit{bug detection}.

\subsubsection{Feedback for Page Pass-Through (RQ2.1)}

\begin{table*}
    \centering
    \scriptsize
    \caption{Effectiveness and overhead summary for feedback-enhanced prompting in RQ2.}
    \label{TAB:result-RQ2-summary}
    \setlength\tabcolsep{1.4mm}
    \begin{tabular}{llrrrcc}
        \hline
        \textbf{Task} & \textbf{Metric} &
        \makecell[c]{\textbf{No-Feedback}\\\textbf{($x$)}} &
        \makecell[c]{\textbf{Feedback}\\\textbf{($y$)}} &
        \makecell[c]{\textbf{Absolute Difference}\\\textbf{($y-x$)}} &
        \makecell[c]{\textbf{Relative Increase}\\\textbf{($(y-x)/x$)}} &
        \textbf{Comparison} \\
        \hline

        \multirow{3}{*}{\makecell[l]{RQ2.1\\Page Pass-Through}}
        & PPTR@3$\uparrow$        & 69.18\%  & 73.78\%   & \up 4.60 pp    & \up 6.65\%     & \ding{52} (<0.001, 0.52) \\
        & Total Tokens$\downarrow$ & 4,944.6  & 5,227.6   & \up 283.0      & \up 5.72\%     & \ding{109} (0.580, 0.46) \\
        & Latency (s)$\downarrow$  & 24.6     & 34.1      & \up 9.5        & \up 38.62\%    & \ding{109} (0.189, 0.40) \\
        \hline

        \multirow{3}{*}{\makecell[l]{RQ2.2\\Bug Detection}}
        & BDR@30$\uparrow$        & 50.95\%  & 64.46\%   & \up 13.51 pp   & \up 26.52\%    & \ding{52} (<0.001, 0.57) \\
        & Total Tokens$\downarrow$ & 10,564.1 & 42,042.0  & \up 31,477.9   & \up 297.98\%   & \ding{54} (<0.001, 0.09) \\
        & Latency (s)$\downarrow$  & 35.4     & 506.1     & \up 470.7      & \up 1329.66\%  & \ding{54} (<0.001, 0.05) \\
        \hline
    \end{tabular}
\end{table*}

\begin{table*}
    \centering
    \scriptsize
    \captionsetup{skip=2pt}
    \caption{Feedback (FB) improvements over No-Feedback (No-FB) for RQ2.}
    \label{TAB:result-RQ2-feedback-gain}
    \setlength\tabcolsep{0.75mm}
    \begin{tabular}{lccccc c ccccc}
        \hline
        \multirow{3}{*}{\textbf{LLM}} & \multicolumn{5}{c}{\textbf{RQ2.1 Page Pass-Through}} & & \multicolumn{5}{c}{\textbf{RQ2.2 Bug Detection}} \\
        \cline{2-6}\cline{8-12}
        & \makecell[c]{\textbf{No-FB}\\\textbf{($x$)}} 
        & \makecell[c]{\textbf{FB}\\\textbf{($y$)}} 
        & \makecell[c]{\textbf{Gain}\\\textbf{($y-x$)}} 
        & \makecell[c]{\textbf{Improvement}\\\textbf{($(y-x)/x$)}} 
        & \textbf{Comparison} 
        & 
        & \makecell[c]{\textbf{No-FB}\\\textbf{($x$)}} 
        & \makecell[c]{\textbf{FB}\\\textbf{($y$)}} 
        & \makecell[c]{\textbf{Gain}\\\textbf{($y-x$)}} 
        & \makecell[c]{\textbf{Improvement}\\\textbf{($(y-x)/x$)}} 
        & \textbf{Comparison} \\ \hline

        \GeminiM & 72.75\% & 74.49\% & \up 1.74 pp & \up 2.39\% & \ding{109} (0.605, 0.51) && 49.55\% & 66.67\% & \up 17.12 pp & \up 34.55\% & \ding{52} (0.010, 0.59) \\
        \ClaudeM & 70.14\% & 75.36\% & \up 5.22 pp & \up 7.44\% & \ding{109} (0.124, 0.53) && 54.05\% & 71.17\% & \up 17.12 pp & \up 31.67\% & \ding{52} (0.009, 0.59) \\
        \GrokM & 71.59\% & 72.17\% & \up 0.58 pp & \up 0.81\% & \ding{109} (0.866, 0.50) && 50.45\% & 60.36\% & \up 9.91 pp & \up 19.64\% & \ding{109} (0.139, 0.55) \\
        \GPTM & 72.46\% & 75.36\% & \up 2.90 pp & \up 4.00\% & \ding{109} (0.386, 0.51) && 59.46\% & 65.77\% & \up 6.31 pp & \up 10.61\% & \ding{109} (0.333, 0.53) \\
        \DoubaoM & 70.72\% & 74.20\% & \up 3.48 pp & \up 4.92\% & \ding{109} (0.307, 0.52) && 51.35\% & 71.17\% & \up 19.82 pp & \up 38.60\% & \ding{52} (0.003, 0.60) \\
        \QwenM & 73.04\% & 75.07\% & \up 2.03 pp & \up 2.78\% & \ding{109} (0.544, 0.51) && 45.95\% & 64.86\% & \up 18.91 pp & \up 41.15\% & \ding{52} (0.005, 0.59) \\
        \ERNIEM & 70.72\% & 73.04\% & \up 2.32 pp & \up 3.28\% & \ding{109} (0.499, 0.51) && 53.15\% & 63.06\% & \up 9.91 pp & \up 18.65\% & \ding{109} (0.136, 0.55) \\
        \KimiM & 72.75\% & 75.07\% & \up 2.32 pp & \up 3.19\% & \ding{109} (0.488, 0.51) && 47.75\% & 71.17\% & \up 23.42 pp & \up 49.05\% & \ding{52} (<0.001, 0.62) \\
        \MistralM & 48.41\% & 69.28\% & \up 20.87 pp & \up 43.11\% & \ding{52} (<0.001, 0.60) && 46.85\% & 53.13\% & \up 6.28 pp & \up 13.40\% & \ding{109} (0.394, 0.53) \\

        \textbf{Avg.} & 69.18\% & 73.78\% & \up 4.60 pp & \up 6.65\% & \ding{52} (<0.001, 0.52) && 50.95\% & 64.46\% & \up 13.51 pp & \up 26.52\% & \ding{52} (<0.001, 0.57) \\\hline
    \end{tabular}
\end{table*}

\begin{figure}
	\centering
	\includegraphics[width=0.76\textwidth]{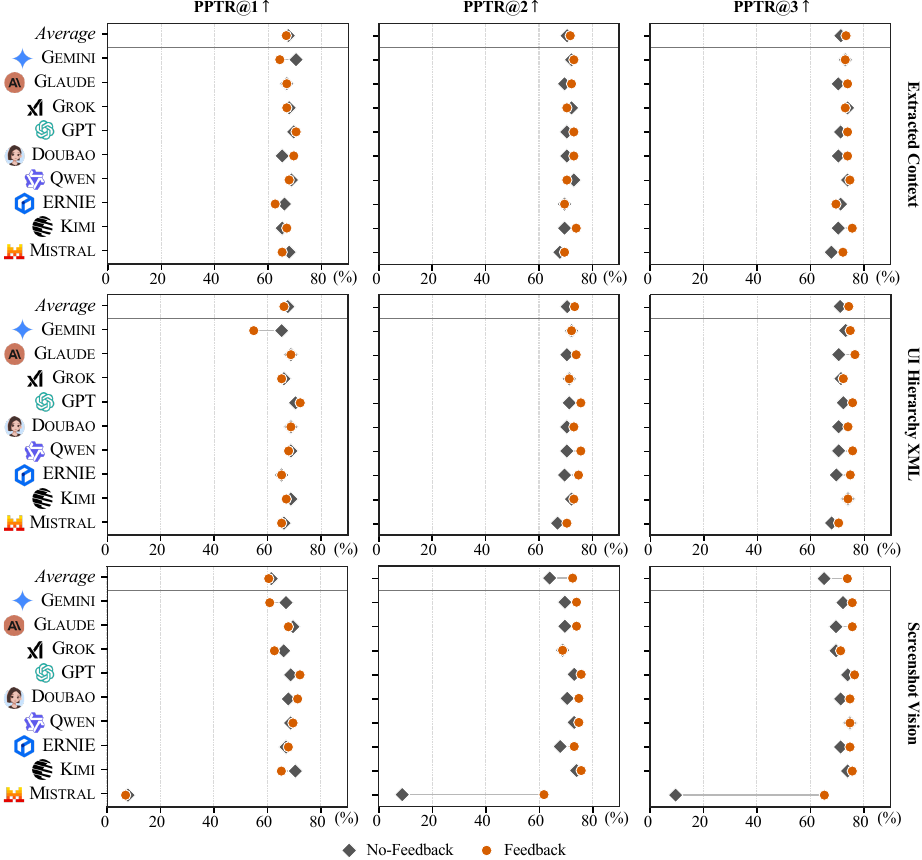}\captionsetup{skip=2pt}
	\caption{Effectiveness comparison between no-feedback and feedback-enhanced prompting, for PPTR.}
	\label{fig:page_pass_no_feedback_feedback_effectiveness}
\end{figure}

Table~\ref{TAB:result-RQ2-summary} and Figure~\ref{fig:page_pass_no_feedback_feedback_effectiveness} present the main page-pass-through results for the feedback protocol.
More detailed per-LLM results and statistical comparisons are shown in Appendix Tables~\ref{TAB:result-RQ2.1} and~\ref{TAB:result-RQ2.1-stat}; 
and the settings-level effectiveness and overhead details are shown in Appendix Figures~\ref{fig:page_pass_feedback_prompt_setting_effectiveness},~\ref{fig:page_pass_feedback_prompt_setting_overhead}, and~\ref{fig:page_pass_no_feedback_feedback_overhead}.
Based on the results, we have the following observations:
\begin{itemize}[leftmargin=2em, topsep=0pt, itemsep=0pt, parsep=0pt, partopsep=0pt]
    \item
    Compared with no-feedback generation, the average PPTR@3 across LLMs increased from 69.18\% to 73.78\%, a 4.61 pp gain and a 6.65\% relative improvement.
    The average comparison was statistically significant ($p<0.001$, $\hat{A}_{12}=0.52$), but the effect size was small, which means feedback is useful for page pass-through, but should not be overstated.

    \item 
    The improvement was not uniformly large.
    In Table~\ref{TAB:result-RQ2-feedback-gain}, \Mistral\ showed the largest PPTR gain, increasing from 48.41\% to 69.28\% (a 20.87 pp gain and a 43.11\% relative improvement), and was the only individual model with a statistically significant page-pass-through improvement.
    For the other models, feedback improved the PPTR@3 by between 0.58 pp and 5.22 pp (corresponding to between 0.81\% and 7.44\% relative improvement), but was statistically indistinguishable from the no-feedback counterpart.

    \item 
    Feedback improved the page pass-through mainly by repairing failed attempts, but the repair opportunity was limited because many pages had already been solved in the first attempt.
    In the feedback-enhanced runs, 2001 of the 2291 successful model-setting-page combinations passed on the first attempt; 
    250 more passed on the second attempt; and 
    the remaining 40 passed on the third attempt.
    Thus, only 290 (12.7\%) of the successful combinations were recovered after at least one failed attempt.
    This explains why the net PPTR@3 gain is positive, but moderate.

    \item 
    Feedback increased the per-attempt cost because the later prompts included both the failure information and the recent attempt history.
    Table~\ref{TAB:result-RQ2-summary} shows that the average total number of tokens per attempt increased from 4944.6 to 5227.6, and the average latency increased from 24.6s to 34.1s.
    These overhead differences were not statistically significant
    (Token: $p=0.580$, $\hat{A}_{12}=0.46$; Latency: $p=0.189$, $\hat{A}_{12}=0.40$, lower-is-better).
    The setting-level details in Appendix Table~\ref{TAB:result-RQ2.1} and Appendix Figure~\ref{fig:page_pass_feedback_prompt_setting_overhead} show that the extracted-context prompting remained the lowest-cost setting, even after adding feedback.
\end{itemize}

A concrete role of the feedback is to convert a failed output into constraints for the next input generation.
As explained in Sections~\ref{SEC:LLM Communication} and~\ref{subsubsec:settings-of-rq2}, the feedback prompt reports whether the previous turn failed because of JSON syntax, incorrect resource IDs, or a semantic page-pass-through failure;
it also includes the required resource IDs, the required JSON schema, and recent failed parsed JSON values.
For the page-pass-through cases, this can help the LLMs to repair two common failures:
(1) missing or mismatched keys for pages with multiple required fields; and 
(2) input values that appear reasonable, but violate page-specific validation rules.
For example, the validation rules for P040 and P042 reject common placeholder-like search or email values, such as generic library names or domains containing \texttt{example}/\texttt{test}:
When such values appear in the failed history, the next prompt explicitly asks the model not to repeat previous failed values.
Therefore, the feedback protocol does not provide additional UI context; instead, 
it narrows the next generation attempt by discouraging the repetition of error patterns and failed values observed in previous attempts.

\begin{tcolorbox}[breakable,colframe=black,arc=1mm,left={1mm},top={0mm},bottom={1mm},right={1mm},boxrule={0.25mm},before upper={\raisebox{-0.25\height}{\includegraphics[scale=0.013]{figures/icons/android}}~}]
    \textit{\textbf{Summary of Answers to RQ2.1:}}
    Execution feedback improves the average page-pass-through effectiveness, but the gain is moderate and model-dependent.
    Feedback is best used for difficult page-pass-through pages, where failed attempts reveal constraints that can guide subsequent input generation.
\end{tcolorbox}

\subsubsection{Feedback for Bug Detection (RQ2.2)}

\begin{figure}
	\centering
	\includegraphics[width=\textwidth]{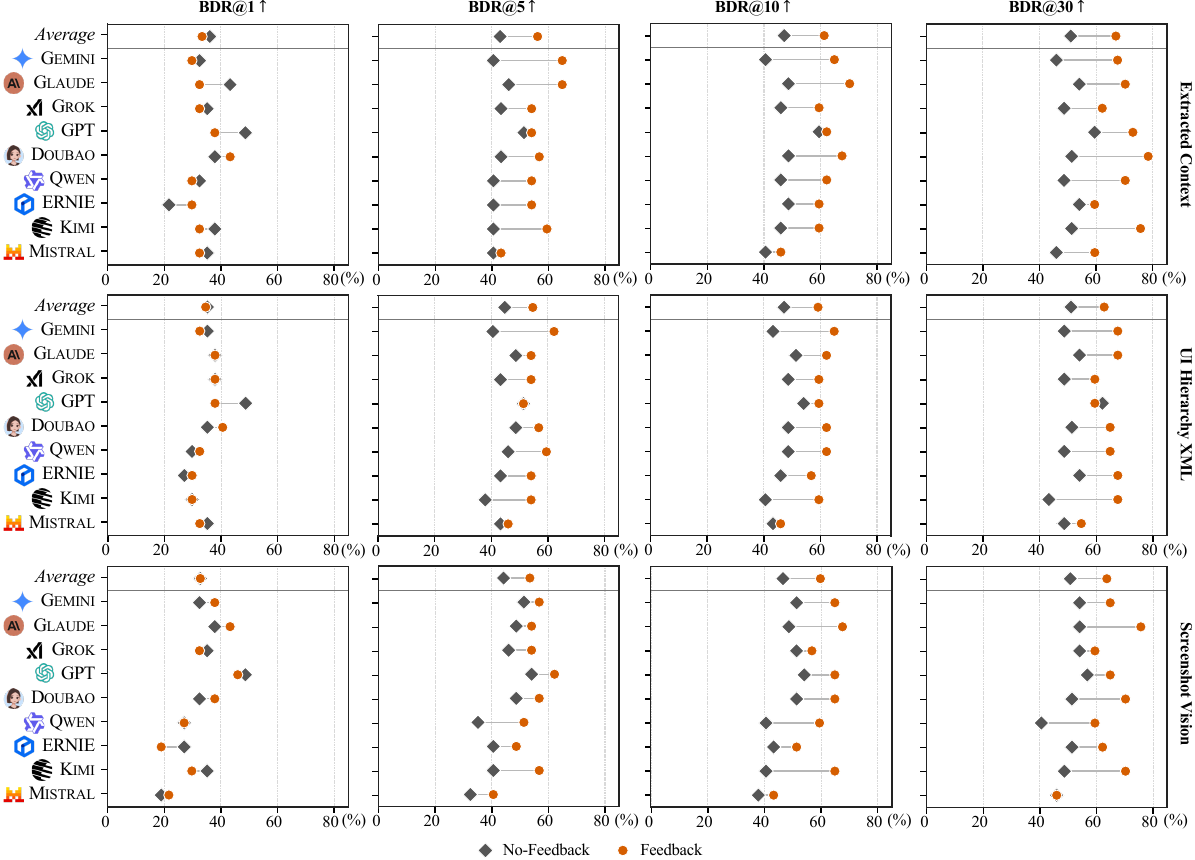}\captionsetup{skip=2pt}
	\caption{Effectiveness comparison between no-feedback and feedback-enhanced prompting, for bug revealing.}
	\label{fig:bug_reveal_no_feedback_feedback_effectiveness}
\end{figure}

Table~\ref{TAB:result-RQ2-summary} and Figure~\ref{fig:bug_reveal_no_feedback_feedback_effectiveness} present the main bug-detection results for the feedback protocol.
More detailed per-model results and statistical comparisons are shown in Appendix Tables~\ref{TAB:result-RQ2.2} and~\ref{TAB:result-RQ2.2-stat}; 
settings-level effectiveness and overhead details are shown in Appendix Figures~\ref{fig:bug_reveal_feedback_prompt_setting_effectiveness},~\ref{fig:bug_reveal_feedback_prompt_setting_overhead}, and~\ref{fig:bug_reveal_no_feedback_feedback_overhead}.
Based on the results, we have the following observations:
\begin{itemize}[leftmargin=2em, topsep=0pt, itemsep=0pt, parsep=0pt, partopsep=0pt]
    \item
    The feedback has a larger aggregate effect on bug detection than on page pass-through.
    Tables \ref{TAB:result-RQ2-summary} and \ref{TAB:result-RQ2-feedback-gain} show that the average BDR@30 increased from 50.95\% without feedback to 64.46\% with feedback, a 13.51 pp gain (a 26.52\% relative improvement).
    The average comparison was statistically significant ($p<0.001$, $\hat{A}_{12}=0.57$).

    \item 
    After feedback, the extracted-context prompting had the highest average BDR@30 value (68.5\%), while the UI-hierarchy-XML and screenshot-vision prompting both reached 63.7\% (Appendix Table~\ref{TAB:result-RQ2.2}).
    The detailed trends in Appendix Table~\ref{TAB:result-RQ2.2} show that much of this improvement occurs by BDR@5 and BDR@10, which suggests that early failure information helps the subsequent invalid-input attempts become more targeted.

    \item 
    The improvement was larger for bug detection because the feedback prompt includes the invalid inputs that failed to trigger the bug in previous attempts, and asks the LLM to generate different invalid inputs in the next attempt.   
    In the feedback-enhanced runs, 335 successful model-setting-bug combinations detected the bug on the first attempt;
    212 additional combinations were detected by attempt 5;
    53 more by attempt 10; and 
    44 more by attempt 30.
    Overall, only 309 of the 644 successful feedback-enhanced combinations (48.0\%) were detected after at least one failed invalid input.

    \item 
    The model-level gains in Table~\ref{TAB:result-RQ2-feedback-gain} were not uniform.
    Five models showed statistically significant BDR@30 improvements:
    \Gemini\ (17.12 pp; 34.55\%);
    \Claude\ (17.12 pp; 31.67\%);
    \Doubao\ (19.82 pp; 38.60\%);
    \Qwen\ (18.91 pp; 41.15\%); and
    \Kimi\ (23.42 pp; 49.05\%).
    The remaining models also relatively improved by between 10.61\% and 19.64\%, but these improvements were not statistically significant.

    \item 
    The gains in effectiveness came with high costs.
    For the bug detection (Table~\ref{TAB:result-RQ2-summary}), the average total number of tokens per attempt increased from 10,564.1 to 42,042.0, and the average latency increased from 35.4s to 506.1s.
    The token and latency increases were statistically significant 
    (Token: $p<0.001$, $\hat{A}_{12}=0.09$; Latency: $p<0.001$, $\hat{A}_{12}=0.05$, lower-is-better).
    Table~\ref{TAB:result-RQ2.2} reports the settings-level feedback overhead: 
    31,840.6, 45,090.1, and 49,195.3 total tokens per attempt for extracted context, XML, and screenshot vision, respectively.
    The average latency was around 500 seconds per attempt.
    Therefore, the feedback-enhanced invalid-input generation was identified as more appropriate when the bug-detection effectiveness is more important than the token or latency budgets.
\end{itemize}

The feedback protocol was especially useful for invalid text-input generation, because many text-input-related bugs were triggered by a narrow range of values, rather than simply by any invalid strings.
The feedback prompt does not reveal the exact triggering value: 
Instead, it reports that the previous candidate failed, preserves the required resource IDs and JSON schema, and lists recent failed parsed JSON values.
This helps the LLMs to avoid simply repeating generic invalid values and encourages them to try other types of invalid inputs, such as empty strings, whitespace-like values, or unusually long strings.
For example, Appendix Table~\ref{TAB:bugs_list} lists several recovered cases whose trigger conditions are highly specific:
Bug 17 in SASAbus is triggered when either the ``From'' or ``To'' text-input component is left empty;
Bug 30 in Health Log requires the ``Enter Hospital ID'' text-input component to be empty;
Bug 34 in Activity Diary is associated with empty character input; and
Bug 5 in NewPipe requires a search query longer than 100 characters.
A no-feedback prompt that asks for an invalid value may generate non-empty malformed text, security-like payloads, or other generic negative examples:
The feedback turns those failed candidates into explicit negative evidence, and pushes later attempts toward untried, field-specific boundaries such as empty strings, whitespace-like values, or long strings.

The feedback can also repair failures caused by output-format issues. 
For example, some models generate descriptions of invalid inputs rather than concrete text values:
Once such outputs are recorded in the failed history, subsequent prompts can guide the model to produce executable inputs. 
Nevertheless, the benefit of feedback is limited for highly app-specific bugs. 
Although failed candidates serve as useful negative evidence, they often do not reveal enough information about the application’s hidden semantic constraints or bug-triggering conditions.

\begin{tcolorbox}[breakable,colframe=black,arc=1mm,left={1mm},top={0mm},bottom={1mm},right={1mm},boxrule={0.25mm},before upper={\raisebox{-0.25\height}{\includegraphics[scale=0.013]{figures/icons/android}}~}]
    \textit{\textbf{Summary of Answers to RQ2.2:}}
    Feedback substantially improved the average BDR@30, especially for several models, with statistically significant gains.
    The improvement occurs because failed invalid inputs become negative examples that guide later attempts toward untried field-specific boundary values.
    However, the protocol is expensive, and should only be used when the bug-detection objective justifies the additional token and latency cost.
\end{tcolorbox}

\subsection{Human Assessment and Modification (RQ3)}

\begin{figure*}
    \centering

    \begin{minipage}{0.48\textwidth}
        \centering
        \includegraphics[width=\linewidth]{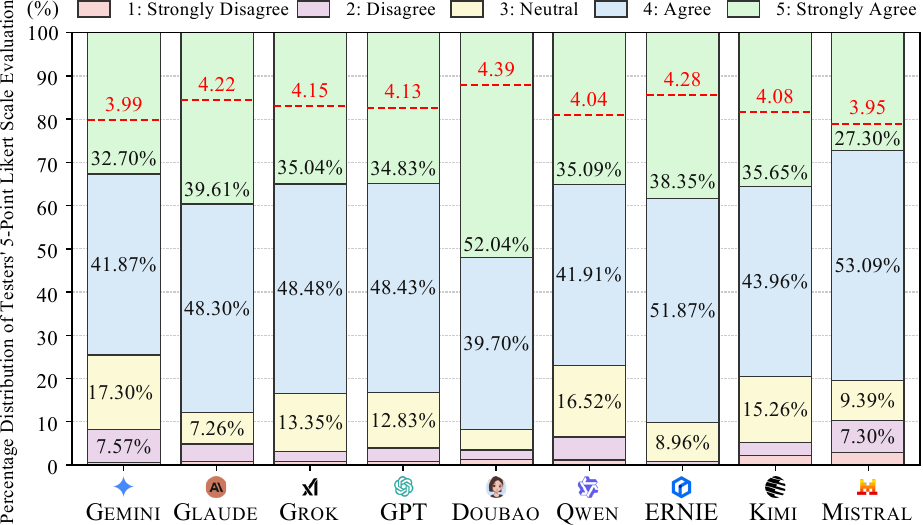}

        {\footnotesize
        (a) Page Pass-Through --- Extracted Context\\[-1pt]
        (Mean = 4.14;\ Agree/Strongly Agree = 83.1\%;\ W = 0.85)}
    \end{minipage}
    \hfill
    \begin{minipage}{0.48\textwidth}
        \centering
        \includegraphics[width=\linewidth]{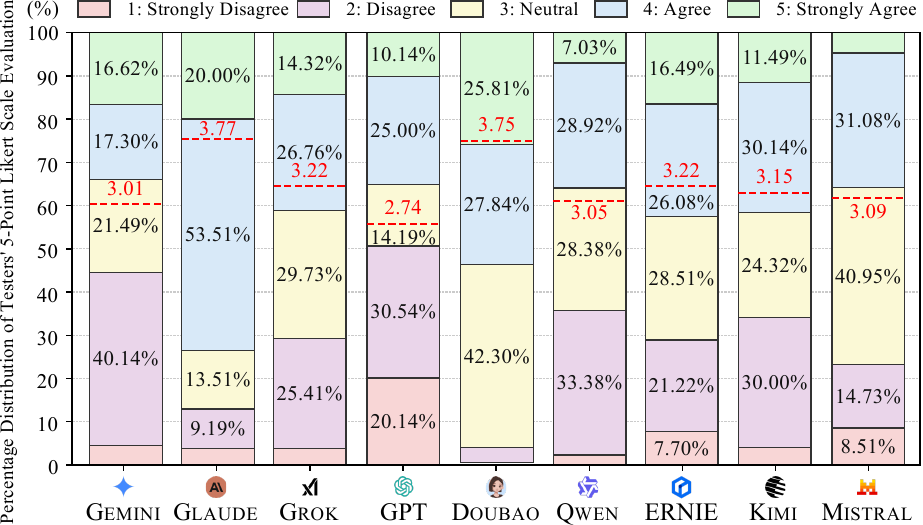}

        {\footnotesize
        (b) Bug Detection --- Extracted Context\\[-1pt]
        (Mean = 3.22;\ Agree/Strongly Agree = 43.7\%;\ W = 0.79)}
    \end{minipage}

    \vspace{1mm}

    \begin{minipage}{0.48\textwidth}
        \centering
        \includegraphics[width=\linewidth]{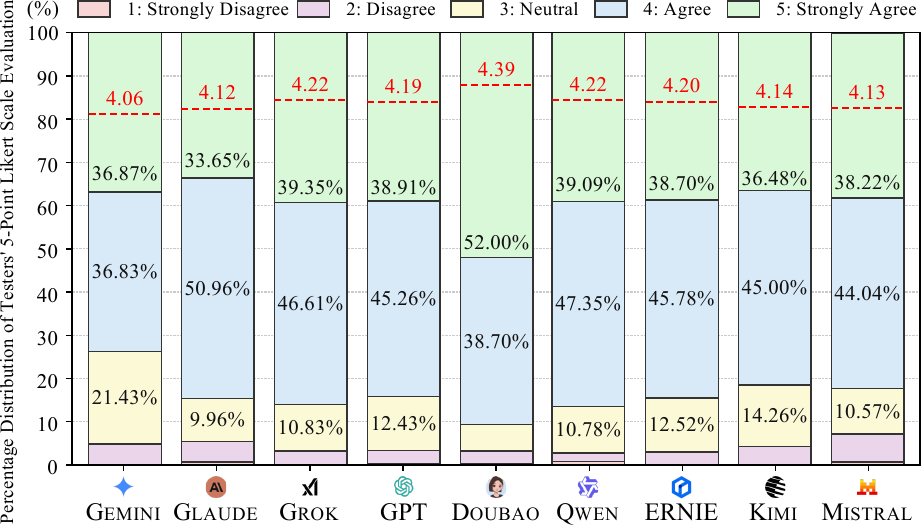}

        {\footnotesize
        (c) Page Pass-Through --- UI-Hierarchy XML\\[-1pt]
        (Mean = 4.19;\ Agree/Strongly Agree = 83.8\%;\ W = 0.76)}
    \end{minipage}
    \hfill
    \begin{minipage}{0.48\textwidth}
        \centering
        \includegraphics[width=\linewidth]{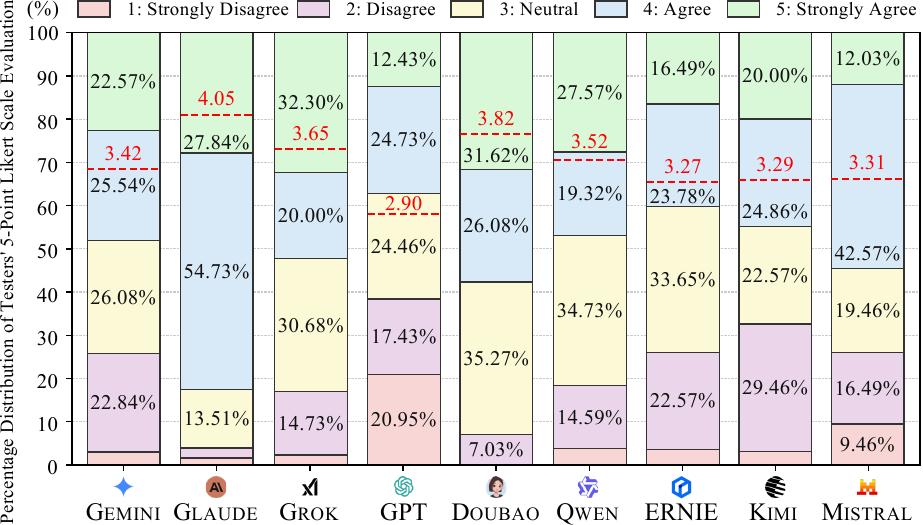}

        {\footnotesize
        (d) Bug Detection --- UI-Hierarchy XML\\[-1pt]
        (Mean = 3.47;\ Agree/Strongly Agree = 51.6\%;\ W = 0.87)}
    \end{minipage}

    \vspace{1mm}

    \begin{minipage}{0.48\textwidth}
        \centering
        \includegraphics[width=\linewidth]{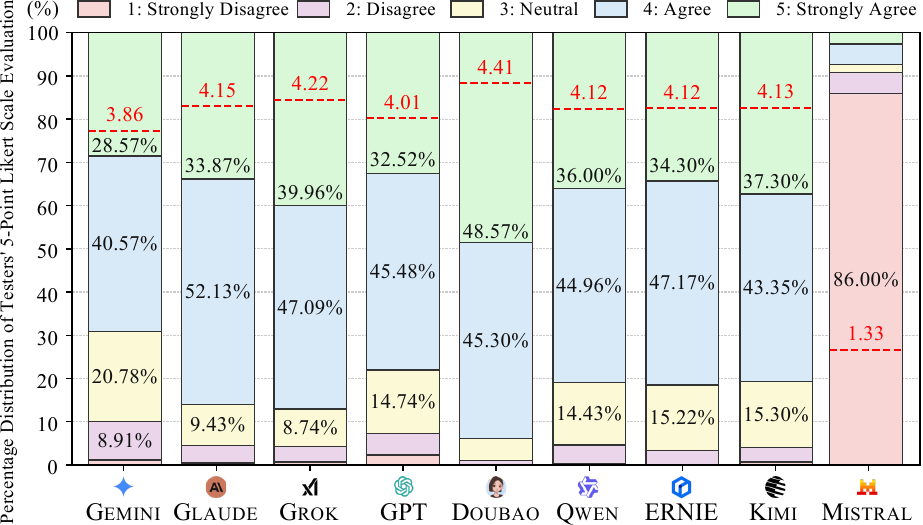}

        {\footnotesize
        (e) Page Pass-Through --- Screenshot Vision\\[-1pt]
        (Mean = 3.82;\ Agree/Strongly Agree = 73.8\%;\ W = 0.85)}
    \end{minipage}
    \hfill
    \begin{minipage}{0.48\textwidth}
        \centering
        \includegraphics[width=\linewidth]{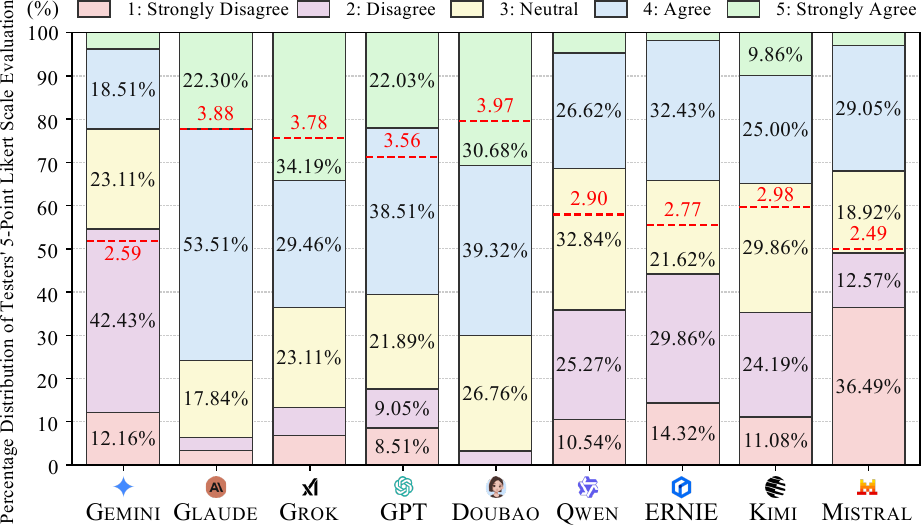}

        {\footnotesize
        (f) Bug Detection --- Screenshot Vision\\[-1pt]
        (Mean = 3.22;\ Agree/Strongly Agree = 47.2\%;\ W = 0.94)}
    \end{minipage}

    \captionsetup{skip=1pt}
    \caption{
    Percentage distribution of testers' 5-point Likert-scale evaluations.
    The red bold markers and dashed lines indicate the mean scores across LLMs.
    }
    \label{FIG:boxplot_ratings}
    \vspace{-1cm}
\end{figure*}

\subsubsection{Human Assessment (RQ3.1)}
\label{sec:Human Assessment (RQ3.1)}
To elicit the testers' assessments of the LLM-generated texts, they were prompted with the question:
\textit{Do you think the following text inputs are suitable for the UI pages? Please assess their quality.}\textsuperscript{\ref{footnot:web}}
Figure~\ref{FIG:boxplot_ratings} shows the testers' Likert-scale assessments across the three UI-context prompting settings and two testing objectives.
Based on these results, we have the following observations:

\begin{itemize}[leftmargin=2em, topsep=0pt, itemsep=0pt, parsep=0pt, partopsep=0pt]
    \item
    The panel-level Kendall's W values~\cite{LiuCWCHHW23, Schmidt1997Delphi} in Figure~\ref{FIG:boxplot_ratings} range from 0.76 to 0.94.
    This indicates substantial agreement among the 20 testers for both page-pass-through and bug-detection assessments.
    The assessment data contain 82,080 score records:
    62,100 records for valid page-pass-through inputs, and 19,980 records for invalid bug-detection inputs.

    \item
    For valid page pass-through, testers generally rated the generated values positively.
    The mean scores were:
    4.14 for extracted context;
    4.19 for UI-hierarchy XML; and 
    3.82 for screenshot vision.
    The proportions of high ratings (scores 4--5) were 83.1\%, 83.8\%, and 73.8\%, respectively.
    This means that testers often found the valid inputs plausible for the visible UI. 
    However, screenshot-vision prompting received a lower mean score and a higher proportion of low ratings than the other two settings, suggesting that screenshots alone did not consistently provide enough textual or structural cues for generating plausible valid inputs.

    \item
    The mean scores were:
    3.22 for extracted context;
    3.47 for UI-hierarchy XML; and 
    3.22 for screenshot vision.
    The high-rating proportions were 43.7\%, 51.6\%, and 47.2\%, respectively (much lower than for the valid-input panels).
    This difference was expected, because an invalid-input set must do more than look malformed:
    It must provide diverse and field-specific values that could reach the bug-relevant validation or execution path.
\end{itemize}

The need for tester assessment comes from a limitation of automatic checks:
A value can be parseable and type-consistent, but may still not be suitable for practical testing.
For instance, a syntactically valid email address can still be a placeholder-like value, and a malformed, invalid string can still miss the target bug branch.
Therefore, we need to evaluate whether or not testers can reconstruct or modify sampled inputs to improve downstream effectiveness:
This motivated the sampled reconstruction and modification study in RQ3.2.

\begin{tcolorbox}[breakable,colframe=black,arc=1mm,left={1mm},top={0mm},bottom={1mm},right={1mm},boxrule={0.25mm},before upper={\raisebox{-0.25\height}{\includegraphics[scale=0.013]{figures/icons/android}}~}]
    \textit{\textbf{Summary of Answers to RQ3.1:}}
    Testers showed substantial agreement when assessing the generated inputs.
    The valid page-pass-through inputs received higher ratings than the invalid bug-detection inputs:
    This was because the valid inputs only really needed to be contextually plausible, whereas the invalid sets needed diversity and bug-triggering potential.
\end{tcolorbox}

\subsubsection{Human Modification (RQ3.2)}
\label{SEC: Testers Evaluation}

\begin{table*}
    \centering
    \scriptsize
    \captionsetup{skip=2pt}
    \caption{Comparison of effectiveness between LLM-generated, tester-generated, and tester-modified text inputs for RQ3.2.}
    \label{TAB:result-RQ3.2-tester-modifiy}
    \setlength\tabcolsep{0.9mm}
    \begin{tabular}{clcccc c cccc}
        \hline
        \multicolumn{2}{c}{\multirow{2}{*}{\textbf{Method}}} &
        \multicolumn{4}{c}{\textbf{Page Pass-Through}} & &
        \multicolumn{4}{c}{\textbf{Bug Detection}} \\
        \cline{3-6}\cline{8-11}
        \multicolumn{2}{c}{} 
        & \makecell[c]{\textbf{PPTR}\\\textbf{($y$)}} 
        & \makecell[c]{\textbf{Gain}\\\textbf{($y-x$)}} 
        & \makecell[c]{\textbf{Improvement}\\\textbf{$(y-x)/x$}} 
        & \textbf{Comparison} 
        & 
        & \makecell[c]{\textbf{BDR}\\\textbf{($y$)}} 
        & \makecell[c]{\textbf{Gain}\\\textbf{($y-x$)}} 
        & \makecell[c]{\textbf{Improvement}\\\textbf{$(y-x)/x$}} 
        & \textbf{Comparison} \\
        \hline

        \multicolumn{2}{c}{LLM-generated ($x$)} 
        & 59.75\% & -- & -- & -- 
        & 
        & 36.10\% & -- & -- & -- \\
        \hline

        \multirow{4}{*}{Tester-generated}
        & \textit{Tester 1} & 90.00\% & \up 30.25 pp & \up 50.63\% & \ding{52} (0.006, 0.65) & & 100.00\% & \up 63.90 pp & \up 177.01\% & \ding{52} (<0.001, 0.82) \\
        & \textit{Tester 2} & 90.00\% & \up 30.25 pp & \up 50.63\% & \ding{52} (0.006, 0.65) & & 95.00\% & \up 58.90 pp & \up 163.16\% & \ding{52} (<0.001, 0.79) \\
        & \textit{Tester 3} & 90.00\% & \up 30.25 pp & \up 50.63\% & \ding{52} (0.006, 0.65) & & 85.00\% & \up 48.90 pp & \up 135.46\% & \ding{52} (<0.001, 0.74) \\
        \cline{2-6}\cline{8-11}
        & \textit{Average} & 90.00\% & \up 30.25 pp & \up 50.63\% & \ding{52} (0.006, 0.65) & & 93.33\% & \up 57.23 pp & \up 158.53\% & \ding{52} (<0.001, 0.79) \\
        \hline

        \multirow{4}{*}{Tester-modified}
        & \textit{Tester 1} & 90.00\% & \up 30.25 pp & \up 50.63\% & \ding{52} (0.006, 0.65) & & 100.00\% & \up 63.90 pp & \up 177.01\% & \ding{52} (<0.001, 0.82) \\
        & \textit{Tester 2} & 90.00\% & \up 30.25 pp & \up 50.63\% & \ding{52} (0.006, 0.65) & & 100.00\% & \up 63.90 pp & \up 177.01\% & \ding{52} (<0.001, 0.82) \\
        & \textit{Tester 3} & 90.00\% & \up 30.25 pp & \up 50.63\% & \ding{52} (0.006, 0.65) & & 100.00\% & \up 63.90 pp & \up 177.01\% & \ding{52} (<0.001, 0.82) \\
        \cline{2-6}\cline{8-11}
        & \textit{Average} & 90.00\% & \up 30.25 pp & \up 50.63\% & \ding{52} (0.006, 0.65) & & 100.00\% & \up 63.90 pp & \up 177.01\% & \ding{52} (<0.001, 0.82) \\
        \hline
    \end{tabular}
\end{table*}

\begin{table*}
    \centering
    \scriptsize
    \captionsetup{skip=2pt}
    \caption{Representative tester-modification examples for RQ3.2.}
    \label{TAB:result-RQ3.2-modification-examples}
    \setlength\tabcolsep{1mm}
    \begin{tabular}{ccp{0.18\linewidth}p{0.18\linewidth}p{0.18\linewidth}p{0.18\linewidth}}
        \hline
        \textbf{Objective} & \textbf{Case} & \textbf{Problem pattern in source LLM outputs} & \textbf{Representative tester modification} & \textbf{Why this improved the sampled case} & \textbf{Boundary} \\
        \hline
        Page pass-through & P014 & Multi-character filter values were too restrictive for a page with starts-with, ends-with, and contains fields. & Use short compatible values, e.g., \texttt{A}, \texttt{a}, and \texttt{an}. & The combined filters still leave a non-empty name intersection, whereas longer prefixes can eliminate all candidates. & Requires knowledge that the page performs conjunctive filtering. \\\hline
        Page pass-through & P031 & Frequently generated search terms such as ``Vintage'' failed even though they looked plausible. & Select \texttt{Bokeh}, the term with broader successful execution evidence across source candidates. & The tester used execution history to select an app-indexed effect term rather than the most frequently generated term. & Works when source history exposes a reliable catalogue term. \\\hline
        Page pass-through & P035 & Alphanumeric identifier variants failed for the GGID field. & Preserve the visible numeric value \texttt{123456}. & The value matches the numeric-only identifier format and the prefilled UI value, suggesting a seeded test credential. & Does not generalize when no visible or known test credential exists. \\\hline
        Bug detection & B002 & Generic invalid amounts, such as alphabetic strings, did not reach the locale-sensitive number parser. & Use \texttt{1 000}, a value above 999 with a space grouping separator. & The value directly matches the locale-specific trigger described for the amount-parsing bug. & Requires the locale or grouping-trigger condition to be known. \\\hline
        Bug detection & B003 & Generic special-character, SQL, XSS, empty, or long strings missed the stated trigger. & Use \texttt{0120345678}, a realistic phone number starting with 0. & The bug description identifies a leading-zero phone number as the trigger, so the tester used that specific input type. & Applies when the bug description exposes a concrete trigger. \\\hline
        Bug detection & B017 & Many candidates used non-empty special-character values for an Empty Char bug. & Set the target field to a strict empty string \texttt{""} and keep the other field non-empty. & The edit isolates the empty-field trigger while preventing early rejection caused by leaving multiple fields empty. & Whitespace and empty strings may behave differently across apps. \\
        \hline
    \end{tabular}
\end{table*}

We invited three experienced Android testers to reconstruct and modify some sampled LLM-generated text inputs.
All three testers had industry experience:
Tester 1 had over 15 years;
Tester 2 had over five years; and 
Tester 3 had over three years.
For valid page pass-through, we sampled 20 UI pages from RQ1.1; and 
for the invalid bug detection, we sampled 20 text-input-related bug cases from RQ1.2.
In the regeneration mode, the testers produced a new input from the UI and task context.
In the modification mode, they also inspected the source candidate pool and its execution outcomes before producing a repaired input.
Table~\ref{TAB:result-RQ3.2-tester-modifiy} shows the effectiveness comparison between LLM-generated inputs and tester-generated or tester-modified inputs.
Table~\ref{TAB:result-RQ3.2-modification-examples} lists some representative case-level modification mechanisms.

\begin{itemize}[leftmargin=2em, topsep=0pt, itemsep=0pt, parsep=0pt, partopsep=0pt]
    \item
    For the sampled page-pass-through cases, the LLM-generated baseline reached 59.75\% PPTR, while both tester-generated and tester-modified inputs reached an average of 90.00\%.
    The improvement was statistically significant (Table~\ref{TAB:result-RQ3.2-tester-modifiy}: $p=0.006$, and $\hat{A}_{12}=0.65$).
    The source baseline was low because it averaged all the source candidates in the sampled pool (generic, overly restrictive, wrong-format, and parse-failing values).
    The testers improved the valid inputs by selecting or constructing values that matched the actual page semantics:
    For example, using short topic keywords for search fields;
    preserving known-good identifiers or prefilled values;
    avoiding fictitious phone-number patterns; and 
    filling multi-field forms with mutually compatible values.

    \item
    For the sampled bug-detection cases, the LLM-generated baseline reached a BDR of 36.10\%, while the tester-generated inputs reached an average of 93.33\%, and the tester-modified inputs reached 100.00\%.
    This improvement was larger than in the valid-input task, because many source invalid inputs were only generically invalid.
    The testers, instead, targeted the bug mechanism:
    They used exact trigger strings from bug descriptions;
    locale-sensitive numeric formats;
    strict empty strings or whitespace-only values for Empty Char bugs;
    malformed query syntax for search-parser bugs; and 
    valid non-target field values to avoid early rejection before reaching the buggy branch.

    \item
    The improvements were large because the testers used information that the original LLM generation did not fully exploit.
    For valid cases, the tester modifications were often selection and normalization:
    Choosing the source value that already matched execution history;
    replacing long phrases with short app-indexed keywords;
    normalizing identifiers to numeric-only or single-string formats; and 
    keeping enough fields filled to satisfy the page-level constraints.
    For invalid cases, tester modifications were more aimed at matching known bug-triggering conditions:
    They converted generic invalid strings into bug-specific inputs, such as the locale-formatted amount ``1 000'' for a space-grouping numeric bug;
    a leading-zero phone number for a Specific Input bug; or 
    an empty target field paired with non-empty non-target fields for an Empty Char bug.
    Table~\ref{TAB:result-RQ3.2-modification-examples} shows some concrete examples.

    \item
    These mechanisms are not universally guaranteed improvements, but they are broadly reusable as editing heuristics when testers can observe enough task-specific clues:
    Such clues may come from UI hints and visible values on the page, resource IDs in the UI hierarchy, or previous execution outcomes.
    However, the exact effect depends on app-specific validation logic, and the available execution signal.
    In the sampled page-pass-through task, two pages, P049 and P104, still failed for all three testers after regeneration and modification, leaving the PPTR at 90.00\% (18/20 pages):
    For P049, the testers supplied validly formatted phone numbers, but tapping the \texttt{Next} button did not trigger a page transition; the app stayed on the same phone-number page, so the case was still counted as a page-pass-through failure.
    For P104, the \texttt{com.yelp.android.biz} page contains two name text-input components that share the same resource ID. 
    Although the testers generated values for both fields, the test script could not reliably map the two values back to the corresponding UI components during input insertion, so the case was still counted as a page-pass-through failure.
    These cases show that tester reasoning cannot always overcome app-specific execution conditions or ambiguous UI-resource information.
    Therefore, RQ3.2 should be read as evidence that tester expertise can repair representative LLM-generated inputs in a controlled sample, not as a claim that the same percentage-point gains will generalize to every app.
\end{itemize}

For page pass-through, the LLMs generated inputs based only on the information available in the prompt. 
However, some pages contained additional requirements that were not fully visible from the UI (such as specific field constraints, dependencies between multiple fields, or predefined values expected by the page). 
As a result, the LLMs generated values that appeared reasonable, but still failed to pass through the page.
For bug detection, the prompt asked the LLMs to generate invalid inputs, but many bugs were not triggered by arbitrary invalid values:
Instead, they required a specific invalid-input form, such as leaving a required field empty, using an over-length search query, entering a locale-formatted number, or using a leading-zero phone number.
Consequently, a generally malformed input may not have been effective for revealing the target bug.
In contrast, the testers were given two additional sources of information: 
The task objective (or bug description); and 
the history of previously-generated candidates (and their outcomes). 
This information helped the testers to identify more suitable inputs for the target page or bug, and thus avoid repeatedly generating ineffective values. 
Table~\ref{TAB:result-RQ3.2-modification-examples} summarizes some representative examples of how the testers modified the generated valid and invalid inputs.

\begin{tcolorbox}[breakable,colframe=black,arc=1mm,left={1mm},top={0mm},bottom={1mm},right={1mm},boxrule={0.25mm},before upper={\raisebox{-0.25\height}{\includegraphics[scale=0.013]{figures/icons/android}}~}]
    \textit{\textbf{Summary of Answers to RQ3.2:}}
    In the sampled controlled comparison, tester-generated and tester-modified inputs substantially improved both the page-pass-through and the bug-detection effectiveness.
    The gains arose because the testers could leverage task objectives, bug descriptions, and execution outcomes to refine the generated inputs.
    Nevertheless, the observed improvements are not guaranteed to generalize to all applications, due to hidden validation logic and app-specific behaviors.
\end{tcolorbox}

\subsection{Practical Significance (RQ4)}

This section presents the results of integrating the LLM-generated text inputs into DroidBot \cite{LiYGC17}.
We also provide some insights and advice for using LLMs for Android testing.

\begin{figure}
    \centering
    \includegraphics[width=0.6\textwidth]{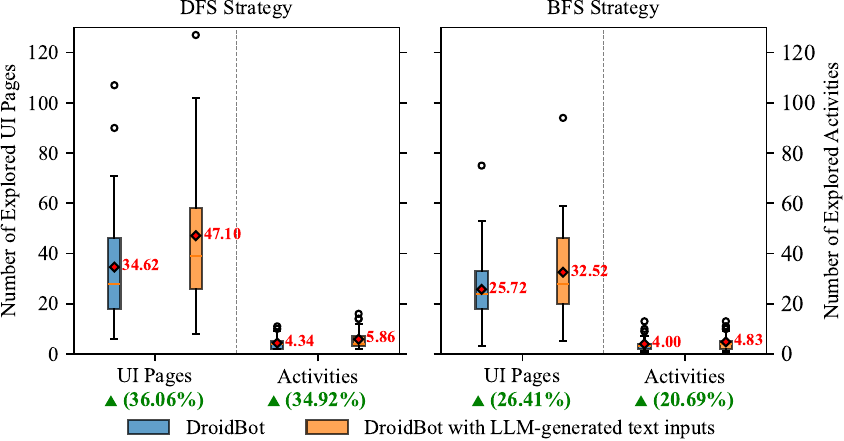}\captionsetup{skip=2pt}
    \caption{Comparison of UI-page and activity exploration capabilities for DroidBot and for DroidBot with LLM-generated text inputs, for both DFS and BFS strategies.
    (The \includegraphics[scale=0.23]{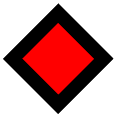} symbol represents the average numbers; and
    the \up~symbols indicate the average percentage relative increase of DroidBot with the LLM-generated text inputs over the original DroidBot.)}
    \label{FIG:droidbot-inmprovement}
\end{figure}

\subsubsection{Tool Improvement (RQ4.1)}\label{subsubsec: tools improvement}

Figure~\ref{FIG:droidbot-inmprovement} presents a comparison of the UI-page and activity-exploration capabilities of DroidBot and DroidBot with LLM-generated text inputs, for both the DFS and the BFS strategies.
Based on these results, we have the following observations:

\begin{itemize}[leftmargin=2em, topsep=0pt, itemsep=0pt, parsep=0pt, partopsep=0pt]
    \item
    For both the DFS and the BFS strategies, DroidBot integrated with LLM-generated text inputs (LLM-integrated DroidBot) was able to explore significantly more UI pages than the original DroidBot:
    With the DFS strategy, the LLM-integrated DroidBot explored an average of 47.10 UI pages, a 36.06\% relative increase over the original DroidBot's 34.62 UI pages.
    With the BFS strategy, the LLM-integrated DroidBot explored an average of 32.52 UI pages, a 26.41\% relative increase over the original 25.72 UI pages.

    \item
    The LLM-integrated DroidBot also improved the number of activities explored, for both DFS and BFS strategies:
    With DFS, the LLM-integrated DroidBot explored an average of 5.86 activities, up from 4.34, representing a 34.92\% relative improvement over the original DroidBot.
    With BFS, the LLM-integrated DroidBot explored an average number that increased from 4.00 to 4.83 activities, representing a 20.69\% relative improvement over the original version.

    \item
    For both UI-page and activity explorations, the LLM-related relative improvement was greater for DFS (36.06\% for UI page, and 34.92\% for activity) than for BFS (26.41\% for UI page, and 20.69\% for activity).
\end{itemize}

The results show that integrating LLM-generated text inputs into DroidBot can significantly enhance exploration capabilities for Android app UI pages and activities, thereby improving test coverage.
The key reason for the improvement lies in overcoming the limitations of the original DroidBot:
When encountering text-input components, the original DroidBot typically fills them with predefined strings (e.g., ``Hello World'') without regard for the specific context of the UI pages.
However, many UI transitions depend on inputting texts that meet specific semantic or format requirements (e.g., a valid email address):
Entering invalid or irrelevant text often fails to trigger UI transitions, preventing exploration of the UI.
LLMs can analyze information from the currently-displayed UI page to understand the required semantics and format.
Based on this understanding, these LLM-generated texts are more likely to satisfy the app's logic, and to trigger valid page transitions:
This enables the testing tool to more effectively explore new UI pages and activities that require valid input to be reached.

\begin{tcolorbox}[breakable,colframe=black,arc=1mm,left={1mm},top={0mm},bottom={1mm},right={1mm},boxrule={0.25mm},before upper={\raisebox{-0.25\height}{\includegraphics[scale=0.013]{figures/icons/android}}~}]
    \textit{\textbf{Summary of Answers to RQ4.1:}}
    LLM-generated context-aware text inputs can effectively enhance UI-page exploration when testing Android apps.
    They are particularly well-suited for deep or complex UI pages that rely on specific text inputs.
    By addressing the original DroidBot's shortcomings in text-input generation, LLM-generated text inputs can significantly improve the ability to explore more UI states.
\end{tcolorbox}

\subsubsection{Testing Insights (RQ4.2)}
\label{SEC:Insights}

Based on the observations and conclusions above, we present the following insights into the use of LLMs for testing Android apps.
\begin{itemize}[leftmargin=2em, topsep=0pt, itemsep=0pt, parsep=0pt, partopsep=0pt]
    \item
    \textit{\textbf{Insight 1: Select the UI-context representation by effectiveness-cost tradeoff.}}
    RQ1 shows that extracted-context and XML prompting achieve similar PPTR values, but that the extracted context has much lower token overheads.
    The screenshot vision can expose visual cues, but in our evaluated page-pass-through and bug-detection tasks, it was not automatically better than textual UI-context settings.

    \item
    \textbf{\textit{Insight 2: Use feedback when the testing objective justifies the overhead.}}
    RQ2 shows that feedback improves both page-pass-through and bug-detection effectiveness, with a larger effect for bug detection.
    The same results also show high token and latency overheads, so feedback is most suitable for difficult pages or bug-detection tasks where failed attempts provide useful diagnostic information.

    \item
    \textbf{\textit{Insight 3: Avoid over-interpreting small model-level differences.}}
    The detailed statistical tables show that many pairwise LLM differences were not statistically distinguishable for the core effectiveness metrics.
    Model choice should therefore consider availability, modality support, latency, and cost, in addition to measured effectiveness on the target task.

    \item
    \textbf{\textit{Insight 4: Combine LLM generation with human or algorithmic refinement for difficult cases.}}
    RQ3.2 shows that professional testers can substantially improve sampled LLM-generated inputs by replacing placeholders, normalizing formats, and repairing off-target values.
    This supports a hybrid workflow in which LLMs provide candidate inputs and human testers (or testing algorithms) refine them when app-specific constraints matter.

    \item
    \textbf{\textit{Insight 5: Treat component criticality as a diagnostic signal.}}
    The prompt-ablation and multi-component analyses (in the Appendices) show that component context and filled-component combinations affect PPTR.
    Criticality should be used to diagnose which fields may block progress, but it should not be treated as a universal rule that one component alone always determines the page outcome.

    \item
    \textbf{\textit{Insight 6: Match the protocol and metric to the testing objective.}}
    PPTR measures whether or not generated values move a page forward, whereas BDR measures whether or not invalid-input attempts detect known bugs within a specific budget.
    Neither metric alone measures full functional correctness, security, or business-logic coverage:
    App-specific invalid-input generation may require additional targeted mutation rules or app-specific checks~\cite{abs-2310-15657}.
\end{itemize}

\begin{tcolorbox}[breakable,colframe=black,arc=1mm,left={1mm},top={0mm},bottom={1mm},right={1mm},boxrule={0.25mm},before upper={\raisebox{-0.25\height}{\includegraphics[scale=0.013]{figures/icons/android}}~}]
    \textit{\textbf{Summary of Answers to RQ4.2:}}
    We have presented six insights for Android testing with LLMs.
    The practical insights emphasize setting selection;
    feedback use;
    cautious model comparison;
    hybrid refinement;
    component-level diagnostics; and 
    metric-aware protocol design.
    We hope that these insights will benefit the Android testing community.    
\end{tcolorbox}

\subsection{Threats to Validity
\label{SEC: Threats to Validity}}

This section examines some potential threats to the validity of our study.

The first threat to validity concerns the selection of apps, bugs, and LLMs.
Our experiments use 115 Google Play UI pages, 37 F-Droid text-input-related bugs, 30 Google Play apps for tool-improvement experiments, and sampled cases for RQ3.2.
These samples cover multiple app categories and are drawn from widely-used Android app sources.
However, they cannot represent all Android apps, custom widgets, WebViews, dynamically-rendered forms, or business-logic constraints.
Similarly, although the nine evaluated LLMs were selected according to availability, recency, leaderboard evidence, and text/vision support, commercial and open-weight models evolve rapidly, and model versions, API behavior, pricing, and availability may change over time.
Therefore, our findings should be interpreted as empirical evidence from the evaluated samples and models rather than universal conclusions about all Android applications or future LLMs.

The second threat concerns the baselines and evaluation protocols.
\texttt{QTypist}~\cite{LiuCWCHHW23} influenced our extracted-context setting, but direct reproduction is difficult because its GPT-3 dependency has been deprecated, and its fine-tuning setup differs from our evaluation of current pre-trained LLMs.
Other approaches, such as TextExerciser~\cite{HeZYCLLYZYZD20} and mutation-based methods~\cite{abs-2310-15657}, are based on different assumptions and objectives.
In addition, our feedback protocol is an empirical iterative-generation strategy and should not be viewed as a complete substitute for targeted mutation rules, app-specific checks, or systematic test-case-generation techniques.

The third threat concerns the limitations of the evaluation metrics.
PPTR measures whether or not generated text enables a page transition within a specific number of attempts, while BDR@i measures whether or not known text-input-related bugs are detected within a specific invalid-input budget.
Although these metrics are appropriate for the objectives of this study, they do not capture broader software qualities, such as functional correctness, security properties, accessibility, privacy, or complete business-logic coverage.
For closed-source apps, our evaluation relies on externally observable behavior and available bug information, which may miss internal state changes or latent faults.

The fourth threat concerns the use of human evaluation.
RQ3.1 relies on ratings from 20 professional testers, with Kendall's W indicating substantial agreement among them.
Nevertheless, Likert-scale ratings remain subjective, and the reconstruction/modification experiments in RQ3.2 were based on sampled cases.
Therefore, the observed improvements should be interpreted as evidence that human expertise can improve representative LLM-generated inputs, rather than as exact estimates of improvement for all apps and pages.

The fifth threat concerns privacy and security considerations.
The LLM-based text-input generation may interact with personal data, authentication fields, or proprietary business logic in real applications.
Although our study focused on controlled testing artifacts and public app samples, practical deployment should avoid sending sensitive user data or confidential application information to third-party LLM services, unless appropriate privacy, security, and compliance controls are in place.

\section{Related Work
\label{SEC: Related Work}}

This section examines related work on automated GUI testing for mobile apps, and on LLM use in software testing.

\subsection{Automated GUI Testing for Mobile Apps
\label{subsec:automated-gui-testing-for-mobile-apps}}

Research into automated mobile-app GUI testing includes methods to explore additional activities or pages.
Automated testing tools can simulate user actions, such as clicking and swiping on UI pages~\cite{LiY0C19, SuMCWYYPLS17, WangLYCZDX18, LiYGC17, BorgesHZ18, WhiteFB19, DegottBZ19, MoranTBVBVPP18, CruzA21}.
Many of these methods have used a random testing approach, triggering UI actions at random.
However, such approaches may not properly handle complex operations, including those that require text inputs with contextual information and constraints, resulting in incomplete exploration of the UI page.
Because text inputs can be regarded as strings, various string-generation methods~\cite{ChenFHHHKLRW22, ChenHHHLRW20, HolikJLRV18, KiezunGAGHE12} have used machine-learning-based approaches to ensure that the generated strings are appropriate for the contextual information.
Although this may work for some string-generation tasks, it may not suffice for testing mobile apps:
Web-sourced inputs~\cite{mcminn2012search} and improved string-readability~\cite{afshan2013evolving} could provide some of the needed enhancements.
Liu et al.~\cite{LiuCWCHHW23} used GPT-3 to generate text inputs based on contextual information in hierarchy files.
Liu et al.~\cite{liu2024make} proposed GPTDroid, enhancing coverage and bug detection with a Q\&A framework.
Yoon et al.~\cite{yoon2024intent} introduced DroidAgent for intent-driven, autonomous testing.
Liu et al.~\cite{liu2024vision} advanced this further with VisionDroid, using multimodal LLMs to detect non-crash bugs through GUI text and visual semantics.
Huang et al.~\cite{HuangQZ25} used inner-app mining and GPT recommendations to generate query-type text inputs for Android apps.

GUI test-case migration \cite{qin2019testmig} is an important part of automated GUI testing.
It involves transferring test cases between apps or platforms, and addresses challenges such as widget-matching and functional similarity \cite{behrang2019test}.
Recent work in this area \cite{talebipour2021ui,hu2018appflow,mariani2021evolutionary,zhang2024learning} makes use of machine learning, NLP, and evolutionary testing to enable cross-platform and cross-app migration.
MigratePro~\cite{zhang2024synthesis}, for example, enhances test-case migration by synthesizing new executable test cases from multiple migrated versions, significantly improving functional coverage, without manual intervention.
Similarly, TRASM~\cite{liu2022test} proposes adaptive semantic-matching strategies to improve GUI event reuse across semantically similar Android apps.
Recent work on UI automation has used LLMs not only for input generation, but also for end-to-end task automation:
AutoDroid~\cite{wen2024autodroid} uses dynamic analysis to inject domain-specific knowledge into LLMs, and to generate high-precision UI action sequences for arbitrary Android tasks, achieving over 70\% end-to-end task completion.
AxNav~\cite{taeb2024axnav} explored accessibility testing using natural language, replaying spoken test instructions using LLMs and pixel-level UI understanding:
This provides a scalable alternative to manual accessibility testing.
This work indicates a shift toward model-centric UI testing workflows, where high-level intents are translated into executable GUI actions.

\subsection{LLMs for Software Testing}

LLMs can be used to generate text inputs, reducing manual GUI analysis costs and the randomness associated with testing tools.
They can analyze complex testing tasks and scenarios.
Among other things in software testing, LLMs have been used for 
fault localization~\cite{CiborowskaD22, abs-2308-15276, KangAY24, ZhuWL21};
unit-test generation~\cite{TufanoDSS22, ChenHZHDY24, abs-2307-00588, LemieuxILS23};
test-oracle generation~\cite{TufanoDSS22, DBLP:conf/icse/NashidSM23,DBLP:conf/icse/NieBLMG23};
test-suite minimization~\cite{PanGB24};
bug analysis~\cite{DBLP:conf/sigsoft/ZhangITH0J22,DBLP:journals/corr/abs-2306-01987}; and
mutation testing~\cite{abs-2301-03543, abs-2310-02407}.

In recent years, the performance of system test-input generation has been widely discussed~\cite{LiuCWCHHW23,0010ZPZMZ17,DBLP:conf/pldi/YeTTHFSBW021,ZhangZBLMXLSL24,XiaPTP024,DBLP:journals/corr/abs-2210-02506,DBLP:conf/icst/ZimmermannK23,abs-2402-00950}.
Studies~\cite{abs-2402-00950,Li2025} have shown that LLMs can guide effective text-input generation for the testing of web apps.
LLMs can also be combined with fuzz testing~\cite{abs-2310-15657} to trigger unexpected behavior and reveal bugs in AUTs.
However, there has not yet been any empirical study of the effectiveness of different LLMs in generating text inputs for Android GUI testing.
Similarly, provision of advice or insights for how testers may best use LLMs for this task is also lacking:
This paper attempts to address this gap in the literature.

\section{Conclusions and Future Work
\label{SEC: Conclusions and Future Work}}

This paper has reported on an empirical study of LLM-based text-input generation for Android GUI testing:
It examined different UI-context prompting settings, feedback protocols, and human-refinement settings.
We evaluated nine recent LLMs on 115 Google Play UI pages, with 183 text-input components and on 37 real-world text-input-related bugs.
The results showed that extracted-context prompting and UI-hierarchy-XML prompting had similar page-pass-through effectiveness, while extracted-context prompting was substantially cheaper; 
screenshot-vision prompting provided visual information, but was not automatically better for the evaluated page-pass-through and bug-detection tasks.
Feedback improved both the page-pass-through and (especially) bug-detection effectiveness, but introduced substantial token and latency overheads.
The human-assessment and human-modification results showed that tester expertise can identify and repair practical weaknesses in selected LLM-generated inputs.
Finally, integrating LLM-generated text inputs into DroidBot improves the UI-page and activity exploration on the sampled apps.
We have also provided six testing insights based on these results.

Our future work will include further investigation of the following:
\begin{itemize}[leftmargin=2em, topsep=0pt, itemsep=0pt, parsep=0pt, partopsep=0pt]
    \item
    \textit{\textbf{Extracting task-relevant contextual information for prompt construction.}}
    Our results indicate that exposing more context is not automatically better: 
    A useful direction is to extract context that is relevant to the target component and objective, while controlling the token cost.
    In this study, we only extracted some of the context information from the UI pages:
    Other information
    ---
    such as the optional lists of ListViews or Spinners, or scrollable lists that allow users to choose from multiple options
    ---
    could also be extracted and used.
    In the future, we will explore extracting and using such additional contextual information to construct the prompt.

    \item
    \textit{\textbf{Using multimodal information to assist with the Android GUI testing process.}}
    Although this study included screenshot-vision prompting, future work should investigate when visual signals are genuinely needed, such as for custom views, WebViews, image-only labels, and dynamically rendered interfaces.
    UI pages can be considered a series of images (screenshots), and thus can be used to represent the contextual information of the currently-displayed UI page.
    They can then be used with certain multimodal LLMs~\cite{abs-2303-18223} designed to handle both text and images.

    \item
    \textit{\textbf{Evaluating more LLMs and UI components.}}
    In this study, we explored nine LLMs' ability to generate text inputs for Android GUIs:
    In the future, we plan to identify and examine other available LLMs.
    This study focused on text input for EditText, AutoCompleteTextView, and MultiAutoCompleteTextView, which are the most common text-input components in Android GUIs.
    However, there are other UI components (such as Spinners and ListViews) that also process text by offering users choices among candidate selections:
    We will explore using LLMs to select the most appropriate (or test-impactful) candidates for such components.

    \item
    \textbf{\textit{Enhancing app-specific invalid-input generation for effective bug detection.}}
    The current LLMs' ability to generate invalid text inputs that detect app-specific bugs remains limited (aligned with Insight 6 in Section \ref{SEC:Insights}).
    We look forward to designing and implementing other methods, and combining them with LLMs to generate more effective invalid texts in the future.
\end{itemize}

We believe that the findings of this study will be of interest and beneficial to the Android quality-assurance community.
We also look forward to sharing the results of our future work with the community.

\section*{Acknowledgment}
We would like to thank the various testers from enterprises and research institutions for their participation in our study. This work is supported by the Science and Technology Development Fund of Macau, Macao SAR, under Grant Nos. 0069/2025/RIB2 and 0021/2023/RIA1.

{
\bibliographystyle{ACM-Reference-Format}
\bibliography{tosem_major.bib}
}

\appendix
\clearpage

\newpage
\setcounter{page}{1} 
\pagenumbering{arabic} 
\section*{Appendices} 
\addcontentsline{toc}{section}{Appendices} 
These appendices contain detailed information for the paper ``Large Language Models for Mobile GUI Text Input Generation: An Empirical Study'' and provide a deeper understanding.

\setcounter{figure}{0}
\renewcommand{\thefigure}{\thesection.\arabic{figure}}

\section{An Example of Context Extraction}
\label{ASEC: Contextual Integration}

\begin{figure*}[!b]
	\centering
	\includegraphics[width=\textwidth]{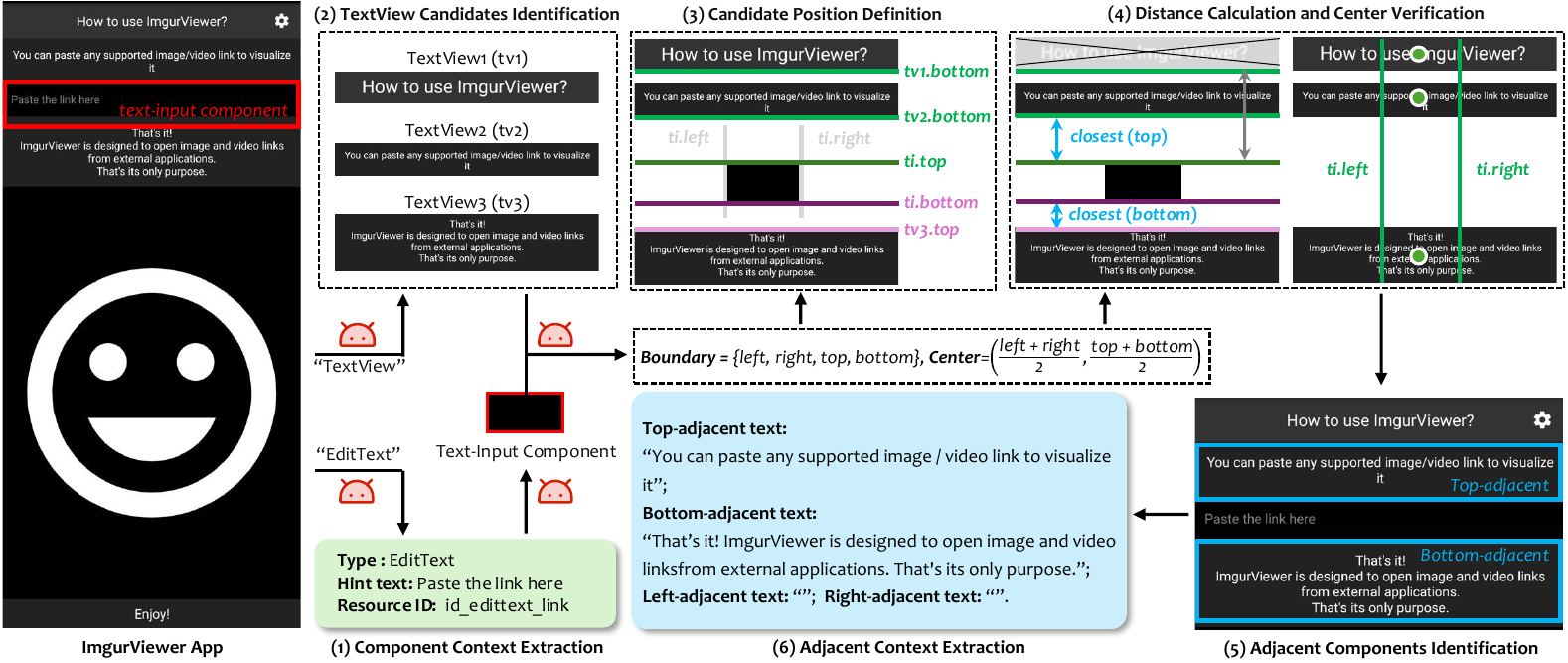}
	\caption{Example of the extraction of the component context and adjacent context from ImgurViewer.
	(The symbol \includegraphics[scale=0.06]{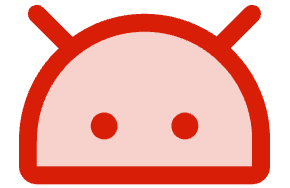} represents UIAutomator, and
	{\color{deepgreen}\CIRCLE\/} represents the center coordinate of the candidate TextViews.)}
	\label{FIG: com_adj_example}
\end{figure*}

Figure~\ref{FIG: com_adj_example} presents an example of extracting the component and adjacent contexts from ImgurViewer\footnote{\url{https://play.google.com/store/apps/details?id=com.ensoft.imgurviewer.}}.

\begin{itemize}
    \item
    The \textit{component context} refers to the essential information about text-input components, including the component type, the hint text displayed within the component (which often provides guidance or a description of what should be entered), and the resource ID (which is a unique identifier to match and position the text-input component on the UI page).
    As shown in Figure~\ref{FIG: com_adj_example}, in Step (1), UIAutomator is used to extract the component context from the text-input component.

    \item
    The \textit{adjacent context} refers to the texts adjacent to the text-input components.
    We defined four adjacent positions around the text-input component: 
    top;
    bottom;
    left; and 
    right.
    The text labels at these four positions are referred to as the adjacent context.
    The adjacent context can help interpret the function and purpose of the text-input components.
    As shown in Figure~\ref{FIG: com_adj_example}, in Step (2), UIAutomator is used to identify the text-input component based on the resource ID from the corresponding extracted component context (in Step (1)), and all candidate TextViews.
    In Step (3), we use UIAutomator to locate the positions of the candidate TextViews and the text-input component.
    Specifically, we used the \textit{boundary} and \textit{center} attributes to determine their relative positions
    ---
    for example, if the bottom boundary of a candidate TextView is higher than the top boundary of the text-input component, then this candidate TextView is considered to be on top of the text-input component.
    (The other three positions (bottom, left, and right), are determined similarly.)
    After determining the relative positions, all candidate TextViews are classified into one of the four categories:
    top;
    bottom;
    left; or 
    right.    
    There are two TextViews in the top position of Step (3) in Figure~\ref{FIG: com_adj_example};
    one in the bottom position; and
    none in either the left or right positions.
    In Step (4),
    the nearest TextViews for each of the four positions are identified, following two constraints:
    the candidate TextView has the shortest distance to the text-input component; and
    the center of the candidate TextViews is within the boundaries of the text-input component.
    After calculating the distances, in Step (5), the nearest-neighbor components for each of the four positions are identified.
    Finally, in Step (6), we use UIAutomator to extract the text labels of the adjacent components as adjacent context.
\end{itemize}

\section{Details of UI Pages}\label{ASEC:Details of UI Pages}

    Table \ref{ATAB:UIPAGESDETAILS} details the 115 UI pages utilized in our experiments, including:
    application identifiers (App ID),
    download counts (Down., collected on April 17, 2025),
    number of text-input components ($d$), and
    UI actions trigger GUI transitions to subsequent pages.
    Entries marked with $\bigstar$ denote 20 randomly selected applications that were manually modified and generated by three testers.
    Applications with $d \geq 2$ were employed in the multiple text-input component ablation study.

\setcounter{table}{0}
\renewcommand{\thetable}{\thesection.\arabic{table}}

\begin{scriptsize}
\renewcommand{\arraystretch}{1}

\setlength{\extrarowheight}{1pt}

\begin{longtable}{cllcl}
\captionsetup{skip=2pt}
\caption{DETAILS OF UI PAGES.}
\label{ATAB:UIPAGESDETAILS}\\
\hline
\textbf{No.} & \textbf{App ID} & \textbf{Download} & \textbf{$d$} & \textbf{UI Actions} \\\hline
\endfirsthead

\hline
\textbf{No.} & \textbf{App ID} & \textbf{Download} & \textbf{$d$} & \textbf{UI Actions} \\\hline
\endhead

\hline
    \multicolumn{5}{r}{Table \ref{ATAB:UIPAGESDETAILS} (Continued on next page)}\\

    \endfoot

\hline
\multicolumn{3}{c}{\textbf{\textit{Total Text-input Component Number}}} & 183 & \\\hline
\endlastfoot
1 & ai.chat.gpt.bot & 10M+ & 1 & Click the button with text ``Continue'' \\
2 & ai.moises & 10M+ & 1 & Click the button with text ``Create'' \\
3 & ai.photo.enhancer.photoclear & 50M+ & 1 & Click the button with text ``Send'' \\
4 & ai.socialapps.speakmaster & 10M+ & 1 & Click the button with coordinate \\
5 & app.hallow.android & 5M+ & 2 & Click the button with text ``Save'' \\
6 & app.kids360.kid & 1M+ & 1 & Click the button with text ``Connect'' \\
7 & app.zyng.app & 1M+ & 1 & Click the button with text ``Continue'' \\
8 & audioeditor.musiceditor.soundeditor.songeditor & 10M+ & 1 & Type an ``Enter'' key \\
9 & bible.offline.kingjames.holy.kjv.bible.daily.verse & 1M+ & 1 & Click the button with coordinate \\
10 & co.fable.fable & 500K+ & 2 & Click the button with text ``Done'' \\
11 & com.BabyName.start ($\bigstar$) & 1M+ & 3 & Click the button with coordinate \\
12 & com.aiphone.secondphonenumber & 100K+ & 4 & Click the button with text ``Save'' \\
13 & com.applabstudios.ai.mail.homescreen.inbox ($\bigstar$) & 1M+ & 5 & Click the button with coordinate \\
14 & com.aspiro.tidal & 10M+ & 1 & Type an ``Enter'' key \\
15 & com.beint.zangi & 10M+ & 2 & Click the button with text ``Continue'' \\
16 & com.brandingbrand.reactnative.and.bjs & 1M+ & 1 & Type an ``Enter'' key \\
17 & com.cars.android & 10M+ & 5 & Type an ``Enter'' key \\
18 & com.chewy.android & 10M+ & 1 & Click the button with coordinate \\
19 & com.citymapper.app.release & 10M+ & 1 & Click the button with coordinate \\
20 & com.colt ($\bigstar$) & 5M+ & 1 & Click the button with coordinate \\
21 & com.contorra.golfpad & 5M+ & 2 & Click the button with text ``Done'' \\
22 & com.deepstash ($\bigstar$) & 1M+ & 1 & Click the button with coordinate \\
23 & com.duolingo & 500M+ & 2 & Click the button with text ``\begin{CJK*}{UTF8}{gbsn}登录\end{CJK*}'' \\
24 & com.economist.lamarr ($\bigstar$) & 1M+ & 1 & Click the button with coordinate \\
25 & com.education.android.h.intelligence & 50M+ & 1 & Click the button with text ``Save'' \\
26 & com.family.locator.find.my.kids & 5M+ & 1 & Click the button with text ``Submit'' \\
27 & com.fandango & 10M+ & 1 & Type an ``Enter'' key \\
28 & com.fly.hong.pdfreader & 1M+ & 1 & Click the button with coordinate \\
29 & com.forecasts.noaa.live.weather & 100K+ & 1 & Click the button with coordinate \\
30 & com.foxnews.android & 10M+ & 1 & Click the button with coordinate \\
31 & com.frameme.photoeditor.collagemaker.effects ($\bigstar$) & 10M+ & 1 & Click the button with coordinate \\
32 & com.gametime.gametime & 5M+ & 1 & Click the button with coordinate \\
33 & com.getyourguide.android & 10M+ & 5 & Click the button with text ``Save'' \\
34 & com.goeuro.rosie & 10M+ & 1 & Click the button with coordinate \\
35 & com.golftripgenius.android.leaguegenius ($\bigstar$) & 500K+ & 1 & Click the button with text ``SIGN IN'' \\
36 & com.gowish.app & 1M+ & 4 & Click the button with text ``ADD WISH'' \\
37 & com.greyhound.mobile.consumer & 1M+ & 1 & Click the button with coordinate \\
38 & com.hellofresh.androidapp & 10M+ & 1 & Click the button with text ``Continue'' \\
39 & com.hometogo & 1M+ & 1 & Click the button with coordinate \\
40 & com.hoopladigital.android ($\bigstar$) & 5M+ & 1 & Click the button with text ``Next'' \\
41 & com.ibragunduz.applockpro & 100M+ & 1 & Click the button with coordinate \\
42 & com.immediasemi.android.blink & 10M+ & 1 & Click the button with text ``Next'' \\
43 & com.isharing.isharing & 10M+ & 1 & Click the button with text ``Save'' \\
44 & com.jeyluta.timestampcamerafree & 50M+ & 1 & Click the button with text ``Save'' \\
45 & com.kajda.fuelio & 5M+ & 5 & Click the button with text ``Start date'' \\
46 & com.kroger.mobile & 10M+ & 1 & Click the button with coordinate \\
47 & com.kw.literie ($\bigstar$) & 500K+ & 1 & Type an ``Enter'' key \\
48 & com.learnpaladin & 50K+ & 1 & Click the button with text ``SAVE'' \\
49 & com.limebike ($\bigstar$) & 10M+ & 1 & Click the button with text ``Next'' \\
50 & com.michaels.michaelsstores & 5M+ & 1 & Type an ``Enter'' key \\
51 & com.mobile.bizo.inpainting.clothes & 1M+ & 1 & Click the button with text ``OK'' \\
52 & com.mobillium.airalo & 5M+ & 1 & Click the button with coordinate \\
53 & com.musescore.playerlite ($\bigstar$) & 5M+ & 1 & Type an ``Enter'' key \\
54 & com.musicplayer.playermusic & 50M+ & 1 & Click the button with text ``Done'' \\
55 & com.mxtech.videoplayer.ad & 1B+ & 1 & Click the button with text ``Create'' \\
56 & com.myfitnesspal.android & 100M+ & 2 & Click the button with text ``CHANGE PASSWORD'' \\
57 & com.mysugr.android.companion & 5M+ & 1 & Click the button with text ``Next'' \\
58 & com.naraorganics.nara & 100K+ & 3 & Click the button with text ``Create Account'' \\
59 & com.nihaojewelry.nihao & 1M+ & 8 & Click the button with coordinate \\
60 & com.novanews.localnews.en & 5M+ & 1 & Click the button with text ``Post'' \\
61 & com.nytimes.android & 10M+ & 1 & Type an ``Enter'' key \\
62 & com.okta.android.auth & 10M+ & 2 & Click the button with text ``Add Account'' \\
63 & com.olo.ihop & 1M+ & 1 & Type an ``Enter'' key \\
64 & com.omronhealthcare.omronconnect & 1M+ & 3 & Click the button with text ``ADD'' \\
65 & com.onwaterapp & 50K+ & 1 & Click the button with coordinate \\
66 & com.opentable ($\bigstar$) & 10M+ & 1 & Click the button with text ``Continue'' \\
67 & com.peerspace.app & 50K+ & 4 & Click the button with text ``Save'' \\
68 & com.podbean.app.podcast & 10M+ & 1 & Click the button with text ``Save'' \\
69 & com.punicapp.whoosh & 10M+ & 1 & Click the button with text ``Get code'' \\
70 & com.qianfan.aihomework & 10M+ & 1 & Click the button with coordinate \\
71 & com.qvon.novellair & 100K+ & 1 & Type an ``Enter'' key \\
72 & com.radio.pocketfm & 100M+ & 1 & Type an ``Enter'' key \\
73 & com.redfin.android & 5M+ & 1 & Click the button with text ``UPDATE'' \\
74 & com.rent & 1M+ & 1 & Type an ``Enter'' key \\
75 & com.rnovel.android.app & 1M+ & 1 & Click the button with text ``Done'' \\
76 & com.roomthermometer.interior & 1M+ & 1 & Type an ``Enter'' key \\
77 & com.scribd.app.reader0 ($\bigstar$) & 10M+ & 1 & Type an ``Enter'' key \\
78 & com.servicemagic.consumer & 5M+ & 1 & Click the button with text ``Next'' \\
79 & com.shazam.android & 500M+ & 1 & Type an ``Enter'' key \\
80 & com.shiftsmart.workerapp & 1M+ & 1 & Click the button with text ``Continue'' \\
81 & com.shopify.arrive & 10M+ & 3 & Click the button with coordinate \\
82 & com.shopify.mobile & 10M+ & 3 & Click the button with coordinate \\
83 & com.sillens.shapeupclub & 10M+ & 1 & Click the button with coordinate \\
84 & com.skool.skoolcommunities & 1M+ & 4 & Type an ``Enter'' key \\
85 & com.smule.singandroid & 100M+ & 1 & Type an ``Enter'' key \\
86 & com.snowcorp.stickerly.android & 100M+ & 1 & Type an ``Enter'' key \\
87 & com.spotify.music & 1B+ & 2 & Click the button with text ``Save'' \\
88 & com.sunrun.customerapp & 100K+ & 1 & Click the button with text ``Next'' \\
89 & com.teamreach.app ($\bigstar$) & 1M+ & 2 & Click the button with text ``Create'' \\
90 & com.thanx.nothingbundtcakes & 100K+ & 3 & Click the button with text ``Next'' \\
91 & com.thedyrt.wayfinder & 1M+ & 3 & Click the button with text ``Next'' \\
92 & com.thumbtack.consumer & 5M+ & 2 & Type an ``Enter'' key \\
93 & com.tickpickllc.ceobrien.tickpick ($\bigstar$) & 1M+ & 1 & Click the button with text ``Save \& Continue'' \\
94 & com.touchtunes.android & 5M+ & 1 & Type an ``Enter'' key \\
95 & com.trubeacon.scooters\_mobile\_android ($\bigstar$) & 500K+ & 1 & Type an ``Enter'' key \\
96 & com.ultimateguitar.tabs ($\bigstar$) & 10M+ & 1 & Type an ``Enter'' key \\
97 & com.virtual.video.i18n & 1M+ & 1 & Click the button with text ``Confirm'' \\
98 & com.vkontakte.android & 100M+ & 1 & Click the button with text ``Continue'' \\
99 & com.wendys.nutritiontool & 10M+ & 1 & Click the button with coordinate \\
100 & com.wizards.winter\_orb ($\bigstar$) & 1M+ & 1 & Type an ``Enter'' key \\
101 & com.wonder & 10M+ & 1 & Click the button with text ``OK'' \\
102 & com.wondershare.famisafe & 5M+ & 1 & Click the button with text ``Next'' \\
103 & com.xogrp.planner & 1M+ & 4 & Click the button with text ``Save'' \\
104 & com.yelp.android.biz ($\bigstar$) & 1M+ & 2 & Click the button with text ``SAVE'' \\
105 & com.zennioptical.app & 1M+ & 1 & Type an ``Enter'' key \\
106 & com.zillow.android.zillowmap & 50M+ & 1 & Type an ``Enter'' key \\
107 & heartratemonitor.heartrate.pulse.pulseapp & 10M+ & 3 & Click the button with text ``Save'' \\
108 & net.wordbit.enes & 5M+ & 1 & Click the button with coordinate \\
109 & notes.notepad.checklist.calendar.todolist.notebook & 10M+ & 2 & Click the button with coordinate \\
110 & org.peopleready.jobstack & 100K+ & 2 & Click the button with coordinate \\
111 & org.wikipedia ($\bigstar$) & 50M+ & 1 & Type an ``Enter'' key \\
112 & sanity.podcast.freak & 1M+ & 1 & Type an ``Enter'' key \\
113 & tunein.player & 100M+ & 1 & Click the button with text ``SAVE'' \\
114 & tv.telepathic.hooked & 10M+ & 1 & Click the button with coordinate \\
115 & us.mitene & 10M+ & 2 & Click the button with text ``DONE'' \\
\end{longtable}
 \end{scriptsize}

\section{Details of Bugs}\label{ASEC:Details of Bugs}
Table \ref{TAB:bugs_list} presents the 37 buggy UI pages, including their app names, bug descriptions, bug categories, and corresponding issue report links.

\setcounter{table}{0}
\renewcommand{\thetable}{\thesection.\arabic{table}}

\begin{scriptsize}
\renewcommand{\arraystretch}{1.25}

\setlength{\extrarowheight}{1.5pt}
    \begin{longtable}{@{}c|m{2cm}|m{4.8cm}|c|m{3.8cm}@{}}
    \caption{A Comprehensive Overview of Bugs.}
    \label{TAB:bugs_list}\\  
    \hline
    \textbf{No.} & \multicolumn{1}{c|}{\textbf{App Name}} & \multicolumn{1}{c|}{\textbf{Bug Description}} & \textbf{Category} & \textbf{Issue Report Link} \\
    \hline
    \endfirsthead  

    \hline
    \textbf{No.} & \multicolumn{1}{c|}{\textbf{App Name}} & \multicolumn{1}{c|}{\textbf{Bug Description}} & \textbf{Category} & \textbf{Issue Report Link} \\
    \hline
    \endhead

    \hline
    \multicolumn{5}{r}{Table \ref{TAB:bugs_list} (Continued on next page)}\\
    \endfoot

    \hline 
    \endlastfoot

    1 & Unser Hörsaal & The application crashes when the input field is empty. & Empty Char
    & \url{https://github.com/ASE-Projekte-WS-2021/ase-ws-21-unser-horsaal/issues/239}\\

    2 & Debitum & The application crashes when entering amounts > 999 in locales with a space as the grouping character. & Constraint
    & \url{https://github.com/Marmo/debitum/issues/86}\\

    3 & Mozc & A phone number starting with ``0'' causes an issue. & Specific Input
    & \url{https://github.com/google/mozc/issues/375} \\

    4 & Mozc & A user-dictionary email entry is not suggested when typing its registered reading U+3081 (Japanese Hiragana ``me'').
    & Specific Input
    & \url{https://github.com/google/mozc/issues/376}                                  \\

    5 & NewPipe & The application shows an error snackbar when retrieving YouTube search suggestions for a query longer than 100 characters, with spaces counted as characters.
    & Constraint & \url{https://github.com/TeamNewPipe/NewPipe/issues/5588}\\

    6 & SORMAS & Entering non-numeric characters in the immunization number-of-doses field causes the app to crash. & Constraint
    & \url{https://github.com/hzi-braunschweig/SORMAS-Project/issues/6994}\\

    7 & Lab1-Calc & The application crashes when invalid input is provided (non-numeric characters). & Special Char
    & \url{https://github.com/Krow10/Lab1-Calc/issues/7}\\

    8 & AppManager & Entering a non-numeric app-op name, such as ``WRITE\_SETTINGS'', in the profile app-ops configuration causes a NumberFormatException and crashes the app. & Specific Input
    & \url{https://github.com/MuntashirAkon/AppManager/issues/504}               \\

    9 & AnySoft-Keyboard & Inputting ``:'' (contains) causes the application to crash. & Special Char
    & \url{https://github.com/AnySoftKeyboard/AnySoftKeyboard/issues/2785}\\

    10 & KISS & After typing ``a'' or ``A'', entering any additional ordinary character causes app to crash. & Constraint
    & \url{https://github.com/Neamar/KISS/issues/1687}\\

    11 & KISS & The input contains any letter other than ``a'' (uppercase or lowercase) or any special symbols (\%\&\texteuro\textasciicircum¡¿\textasciitilde®©™¢¥\$¦¬°¶§×, excluding accented letters and common punctuation) will trigger the bug. & Special Char
    & \url{https://github.com/Neamar/KISS/issues/1687}\\

    12 & FlorisBpard & When any input field contains a word starting with lowercase ``auto'' followed by a space or symbol, FlorisBoard incorrectly appends an extra ``0''. & Specific Input
    & \url{https://github.com/florisboard/florisboard/issues/2041}\\

    13 & FlorisBpard & When any input field contains ``auto0'' followed by another ``0'', a space, or a symbol, FlorisBoard incorrectly appends additional zeros. & Specific Input
    & \url{https://github.com/florisboard/florisboard/issues/2041} \\

    14 & wallabag & Entering a malformed full-text search query, such as ``"'' or ``-foo'', in the article search box causes the wallabag app to crash. & Special Char
    & \url{https://github.com/wallabag/android-app/issues/1125}\\

    15 & libgdx & If the input string is just ``['', the BitmapFont markup causes a crash. & Specific Input
    & \url{https://github.com/libgdx/libgdx/issues/2886}\\

    16 & Kontalk & Detects non-emoji textual contexts that may accidentally form smiley-like sequences and be converted into graphics. & Specific Input
    & \url{https://github.com/kontalk/androidclient/issues/1082}\\

    17 & SASAbus & When either the From or To departure/arrival EditText is empty, tapping refresh triggers an error, and subsequently tapping ``Search ...'' causes the app to crash. & Empty Char
    & \url{https://github.com/SASAbus/SASAbus/issues/17}\\

    18 & Masked-Edittext & If non-digit symbols are entered, the EditText removes mask characters and then crashes. & Special Char
    & \url{https://github.com/pinball83/Masked-Edittext/issues/7}\\

    19 & Riot &  After entering any non-empty text and pressing Enter, the app crashes. & Constraint
    & \url{https://github.com/vector-im/riot-android/issues/634}\\

    20 & Quote (Anecdote) & Entering only whitespace in the DOM selector field when adding a custom website causes the app to crash. & Empty Char
    & \url{https://github.com/HugoGresse/Anecdote/issues/27}\\

    21 & Quote (Anecdote) & Entering a malformed URL in the website URL field when adding a custom website causes the app to crash. & Constraint
    & \url{https://github.com/HugoGresse/Anecdote/issues/27}\\

    22 & Landfill E-Forms & Inputting a number starting with ``.'' (e.g., ``.67'') for barometric pressure or methane fields causes the application to become unresponsive and crash. & Specific Input
    & \url{https://github.com/grantkang/LandfillE-Forms/issues/21}\\

    23 & Nextcloud Notes & Previewing a note whose content is an empty two-digit numbered-list item, e.g., ``10.'', causes the application to crash. & Specific Input
    & \url{https://github.com/stefan-niedermann/nextcloud-notes/issues/668}\\

    24 & Nextcloud Notes & The application becomes sluggish and crashes after rapidly entering a long text. & Constraint
    & \url{https://github.com/stefan-niedermann/nextcloud-notes/issues/267}\\

    25 & MinTone & Clicking ``Set Freq Manually'' with an empty text field causes the application to crash. & Empty Char
    & \url{https://github.com/dgets/MinTone/issues/22}\\

    26 & K-9 Mail & An invalid recipient address leads to a crash. & Constraint
    & \url{https://github.com/k9mail/k-9/issues/5991} \\

    27 & CommuniDog & Clicking the Login button with missing required input causes the app to crash. & Empty Char
    & \url{https://github.com/IdoSagiv/CommuniDog/issues/103}\\

    28 & WiseTrack & Entering a single space into the search bar causes an issue. & Special Char
    & \url{https://github.com/CMPUT301W21T12/WiseTrack/issues/104}\\

    29 & Sora-editor & In wordwrap mode, entering text spanning multiple lines and then holding backspace may trigger ANR/crash. & Constraint
    & \url{https://github.com/Rosemoe/sora-editor/issues/168}\\

    30 & Health Log & Leaving the ``Enter Hospital ID'' EditText empty causes an issue.       & Empty Char
    & \url{https://github.com/Technical-Hackers/Health-Log/issues/35}\\

    31 & Interval Timer & Entering a minute/second value outside the 0--60 range displays an unfriendly validation error message. & Constraint
    & \url{https://github.com/SecUSo/privacy-friendly-interval-timer/issues/33}\\

    32 & Revolution IRC & Editing a theme color hex field to a standalone ``\#'' causes a crash. & Special Char
    & \url{https://github.com/MCMrARM/revolution-irc/issues/201}                 \\

    33 & Organic Maps & Updating a field name with prohibited characters hides the red error message. & Special Char
    & \url{https://github.com/organicmaps/organicmaps/issues/4343}\\

    34 & Activity Diary & Empty character input is not allowed. & Empty Char
    & \url{https://github.com/ramack/ActivityDiary/issues/285} \\

    35 & Eventyay Attendee & Multi-line input is not allowed. & Constraint
    & \url{https://github.com/fossasia/open-event-attendee-android/issues/2621} \\

    36 & Eventyay Attendee & Invalid email addresses can be entered without restriction. & Constraint
    & \url{https://github.com/fossasia/open-event-attendee-android/issues/865}\\

    37 & Eventyay Organizer & The application crashes when the password and confirm password fields do not match. & Constraint
    & \url{https://github.com/fossasia/open-event-organizer-android/pull/1972} ( \url{https://github.com/fossasia/open-event-organizer-android/issues/1971})\\
\end{longtable}
\end{scriptsize}

\section{Details of Apps for Tool Improvement Experiment}
\label{ASEC:Details of Apps}
Table \ref{ATAB:tool_improve} presents the 30 apps used in the tool-improvement experiments, including application identifiers (App ID) and download counts (collected on July 22, 2025).

\setcounter{table}{0}
\renewcommand{\thetable}{\thesection.\arabic{table}}

\begin{scriptsize}
    \begin{longtable}{@{}c l c | c l c@{}}  
    \captionsetup{skip=2pt}
    \caption{A Comprehensive Overview of Apps Used in Tool Improvement Experiments.}
    \label{ATAB:tool_improve}\\
    \hline
    
    \textbf{No} & \textbf{App ID} & \textbf{Down.} & \textbf{No} & \textbf{App ID} & \textbf{Down.} \\
    \hline
    \endfirsthead

    \hline
    
    \textbf{No} & \textbf{App ID} & \textbf{Down.} & \textbf{No} & \textbf{App ID} & \textbf{Down.} \\
    \hline
    \endhead

    \hline
    \multicolumn{6}{r}{Table \ref{ATAB:tool_improve} (Continued on next page)}\\
    \hline
    \endfoot

    \hline
    \endlastfoot

    \rowcolors{1}{white}{black!10} 
    \tiny 
    \setlength\tabcolsep{1.5mm}   
    \renewcommand{\arraystretch}{1.3}

    1  & com.abaenglish.videoclass                  & 10M+               & 16 & com.kimcy929.secretvideorecorder           & 10M+               \\
    2  & com.ada.app                                & 10M+               & 17 & com.mi.health                              & 5M+                \\
    3  & com.amazon.cosmos                          & 100K+              & 18 & com.originatorkids.EndlessAlphabet         & 1M+                \\
    4  & com.apple.android.music                    & 100M+              & 19 & com.p1.mobile.putong                       & 5M+                \\
    5  & com.barnesandnoble.app                     & 500K+              & 20 & com.qooapp.qoohelper                       & 1M+                \\
    6  & com.bk.pt                                  & 1M+                & 21 & com.ss.android.article.news                & 1M+                \\
    7  & com.brother.ph.brcamera                    & 50K+               & 22 & com.teamsnap.teamsnap                      & 5M+                \\
    8  & com.disney.datg.videoplatforms.android.abc & 100M+              & 23 & com.telangana.twallet                      & 1M+                \\
    9  & com.eg.android.AlipayGphone                & 10M+               & 24 & com.valvesoftware.android.steam.community  & 100M+              \\
    10 & com.github.kr328.clash.foss                & 10K+               & 25 & com.zillow.android.zillowmap               & 50M+               \\
    11 & com.google.android.apps.translate          & 1B+                & 26 & enterprises.dating.boo                     & 10M+               \\
    12 & com.huawei.maps.app                        & 100M+              & 27 & org.quantumbadger.redreader                & 100K+              \\
    13 & com.iconology.comics                       & 10M+               & 28 & pi.browser                                 & 10M+               \\
    14 & com.iloen.melon                            & 50M+               & 29 & tv.danmaku.bili                            & 1M+                \\
    15 & com.iqiyi.i18n                             & 100M+              & 30 & uk.co.bbc.bbc\_plus                        & 10K+               \\
\end{longtable}
\end{scriptsize}

\section{Supplementary Ablation Analysis}\label{ASEC:Supplementary Ablation Analysis}
This appendix reports the findings of the two supplementary ablation experiments, focusing on two key aspects that help interpret RQ1 page-pass-through results: the \textit{sub-prompts}, and the influence of \textit{multiple text-input components} on a single UI page.

\subsection{Sub-Prompt Ablation Settings}\label{ASEC:Prompt-Ablation Settings}

The original prompt-component and multi-component ablation experiments have been retained as supplementary analyses for interpreting the RQ1 page-pass-through results.
For the prompt-component ablation, we investigated the contribution of each sub-prompt in Section~\ref{SEC: Prompt Construction}.
The restrictive sub-prompt was retained in all variants because preliminary attempts without it prevented reliable extraction and parsing of LLM outputs.
The prompt variants, therefore, removed one contextual or guiding sub-prompt at a time, including the global sub-prompt, component sub-prompt, adjacent sub-prompt, and guiding sub-prompt, while keeping the restrictive sub-prompt fixed.

\subsection{Multi-Component Ablation Settings}\label{ASEC:Multi-Component Ablation Settings}
For the multi-component ablation, we used systematic input masking to assess the contribution of individual text-input components on UI pages with more than one text-input component.
We identified 32 UI pages containing multiple text-input components among the 115 pages analyzed in RQ1.1.
Table~\ref{tab:ablation_matrix} presents the combinatorial test scenario matrix for these applications
($d$ denotes the number of text-input components per app, and $\tau$ denotes the number of actively-filled text-input components in each test scenario).
The combinatorial ablation design systematically varied the set of $\tau$-filled text-input components while keeping others empty, enabling isolation of component contributions.
For each app with $d$ components, we generated all possible combinations of filled text-input components at strength levels $\tau = 0$ to $\tau = d$, resulting in a total of 584 test scenarios across the 32 apps.

\begin{table*}
\centering
\scriptsize
\captionsetup{skip=2pt}
\caption{Combinatorial Test Scenario Matrix of Selected 32 Apps.}\label{tab:ablation_matrix}
	\setlength\tabcolsep{1.2mm}

\begin{tabular}{cccccccccccccc}
\hline
\multicolumn{1}{c}{\multirow{2}{*}{\shortstack{\textbf{No. of text-input} \\ \textbf{components} ($d$)}}}
    & \multicolumn{9}{c}{\textbf{No. of actively-filled components by combinatorial strength} ($\tau$)}
    && \multirow{2}{*}{\shortstack{\textbf{No. of test} \\ \textbf{scenarios per app}}}
    & \multirow{2}{*}{\shortstack{\textbf{No. of} \\ \textbf{apps}}}
    & \multirow{2}{*}{\shortstack{\textbf{Total} \\ \textbf{scenarios}}} \\
\cline{2-10}
& $\tau=0$ & $\tau=1$ & $\tau=2$ & $\tau=3$ & $\tau=4$ & $\tau=5$ & $\tau=6$ & $\tau=7$ & $\tau=8$ & & & \\
\hline
$d=2$ & 1 & 2 & 1 & -- & -- & -- & -- & -- & -- && 4 & 14 & 56 \\
$d=3$ & 1 & 3 & 3 & 1 & -- & -- & -- & -- & -- && 8 & 8 & 64 \\
$d=4$ & 1 & 4 & 6 & 4 & 1 & -- & -- & -- & -- && 16 & 5 & 80 \\
$d=5$ & 1 & 5 & 10 & 10 & 5 & 1 & -- & -- & -- && 32 & 4 & 128 \\
$d=8$ & 1 & 8 & 28 & 56 & 70 & 56 & 28 & 8 & 1 && 256 & 1 & 256 \\
\hline
\multicolumn{12}{c}{\textbf{Grand Total}}& 32 & 584 \\
\hline
\end{tabular}
\end{table*}

We used a \textit{criticality score} to quantify the contribution of individual text-input components to the PPTR results:
$\Psi(c_i)$, for each component $c_i~(i=1,2,\dots,d)$, where $d$ denotes the total number of text-input components on a given UI page.
This metric captures the average contribution of $c_i$ to PPTRs across various combinations of actively-filled text-input components.
It is computed as:
\begin{equation}
    \Psi(c_i) = \frac{1}{|S_i|} \sum_{s \in S_i} \left[
        I(s_{\text{full}}) - I(s_{\text{full}} \setminus \{c_i\})
    \right] \times \omega(\tau_s),
    \label{eq:criticality_score}
\end{equation}
where
$S_i$ denotes the set of all combinations containing text-input component $c_i$;
$s_{\text{full}}$ represents a combination with all $\tau_s$ text-input components filled;
$s_{\text{full}} \setminus \{c_i\}$ is the state excluding $c_i$;
$I(\cdot)$ is the indicator function (1 for successful pass-through; 0 otherwise), averaged over multiple tests; and
$\omega(\tau_s) = \tau_s / d$ weights the contribution by combination size relative to $d$.
The term $I(s_{\text{full}}) - I(s_{\text{full}} \setminus \{c_i\})$ quantifies the isolated impact of $c_i$:
Positive $\Psi(c_i)$ values indicate that including $c_i$ improves PPTRs, and negative values mean that it harms the PPTRs.
The weighting factor $\omega(\tau_s)$ emphasizes scenarios closer to real-world usage:
Larger weights are used when most components are filled ($\tau_s \approx d$), reflecting higher PPTR probability; while
lower weights are used for sparse combinations ($\tau_s \ll d$), where the PPTRs may be more volatile.

Criticality scores are relative metrics.
They reflect the text-input component's importance within a specific UI page:
The values are comparable within pages but not across pages.
For example, on a given page $A$ ($d=2$), $\Psi{(A_1)} =\Psi{(A_2)} = 0.01$ would indicate that components $A_1$ and $A_2$ have equal criticality; and
on a page $B$ ($d=2$), $\Psi{(B_1)} =\Psi{(B_2)} = 1$ would similarly indicate that $B_1$ and $B_2$ have equal criticality.
To alleviate the ambiguity of numerical values, we applied min-max normalization to map criticality scores to the range  $[0.0, 1.0]$:
\begin{equation}
    \Psi_{\text{norm}}(c_i) =
    \begin{cases}
        1, & \text{if } \max(\Psi) = \min(\Psi) \\
        \frac{\Psi(c_i) - \min(\Psi)}{\max(\Psi) - \min(\Psi)}, & \text{otherwise}.
    \end{cases}
    \label{eq:normalization}
\end{equation}
This normalized metric enables the clear identification of text-input components that significantly influence PPTRs within each UI page:
Components with higher $\Psi_{\text{norm}}$ values have greater relative criticality compared to other text-input components on the same page.

\begin{table*}
    \centering
    \scriptsize
    \captionsetup{skip=2pt}
    \caption{Contribution of sub-prompts to PPTR on 115 UI pages with statistical comparisons.
    Values in parentheses report the change from the full prompt to the ablated prompt.}
    \label{TAB: prompt-ablation}
    \setlength\tabcolsep{0.2mm}
    \resizebox{\textwidth}{!}{
    \begin{tabular}{lccccc c cccc}
        \hline
        \multirow{2}{*}{\textbf{LLM}} & \multicolumn{5}{c}{\textbf{PPTR}} & & \multicolumn{4}{c}{\textbf{\hlgrey{$Prompt$} \textit{vs. w/o}}} \\
        \cline{2-6}\cline{8-11}
        & \hlgrey{$Prompt$} & \textbf{\textit{w/o} \hlpink{$GloP$}} & \textbf{\textit{w/o} \hlpurple{$ComP$}} & \textbf{\textit{w/o} \hlgreen{$AdjP$}} & \textbf{\textit{w/o} \hlblue{$GuiP$}} & & \textbf{\hlpink{$GloP$}} & \textbf{\hlpurple{$ComP$}} & \textbf{\hlgreen{$AdjP$}} & \textbf{\hlblue{$GuiP$}} \\
        \hline
        \GeminiM  & 73.0\% & 72.2\% (\down 0.9 pp; \down 1.19\%) & 70.4\% (\down 2.6 pp; \down 3.57\%) & 71.3\% (\down 1.7 pp; \down 2.38\%) & 71.3\% (\down 1.7 pp; \down 2.38\%) & & \ding{109} (0.50) & \ding{109} (0.51) & \ding{109} (0.51) & \ding{109} (0.51) \\
        \ClaudeM  & 70.4\% & 69.6\% (\down 0.9 pp; \down 1.23\%) & 70.4\% (-- 0.0 pp; 0.00\%) & 71.3\% (\up 0.9 pp; \up 1.23\%) & 67.8\% (\down 2.6 pp; \down 3.70\%) & & \ding{109} (0.50) & \ding{109} (0.50) & \ding{109} (0.50) & \ding{109} (0.51) \\
        \GrokM    & 73.9\% & 73.0\% (\down 0.9 pp; \down 1.18\%) & 68.7\% (\down 5.2 pp; \down 7.06\%) & 71.3\% (\down 2.6 pp; \down 3.53\%) & 72.2\% (\down 1.7 pp; \down 2.35\%) & & \ding{109} (0.50) & \ding{52} (0.53) & \ding{109} (0.51) & \ding{109} (0.51) \\
        \GPTM     & 71.3\% & 69.6\% (\down 1.7 pp; \down 2.44\%) & 67.0\% (\down 4.3 pp; \down 6.10\%) & 71.3\% (-- 0.0 pp; 0.00\%) & 70.4\% (\down 0.9 pp; \down 1.22\%) & & \ding{109} (0.51) & \ding{109} (0.52) & \ding{109} (0.50) & \ding{109} (0.50) \\
        \DoubaoM  & 70.4\% & 69.6\% (\down 0.9 pp; \down 1.23\%) & 68.7\% (\down 1.7 pp; \down 2.47\%) & 69.6\% (\down 0.9 pp; \down 1.23\%) & 70.4\% (-- 0.0 pp; 0.00\%) & & \ding{109} (0.50) & \ding{109} (0.51) & \ding{109} (0.50) & \ding{109} (0.50) \\
        \QwenM    & 73.9\% & 70.4\% (\down 3.5 pp; \down 4.71\%) & 72.2\% (\down 1.7 pp; \down 2.35\%) & 72.2\% (\down 1.7 pp; \down 2.35\%) & 69.6\% (\down 4.3 pp; \down 5.88\%) & & \ding{109} (0.52) & \ding{109} (0.51) & \ding{109} (0.51) & \ding{109} (0.52) \\
        \ERNIEM   & 71.3\% & 67.8\% (\down 3.5 pp; \down 4.88\%) & 69.6\% (\down 1.7 pp; \down 2.44\%) & 70.4\% (\down 0.9 pp; \down 1.22\%) & 71.3\% (-- 0.0 pp; 0.00\%) & & \ding{109} (0.52) & \ding{109} (0.51) & \ding{109} (0.50) & \ding{109} (0.50) \\
        \KimiM    & 70.4\% & 72.2\% (\up 1.7 pp; \up 2.47\%) & 71.3\% (\up 0.9 pp; \up 1.23\%) & 70.4\% (-- 0.0 pp; 0.00\%) & 69.6\% (\down 0.9 pp; \down 1.23\%) & & \ding{109} (0.49) & \ding{109} (0.50) & \ding{109} (0.50) & \ding{109} (0.50) \\
        \MistralM & 67.8\% & 66.1\% (\down 1.7 pp; \down 2.56\%) & 63.5\% (\down 4.3 pp; \down 6.41\%) & 70.4\% (\up 2.6 pp; \up 3.85\%) & 68.7\% (\up 0.9 pp; \up 1.28\%) & & \ding{109} (0.51) & \ding{109} (0.52) & \ding{109} (0.49) & \ding{109} (0.50) \\
        \cline{1-6}\cline{8-11}
        \textbf{Avg.} & 71.4\% & 70.0\% (\down 1.4 pp; \down 1.83\%) & 69.1\% (\down 2.3 pp; \down 3.29\%) & 70.9\% (\down 0.5 pp; \down 0.61\%) & 70.1\% (\down 1.3 pp; \down 1.71\%) & & \ding{52} (0.51) & \ding{52} (0.51) & \ding{109} (0.50) & \ding{52} (0.51) \\
        \hline
    \end{tabular}
    }
\end{table*}
\subsection{Prompt Ablation Results}\label{asec:prompt-ablation}
Table~\ref{TAB: prompt-ablation} presents the changes in the PPTR results after the ablation of different sub-prompts, and the statistical-comparison results.
Based on these results, we have the following observations:
\begin{itemize}
    \item
    Compared with the complete prompt, the average PPTR remains close after ablating any single sub-prompt:
    the average decreased by 1.4 pp, 2.3 pp, 0.5 pp, and 1.3 pp when removing the global, component, adjacent, and guiding sub-prompts, respectively, corresponding to relative decreases of 1.83\%, 3.29\%, 0.61\%, and 1.71\%.
    Among the four ablations, removing the component prompt had the largest average impact, whereas removing the adjacent prompt had the smallest average impact.

    \item
    At the individual-LLM level, most ablations lead to small changes.
    For example, removing the global sub-prompt changed PPTR by between a 3.5 pp decrease and a 1.7 pp increase, and removing the guiding sub-prompt changed PPTR by between a 4.3 pp decrease and a 0.9 pp increase.
    
    \item
    The component prompt provides direct information about the target text-input component, and it is therefore the most sensitive part of the prompt structure in our ablation results.
    Removing it caused the largest average decrease and a statistically significant decrease for \Grok.

    \item
    The statistical comparisons further show that the effect sizes are small overall:
    most $\hat{A}_{12}$ values are between 0.49 and 0.52.
    This indicates that no single sub-prompt alone dominates the effectiveness; instead, the full prompt works as a combination of complementary contextual and guiding information.
\end{itemize}

\begin{tcolorbox}[breakable,colframe=black,arc=1mm,left={1mm},top={0mm},bottom={1mm},right={1mm},boxrule={0.25mm},before upper={\raisebox{-0.25\height}{\includegraphics[scale=0.013]{figures/icons/android}}~}]
    \textit{\textbf{Summary of Supplementary Prompt Ablation:}}
    Different sub-prompts can influence the effectiveness of the text inputs generated by the LLMs.
\end{tcolorbox}

\begin{table*}
    \centering
    \scriptsize
    \captionsetup{skip=2pt}
    \caption{Detailed multiple text-input component ablation results for PPTR.}
    \label{TAB:multi-component-ablation}
    \setlength\tabcolsep{2.5mm}
    \begin{tabular}{clcrrrrrrrrrl}
        \hline
        \multirow{2}{*}{\textbf{No.}} & \multirow{2}{*}{\textbf{Case}} & \multirow{2}{*}{$\boldsymbol{d}$} & \multicolumn{9}{c}{\textbf{PPTR by combinatorial strength $\boldsymbol{\tau}$ (\%)}} & \multirow{2}{*}{$\boldsymbol{\Psi_{\text{norm}}}$} \\
        \cline{4-12}
        & & & \textbf{0} & \textbf{1} & \textbf{2} & \textbf{3} & \textbf{4} & \textbf{5} & \textbf{6} & \textbf{7} & \textbf{8} & \\
        \hline
        1 & P005 & 2 & 0.0 & 50.0 & 100.0 & -- & -- & -- & -- & -- & -- & $[0,1]$ \\
        2 & P010 & 2 & 0.0 & 50.0 & 100.0 & -- & -- & -- & -- & -- & -- & $[1,0]$ \\
        3 & P015 & 2 & 0.0 & 0.0 & 100.0 & -- & -- & -- & -- & -- & -- & $[1,1]$ \\
        4 & P021 & 2 & 0.0 & 50.0 & 100.0 & -- & -- & -- & -- & -- & -- & $[1,0]$ \\
        5 & P023 & 2 & 0.0 & 0.0 & 0.0 & -- & -- & -- & -- & -- & -- & $[1,1]$ \\
        6 & P056 & 2 & 0.0 & 0.0 & 100.0 & -- & -- & -- & -- & -- & -- & $[1,1]$ \\
        7 & P062 & 2 & 0.0 & 0.0 & 0.0 & -- & -- & -- & -- & -- & -- & $[1,1]$ \\
        8 & P087 & 2 & 0.0 & 50.0 & 100.0 & -- & -- & -- & -- & -- & -- & $[1,0]$ \\
        9 & P089 & 2 & 0.0 & 0.0 & 96.3 & -- & -- & -- & -- & -- & -- & $[1,1]$ \\
        10 & P092 & 2 & 0.0 & 0.0 & 100.0 & -- & -- & -- & -- & -- & -- & $[1,1]$ \\
        11 & P109 & 2 & 0.0 & 100.0 & 100.0 & -- & -- & -- & -- & -- & -- & $[1,1]$ \\
        12 & P110 & 2 & 0.0 & 0.0 & 0.0 & -- & -- & -- & -- & -- & -- & $[1,1]$ \\
        13 & P115 & 2 & 0.0 & 0.0 & 100.0 & -- & -- & -- & -- & -- & -- & $[1,1]$ \\
        14 & P014 & 3 & 0.0 & 0.0 & 0.0 & 25.9 & -- & -- & -- & -- & -- & $[1,1,1]$ \\
        15 & P058 & 3 & 0.0 & 0.0 & 0.0 & 0.0 & -- & -- & -- & -- & -- & $[1,1,1]$ \\
        16 & P064 & 3 & 0.0 & 0.0 & 0.0 & 100.0 & -- & -- & -- & -- & -- & $[1,1,1]$ \\
        17 & P081 & 3 & 0.0 & 0.0 & 30.9 & 92.6 & -- & -- & -- & -- & -- & $[1,0,1]$ \\
        18 & P082 & 3 & 0.0 & 100.0 & 100.0 & 100.0 & -- & -- & -- & -- & -- & $[1,1,1]$ \\
        19 & P090 & 3 & 100.0 & 100.0 & 100.0 & 100.0 & -- & -- & -- & -- & -- & $[1,1,1]$ \\
        20 & P091 & 3 & 0.0 & 0.0 & 33.3 & 100.0 & -- & -- & -- & -- & -- & $[1,1,0]$ \\
        21 & P107 & 3 & 100.0 & 100.0 & 100.0 & 100.0 & -- & -- & -- & -- & -- & $[1,1,1]$ \\
        22 & P011 & 4 & 0.0 & 0.0 & 16.7 & 50.0 & 100.0 & -- & -- & -- & -- & $[1,0,0,1]$ \\
        23 & P036 & 4 & 0.0 & 0.0 & 16.7 & 50.0 & 100.0 & -- & -- & -- & -- & $[1,0,0,1]$ \\
        24 & P067 & 4 & 0.0 & 0.0 & 0.0 & 0.0 & 100.0 & -- & -- & -- & -- & $[1,1,1,1]$ \\
        25 & P084 & 4 & 0.0 & 0.0 & 0.0 & 0.0 & 0.0 & -- & -- & -- & -- & $[1,1,1,1]$ \\
        26 & P103 & 4 & 100.0 & 100.0 & 100.0 & 100.0 & 100.0 & -- & -- & -- & -- & $[1,1,1,1]$ \\
        27 & P012 & 5 & 0.0 & 0.0 & 0.0 & 9.6 & 38.5 & 96.3 & -- & -- & -- & $[1,1,1,0,0]$ \\
        28 & P017 & 5 & 0.0 & 0.0 & 0.0 & 0.0 & 0.0 & 0.0 & -- & -- & -- & $[1,1,1,1,1]$ \\
        29 & P033 & 5 & 0.0 & 0.0 & 0.0 & 0.0 & 0.0 & 0.0 & -- & -- & -- & $[1,1,1,1,1]$ \\
        30 & P045 & 5 & 0.0 & 19.3 & 38.5 & 57.8 & 77.0 & 96.3 & -- & -- & -- & $[1,0,0,0,0]$ \\
        31 & P059 & 8 & 0.0 & 0.0 & 0.0 & 0.0 & 0.0 & 0.0 & 3.6 & 25.0 & 100.0 & $[1,1,1,0,1,1,0,1]$ \\
        \hline
        \multicolumn{13}{p{0.96\linewidth}}{\noindent\justifying
        \textbf{Note:}
        \textit{Each PPTR value is aggregated over all LLMs, attempts, and component combinations at the corresponding $\tau$. The entries in $\Psi_{\text{norm}}$ follow the text-input component order in each page context.}}
    \end{tabular}
\end{table*}

\subsection{Multi-Component Ablation Results}\label{ASEC:Multi-Component Ablation Results}
Table~\ref{TAB:multi-component-ablation} presents the per-page multi-component ablation results.
The table reports, for each UI page, the number of text-input components ($d$), the PPTR under each combinatorial strength $\tau$, and the normalized criticality vector $\Psi_{\text{norm}}$ for the corresponding text-input components.
Based on the results, we have the following observations:
\begin{itemize}
    \item
    The normalized criticality vectors reveal whether the text-input components on the same page contribute uniformly.
    Among the 31 pages, 20 pages (64.52\%) have a single $\Psi_{\text{norm}}$ value, indicating that all components on those pages have equal relative criticality.
    The remaining 11 pages (35.48\%) have two criticality levels, indicating that some components are more influential than others.

    \item
    The PPTR trends across $\tau$ further separate pages into stable, gradual-increase, and delayed-increase cases.
    Stable pages either remain at 0\% regardless of how many components are filled (e.g., P023, P062, P110, P058, P084, P017, and P033), or remain at 100\% because the page can pass through without depending on the generated values (e.g., P090, P107, and P103).
    Gradual-increase pages improve as more text-input components are filled.
    For example, P045 increases from 0.0\% at $\tau=0$ to 19.3\%, 38.5\%, 57.8\%, 77.0\%, and 96.3\% as $\tau$ increases from 1 to 5.

    \item
    Delayed-increase pages show that all, or nearly all, components must be filled before pass-through becomes likely.
    For example, P015, P056, P092, P115, P064, and P067 remain at 0\% until all components are filled.
    The largest page, P059 with eight components, remains at 0\% through $\tau=5$, then increases to 3.6\% at $\tau=6$, 25.0\% at $\tau=7$, and 100.0\% at $\tau=8$.

    \item
    Pages with two criticality levels explain why the same $\tau$ can lead to different PPTRs depending on which components are filled.
    For example, P005, P010, P021, and P087 each have two components but only one higher-criticality component, yielding 50.0\% PPTR at $\tau=1$ and 100.0\% at $\tau=2$.
    Larger pages, such as P012 and P059, show the same phenomenon at higher $d$: filling lower-criticality components alone is insufficient, while combinations containing higher-criticality components drive later PPTR increases.
\end{itemize}

\begin{tcolorbox}[breakable,colframe=black,arc=1mm,left={1mm},top={0mm},bottom={1mm},right={1mm},boxrule={0.25mm},before upper={\raisebox{-0.25\height}{\includegraphics[scale=0.013]{figures/icons/android}}~}]
    \textit{\textbf{Summary of Supplementary Multi-Component Ablation:}}
    When multiple text-input components are on a single UI page, their relative criticality can significantly affect the PPTR results.
    Overall, the PPTR increases with the number of filled components.
    Moreover, when users first enter more critical values into text-input components, PPTR increases.
    In contrast, prioritizing less critical components reduces PPTR.
\end{tcolorbox}

\newpage

\section{Detailed Results}
\label{asec:detailed-results}
This section provides the detailed results.

\begin{table*}[!h]
    \centering
    \scriptsize
    \captionsetup{skip=2pt}
    \caption{Page-pass-through trends for RQ1.1 under the three prompting settings.}
    \label{TAB:result-RQ1.1}
    \setlength\tabcolsep{0.7mm}
\begin{tabular}{llrrrrrrcrrrr}
        \hline
        \multirow{2}{*}{\textbf{Setting}} &
        \multirow{2}{*}{\textbf{LLM}} &
        \multicolumn{6}{c}{\textbf{Page-Pass-Through (n, \%) $\uparrow$}} &
        &
        \multicolumn{3}{c}{\textbf{Token $\downarrow$}} &
        \multirow{2}{*}{\makecell[r]{\textbf{Latency}\\\textbf{(s) $\downarrow$}}} \\
        \cline{3-8}\cline{10-12}
        & &
        \textbf{$\Delta$@1} & \textbf{@1} &
        \textbf{$\Delta$@2} & \textbf{@2} &
        \textbf{$\Delta$@3} & \textbf{@3} &
        &
        \textbf{Input} & \textbf{Output} & \textbf{Total} & \\
        \hline

        \multirow{10}{*}{\makecell[c]{Extracted\\Context}}
        & \GeminiM & 81 (70.4\%) & 81 (70.4\%) & 2 (1.7\%) & 83 (72.2\%) & 1 (0.9\%) & 84 (73.0\%) & & 242.7 & 589.0 & 831.7 & 12.7 \\
        & \ClaudeM & 77 (67.0\%) & 77 (67.0\%) & 3 (2.6\%) & 80 (69.6\%) & 1 (0.9\%) & 81 (70.4\%) & & 260.3 & 147.2 & 407.5 & 5.4 \\
        & \GrokM & 78 (67.8\%) & 78 (67.8\%) & 5 (4.3\%) & 83 (72.2\%) & 2 (1.7\%) & 85 (73.9\%) & & 335.5 & 1,318.2 & 1,653.7 & 19.1 \\
        & \GPTM & 80 (69.6\%) & 80 (69.6\%) & 1 (0.9\%) & 81 (70.4\%) & 1 (0.9\%) & 82 (71.3\%) & & 214.0 & 80.1 & 294.1 & 4.8 \\
        & \DoubaoM & 75 (65.2\%) & 75 (65.2\%) & 6 (5.2\%) & 81 (70.4\%) & 0 (0.0\%) & 81 (70.4\%) & & 280.1 & 1,244.1 & 1,524.3 & 29.7 \\
        & \QwenM & 79 (68.7\%) & 79 (68.7\%) & 5 (4.3\%) & 84 (73.0\%) & 1 (0.9\%) & 85 (73.9\%) & & 226.7 & 1,742.3 & 1,969.1 & 34.3 \\
\end{tabular}
\end{table*}

\begin{table*}[!h]
    \centering
    \scriptsize
   \captionsetup{skip=2pt}
    \caption{Symbol-based statistical comparisons and overhead for RQ1.1.}
    \label{TAB:result-RQ1.1-stat}
    \setlength\tabcolsep{0.51mm}
    \begin{tabular}{clccc cccc cccc}

        \multicolumn{13}{l}{\textbf{Panel A: LLM comparisons within the same prompting setting}} \\\hline
       \multirow{2}{*}{\textbf{Setting}} & \multirow{2}{*}{\textbf{LLM}} &
        \multicolumn{3}{c}{\textbf{PPTR@3}} & &
        \multicolumn{3}{c}{\textbf{Token}} & &
        \multicolumn{3}{c}{\textbf{Latency}} \\
        \cline{3-5}\cline{7-9}\cline{11-13}
        & & \ding{52}/\ding{109}/\ding{54} & $p$-value range & $\hat{A}_{12}$ range & &
        \ding{52}/\ding{109}/\ding{54} & $p$-value range & $\hat{A}_{12}$ range & &
        \ding{52}/\ding{109}/\ding{54} & $p$-value range & $\hat{A}_{12}$ range \\
        \hline
        \multirow{9}{*}{\makecell[c]{Extracted\\Context}}
        & \GeminiM & 0/8/0 & [0.388, 0.883] & [0.50, 0.53] & & 5/1/2 & [<0.001, 0.050] & [0.04, 0.96] & & 5/0/3 & [<0.001, 0.026] & [0.04, 1.00] \\
        & \ClaudeM & 0/8/0 & [0.558, 1.000] & [0.48, 0.51] & & 7/0/1 & [<0.001, <0.001] & [0.25, 1.00] & & 7/0/1 & [<0.001, <0.001] & [0.28, 1.00] \\
        & \GrokM & 0/8/0 & [0.311, 1.000] & [0.50, 0.53] & & 2/0/6 & [<0.001, 0.001] & [0.00, 0.76] & & 4/0/4 & [<0.001, <0.001] & [0.00, 1.00] \\
        & \GPTM & 0/8/0 & [0.568, 1.000] & [0.49, 0.52] & & 8/0/0 & [<0.001, <0.001] & [0.75, 1.00] & & 8/0/0 & [<0.001, <0.001] & [0.72, 1.00] \\
        & \DoubaoM & 0/8/0 & [0.558, 1.000] & [0.48, 0.51] & & 3/0/5 & [<0.001, 0.026] & [0.00, 0.80] & & 1/2/5 & [<0.001, 0.876] & [0.00, 0.98] \\
\end{tabular}
    \vspace{1mm}

    \setlength\tabcolsep{1.8mm}
        \begin{tabular}{lccc c ccc c ccc}

    \multicolumn{12}{l}{\textbf{Panel B: Prompting-setting comparisons within the same LLM}} \\
\hline
    \multirow{2}{*}{\textbf{LLM}}
    &
    \multicolumn{3}{c}{\textbf{PPTR@3}}
    &&
    \multicolumn{3}{c}{\textbf{Token}}
    &&
    \multicolumn{3}{c}{\textbf{Latency}}
    \\

    \cline{2-4}
    \cline{6-8}
    \cline{10-12}

    & Ext./XML
    & Ext./Vis.
    & XML/Vis.
    &
    & Ext./XML
    & Ext./Vis.
    & XML/Vis.
    &
    & Ext./XML
    & Ext./Vis.
    & XML/Vis.
    \\
    \hline

    \GeminiM
    & \ding{109} (0.50)
    & \ding{109} (0.50)
    & \ding{109} (0.50)
    &
    & \ding{52} (1.00)
    & \ding{52} (0.98)
    & \ding{54} (0.01)
    &
    & \ding{52} (0.58)
    & \ding{52} (0.84)
    & \ding{52} (0.80)
    \\

    \ClaudeM
    & \ding{109} (0.50)
    & \ding{109} (0.50)
    & \ding{109} (0.50)
    &
    & \ding{52} (0.99)
    & \ding{52} (0.97)
    & \ding{54} (0.01)
    &
    & \ding{52} (0.84)
    & \ding{52} (0.99)
    & \ding{52} (0.96)
    \\

    \GrokM
    & \ding{109} (0.51)
    & \ding{109} (0.52)
    & \ding{109} (0.51)
    &
    & \ding{52} (0.99)
    & \ding{52} (0.99)
    & \ding{54} (0.24)
    &
    & \ding{109} (0.48)
    & \ding{52} (0.68)
    & \ding{52} (0.71)
    \\

    \GPTM
    & \ding{109} (0.50)
    & \ding{109} (0.49)
    & \ding{109} (0.49)
    &
    & \ding{52} (1.00)
    & \ding{52} (0.99)
    & \ding{54} (0.00)
    &
    & \ding{52} (0.80)
    & \ding{52} (0.98)
    & \ding{52} (0.93)
    \\

    \DoubaoM
    & \ding{109} (0.50)
    & \ding{109} (0.50)
    & \ding{109} (0.50)
    &
    & \ding{52} (1.00)
    & \ding{52} (0.95)
    & \ding{54} (0.00)
    &
    & \ding{54} (0.39)
    & \ding{54} (0.41)
    & \ding{109} (0.52)
    \\
\end{tabular}
\end{table*}

\begin{table*}[!h]
    \centering
    \fontsize{6.2pt}{8pt}\selectfont
    \captionsetup{skip=2pt}
    \caption{Bug-reveal trends for RQ1.2 under the three prompting settings.}
    \label{TAB:result-RQ1.2}
    \setlength\tabcolsep{0.38mm}
    \begin{tabular}{llrrrrrrrrrrrrr}
        \hline
        \multirow{2}{*}{\textbf{Setting}} & \multirow{2}{*}{\textbf{LLM}} & \multicolumn{8}{c}{\textbf{Bug-Reveal (n, \%) $\uparrow$}} && \multicolumn{3}{c}{\textbf{Token} $\downarrow$} & \multirow{2}{*}{\makecell[r]{\textbf{Latency}\\\textbf{(s) $\downarrow$}}} \\
        \cline{3-10}\cline{12-14}
        & & \textbf{$\Delta$@1} & \textbf{@1} & \textbf{$\Delta$@5} & \textbf{@5} & \textbf{$\Delta$@10} & \textbf{@10} & \textbf{$\Delta$@30} & \textbf{@30} & & \textbf{Input} & \textbf{Output} & \textbf{Total} & \\
        \hline
        \multirow{10}{*}{\makecell[c]{Extracted\\Context}}
        & \GeminiM & 12 (32.4\%) & 12 (32.4\%) & 3 (8.1\%) & 15 (40.5\%) & 0 (0.0\%) & 15 (40.5\%) & 2 (5.4\%) & 17 (45.9\%) & & 324.4 & 1,149.3 & 1,473.7 & 15.9 \\
        & \ClaudeM & 16 (43.2\%) & 16 (43.2\%) & 1 (2.7\%) & 17 (45.9\%) & 1 (2.7\%) & 18 (48.6\%) & 2 (5.4\%) & 20 (54.1\%) & & 340.8 & 560.6 & 901.4 & 12.1 \\
        & \GrokM & 13 (35.1\%) & 13 (35.1\%) & 3 (8.1\%) & 16 (43.2\%) & 1 (2.7\%) & 17 (45.9\%) & 1 (2.7\%) & 18 (48.6\%) & & 393.2 & 2,458.8 & 2,852.0 & 30.6 \\
        & \GPTM & 18 (48.6\%) & 18 (48.6\%) & 1 (2.7\%) & 19 (51.4\%) & 3 (8.1\%) & 22 (59.5\%) & 0 (0.0\%) & 22 (59.5\%) & & 270.5 & 86.5 & 357.0 & 5.6 \\
        & \DoubaoM & 14 (37.8\%) & 14 (37.8\%) & 2 (5.4\%) & 16 (43.2\%) & 2 (5.4\%) & 18 (48.6\%) & 1 (2.7\%) & 19 (51.4\%) & & 363.1 & 1,845.3 & 2,208.4 & 42.3 \\
        & \QwenM & 12 (32.4\%) & 12 (32.4\%) & 3 (8.1\%) & 15 (40.5\%) & 2 (5.4\%) & 17 (45.9\%) & 1 (2.7\%) & 18 (48.6\%) & & 299.9 & 2,652.0 & 2,951.9 & 50.5 \\
\end{tabular}
\end{table*}

\begin{table*}[!h]
    \centering
    \scriptsize
    \captionsetup{skip=2pt}
    \caption{Symbol-based statistical comparisons and overhead for RQ1.2.}
    \label{TAB:result-RQ1.2-stat}
    \setlength\tabcolsep{0.51mm}
    \begin{tabular}{clccc cccc cccc}

        \multicolumn{13}{l}{\textbf{Panel A: LLM comparisons within the same prompting setting}} \\\hline
       \multirow{2}{*}{\textbf{Setting}} & \multirow{2}{*}{\textbf{LLM}} &
        \multicolumn{3}{c}{\textbf{BugReveal@30}} & &
        \multicolumn{3}{c}{\textbf{Token}} & &
        \multicolumn{3}{c}{\textbf{Latency}} \\
        \cline{3-5}\cline{7-9}\cline{11-13}
        & & \ding{52}/\ding{109}/\ding{54} & $p$-value range & $\hat{A}_{12}$ range & &
        \ding{52}/\ding{109}/\ding{54} & $p$-value range & $\hat{A}_{12}$ range & &
        \ding{52}/\ding{109}/\ding{54} & $p$-value range & $\hat{A}_{12}$ range \\
        \hline
        \multirow{9}{*}{\makecell[c]{Extracted\\Context}}
        & \GeminiM & 0/8/0 & [0.250, 1.000] & [0.43, 0.50] & & 5/0/3 & [<0.001, <0.001] & [0.00, 0.96] & & 5/0/3 & [<0.001, <0.001] & [0.00, 1.00] \\
         & \ClaudeM & 0/8/0 & [0.492, 1.000] & [0.47, 0.54] & & 6/1/1 & [<0.001, 0.812] & [0.09, 0.99] & & 6/1/1 & [<0.001, 0.531] & [0.07, 1.00] \\
         & \GrokM & 0/8/0 & [0.357, 1.000] & [0.45, 0.51] & & 0/2/6 & [<0.001, 0.974] & [0.00, 0.50] & & 4/0/4 & [<0.001, <0.001] & [0.00, 0.95] \\
         & \GPTM & 0/8/0 & [0.250, 0.646] & [0.53, 0.57] & & 8/0/0 & [<0.001, <0.001] & [0.91, 1.00] & & 8/0/0 & [<0.001, <0.001] & [0.93, 1.00] \\
         & \DoubaoM & 0/8/0 & [0.490, 1.000] & [0.46, 0.53] & & 3/1/4 & [<0.001, 0.230] & [0.00, 0.76] & & 3/0/5 & [<0.001, 0.038] & [0.00, 0.87] \\
\end{tabular}
    \vspace{1mm}

    \setlength\tabcolsep{1.75mm}
        \begin{tabular}{lccc c ccc c ccc}

    \multicolumn{12}{l}{\textbf{Panel B: Prompting-setting comparisons within the same LLM}} \\
\hline
    \multirow{2}{*}{\textbf{LLM}}
    &
    \multicolumn{3}{c}{\textbf{BugReveal@30}}
    &&
    \multicolumn{3}{c}{\textbf{Token}}
    &&
    \multicolumn{3}{c}{\textbf{Latency}}
    \\

    \cline{2-4}
    \cline{6-8}
    \cline{10-12}

    & Ext./XML
    & Ext./Vis.
    & XML/Vis.
    &
    & Ext./XML
    & Ext./Vis.
    & XML/Vis.
    &
    & Ext./XML
    & Ext./Vis.
    & XML/Vis.
    \\
    \hline
        \GeminiM & \ding{109} (0.49) & \ding{109} (0.46) & \ding{109} (0.47) && \ding{52} (1.00) & \ding{52} (0.97) & \ding{54} (0.00) && \ding{52} (0.66) & \ding{52} (0.84) & \ding{52} (0.71) \\
\ClaudeM & \ding{109} (0.50) & \ding{109} (0.50) & \ding{109} (0.50) && \ding{52} (1.00) & \ding{52} (0.97) & \ding{54} (0.00) && \ding{52} (0.70) & \ding{52} (0.66) & \ding{109} (0.43) \\
\GrokM & \ding{109} (0.50) & \ding{109} (0.47) & \ding{109} (0.47) && \ding{52} (1.00) & \ding{52} (0.99) & \ding{54} (0.03) && \ding{54} (0.35) & \ding{109} (0.53) & \ding{52} (0.66) \\
\GPTM & \ding{109} (0.49) & \ding{109} (0.51) & \ding{109} (0.53) && \ding{52} (1.00) & \ding{52} (1.00) & \ding{54} (0.00) && \ding{52} (0.90) & \ding{52} (0.96) & \ding{52} (0.72) \\
\DoubaoM & \ding{109} (0.50) & \ding{109} (0.50) & \ding{109} (0.50) && \ding{52} (1.00) & \ding{52} (0.87) & \ding{54} (0.00) && \ding{109} (0.40) & \ding{109} (0.43) & \ding{109} (0.52) \\
\end{tabular}
\end{table*}

\begin{table*}[!h]
    \centering
    \scriptsize
    \captionsetup{skip=2pt}
    \caption{Feedback-enhanced page-pass-through trends for RQ2.1 under the three prompting settings.}
    \label{TAB:result-RQ2.1}
    \setlength\tabcolsep{0.22mm}
    \begin{tabular}{llrrrrrrcrrrr}
        \hline
        \multirow{2}{*}{\textbf{Setting}} &
        \multirow{2}{*}{\textbf{LLM}} &
        \multicolumn{6}{c}{\textbf{Page-Pass-Through (n, \%) $\uparrow$}}
        &
        & \multicolumn{3}{c}{\textbf{Token $\downarrow$}} &
        \multirow{2}{*}{\makecell[r]{\textbf{Latency}\\\textbf{(s) $\downarrow$}}} \\
        \cline{3-8}\cline{10-12}
        &  &
        \textbf{$\Delta$@1} & \textbf{@1} &
        \textbf{$\Delta$@2} & \textbf{@2} &
        \textbf{$\Delta$@3} & \textbf{@3} &
        &
        \textbf{Input} & \textbf{Output} & \textbf{Total} & \\
        \hline
        \multirow{10}{*}{\makecell[c]{Extracted\\Context}}
        & \GeminiM & 74 (64.3\%) & 74 (64.3\%) & 10 (8.7\%) & 84 (73.0\%) & 0 (0.0\%) & 84 (73.0\%) & & 479.6 & 853.0 & 1,332.5 & 13.6 \\
         & \ClaudeM & 77 (67.0\%) & 77 (67.0\%) & 6 (5.2\%) & 83 (72.2\%) & 2 (1.7\%) & 85 (73.9\%) & & 519.2 & 157.9 & 677.0 & 7.4 \\
         & \GrokM & 77 (67.0\%) & 77 (67.0\%) & 4 (3.5\%) & 81 (70.4\%) & 3 (2.6\%) & 84 (73.0\%) & & 565.4 & 2,359.9 & 2,925.3 & 41.5 \\
         & \GPTM & 81 (70.4\%) & 81 (70.4\%) & 3 (2.6\%) & 84 (73.0\%) & 1 (0.9\%) & 85 (73.9\%) & & 423.6 & 94.4 & 518.0 & 5.3 \\
         & \DoubaoM & 80 (69.6\%) & 80 (69.6\%) & 4 (3.5\%) & 84 (73.0\%) & 1 (0.9\%) & 85 (73.9\%) & & 666.1 & 1,558.4 & 2,224.5 & 36.1 \\
         & \QwenM & 78 (67.8\%) & 78 (67.8\%) & 3 (2.6\%) & 81 (70.4\%) & 5 (4.3\%) & 86 (74.8\%) & & 385.2 & 4,005.5 & 4,390.7 & 78.3 \\
\end{tabular}
\end{table*}

\begin{table*}[!h]
    \centering
    \scriptsize
    \captionsetup{skip=2pt}
    \caption{Symbol-based statistical comparisons and overhead for RQ2.1.}
    \label{TAB:result-RQ2.1-stat}
    \setlength\tabcolsep{0.58mm}
    \begin{tabular}{clccc cccc cccc}

        \multicolumn{13}{l}{\textbf{Panel A: LLM comparisons within the same prompting setting}} \\\hline
       \multirow{2}{*}{\textbf{Setting}} & \multirow{2}{*}{\textbf{LLM}} &
        \multicolumn{3}{c}{\textbf{PPTR@3}} & &
        \multicolumn{3}{c}{\textbf{Token}} & &
        \multicolumn{3}{c}{\textbf{Latency}} \\
        \cline{3-5}\cline{7-9}\cline{11-13}
        & & \ding{52}/\ding{109}/\ding{54} & $p$-value range & $\hat{A}_{12}$ range & &
        \ding{52}/\ding{109}/\ding{54} & $p$-value range & $\hat{A}_{12}$ range & &
        \ding{52}/\ding{109}/\ding{54} & $p$-value range & $\hat{A}_{12}$ range \\
        \hline
        \multirow{9}{*}{\makecell[c]{Extracted\\Context}}
        & \GeminiM & 0/8/0 & [0.562, 1.000] & [0.49, 0.52] && 5/0/3 & [<0.001, 0.022] & [0.11, 0.87] && 5/1/2 & [<0.001, 0.215] & [0.07, 0.97] \\
         & \ClaudeM & 0/8/0 & [0.466, 1.000] & [0.49, 0.52] && 7/0/1 & [<0.001, <0.001] & [0.34, 0.98] && 7/0/1 & [<0.001, <0.001] & [0.28, 0.99] \\
         & \GrokM & 0/8/0 & [0.562, 1.000] & [0.49, 0.52] && 1/1/6 & [<0.001, 0.064] & [0.02, 0.59] && 3/1/4 & [<0.001, 0.339] & [0.00, 0.74] \\
         & \GPTM & 0/8/0 & [0.466, 1.000] & [0.49, 0.52] && 8/0/0 & [<0.001, <0.001] & [0.66, 0.98] && 8/0/0 & [<0.001, <0.001] & [0.72, 1.00] \\
         & \DoubaoM & 0/8/0 & [0.466, 1.000] & [0.49, 0.52] && 1/2/5 & [<0.001, 0.622] & [0.02, 0.66] && 2/0/6 & [<0.001, 0.011] & [0.00, 0.71] \\
\end{tabular}
    \vspace{1mm}

    \setlength\tabcolsep{1.8mm}
        \begin{tabular}{lccc c ccc c ccc}

    \multicolumn{12}{l}{\textbf{Panel B: Prompting-setting comparisons within the same LLM}} \\
\hline
    \multirow{2}{*}{\textbf{LLM}}
    &
    \multicolumn{3}{c}{\textbf{PPTR@3}}
    &&
    \multicolumn{3}{c}{\textbf{Token}}
    &&
    \multicolumn{3}{c}{\textbf{Latency}}
    \\

    \cline{2-4}
    \cline{6-8}
    \cline{10-12}

    & Ext./XML
    & Ext./Vis.
    & XML/Vis.
    &
    & Ext./XML
    & Ext./Vis.
    & XML/Vis.
    &
    & Ext./XML
    & Ext./Vis.
    & XML/Vis.
    \\
    \hline
        \GeminiM & \ding{109} (0.49) & \ding{109} (0.49) & \ding{109} (0.50) && \ding{52} (0.86) & \ding{52} (0.86) & \ding{54} (0.20) && \ding{52} (0.58) & \ding{52} (0.85) & \ding{52} (0.73) \\
\ClaudeM & \ding{109} (0.49) & \ding{109} (0.49) & \ding{109} (0.50) && \ding{52} (0.94) & \ding{52} (0.92) & \ding{54} (0.11) && \ding{52} (0.73) & \ding{52} (0.94) & \ding{52} (0.86) \\
\GrokM & \ding{109} (0.50) & \ding{109} (0.51) & \ding{109} (0.50) && \ding{52} (0.93) & \ding{52} (0.86) & \ding{54} (0.25) && \ding{109} (0.49) & \ding{52} (0.66) & \ding{52} (0.68) \\
\GPTM & \ding{109} (0.49) & \ding{109} (0.49) & \ding{109} (0.50) && \ding{52} (1.00) & \ding{52} (0.98) & \ding{54} (0.01) && \ding{52} (0.67) & \ding{52} (0.93) & \ding{52} (0.88) \\
\DoubaoM & \ding{109} (0.50) & \ding{109} (0.50) & \ding{109} (0.50) && \ding{52} (0.99) & \ding{52} (0.83) & \ding{54} (0.04) && \ding{54} (0.41) & \ding{109} (0.43) & \ding{109} (0.53) \\
\end{tabular}
\end{table*}

\begin{table*}[!h]
    \centering
    \scriptsize
    \captionsetup{skip=2pt}
    \caption{Feedback-enhanced bug-reveal trends for RQ2.2 under the three prompting settings.}
    \label{TAB:result-RQ2.2}
    \setlength\tabcolsep{0.01mm}
    \begin{tabular}{llrrrrrrrrcrrrr}
        \hline
        \multirow{2}{*}{\textbf{Setting}} & \multirow{2}{*}{\textbf{LLM}} & \multicolumn{8}{c}{\textbf{Bug-Reveal (n, \%)} $\uparrow$} && \multicolumn{3}{c}{\textbf{Token} $\downarrow$} & \multirow{2}{*}{\makecell[r]{\textbf{Latency}\\\textbf{(s)} $\downarrow$}} \\
        \cline{3-10}\cline{12-14}
        & & \textbf{$\Delta$@1} & \textbf{@1} & \textbf{$\Delta$@5} & \textbf{@5} & \textbf{$\Delta$@10} & \textbf{@10} & \textbf{$\Delta$@30} & \textbf{@30} & & \textbf{Input} & \textbf{Output} & \textbf{Total} & \\
        \hline
        \multirow{10}{*}{\makecell[c]{Extracted\\Context}}
        & \GeminiM & 11 (29.7\%) & 11 (29.7\%) & 13 (35.1\%) & 24 (64.9\%) & 0 (0.0\%) & 24 (64.9\%) & 1 (2.7\%) & 25 (67.6\%) & & 9,280.2 & 11,936.3 & 21,216.5 & 192.9 \\
         & \ClaudeM & 12 (32.4\%) & 12 (32.4\%) & 12 (32.4\%) & 24 (64.9\%) & 2 (5.4\%) & 26 (70.3\%) & 0 (0.0\%) & 26 (70.3\%) & & 7,770.5 & 2,596.0 & 10,366.5 & 62.9 \\
         & \GrokM & 12 (32.4\%) & 12 (32.4\%) & 8 (21.6\%) & 20 (54.1\%) & 2 (5.4\%) & 22 (59.5\%) & 1 (2.7\%) & 23 (62.2\%) & & 9,657.5 & 44,684.6 & 54,342.1 & 668.3 \\
         & \GPTM & 14 (37.8\%) & 14 (37.8\%) & 6 (16.2\%) & 20 (54.1\%) & 3 (8.1\%) & 23 (62.2\%) & 4 (10.8\%) & 27 (73.0\%) & & 5,120.1 & 508.5 & 5,628.6 & 57.0 \\
         & \DoubaoM & 16 (43.2\%) & 16 (43.2\%) & 5 (13.5\%) & 21 (56.8\%) & 4 (10.8\%) & 25 (67.6\%) & 4 (10.8\%) & 29 (78.4\%) & & 10,450.0 & 14,343.7 & 24,793.7 & 364.6 \\
         & \QwenM & 11 (29.7\%) & 11 (29.7\%) & 9 (24.3\%) & 20 (54.1\%) & 3 (8.1\%) & 23 (62.2\%) & 3 (8.1\%) & 26 (70.3\%) & & 7,979.9 & 64,253.7 & 72,233.5 & 1,178.9 \\
\hline
    \end{tabular}
\end{table*}

\begin{table*}[!h]
    \centering
    \scriptsize
    \captionsetup{skip=2pt}
    \caption{Symbol-based statistical comparisons and overhead for RQ2.2.}
    \label{TAB:result-RQ2.2-stat}
    \setlength\tabcolsep{0.58mm}
    \begin{tabular}{clccc cccc cccc}
        \multicolumn{13}{l}{\textbf{Panel A: LLM comparisons within the same prompting setting}} \\\hline
        \multirow{2}{*}{\textbf{Setting}} & \multirow{2}{*}{\textbf{LLM}} &
        \multicolumn{3}{c}{\textbf{BugReveal@30}} &&
        \multicolumn{3}{c}{\textbf{Token}} &&
        \multicolumn{3}{c}{\textbf{Latency}} \\
        \cline{3-5}\cline{7-9}\cline{11-13}
        & & \ding{52}/\ding{109}/\ding{54} & $p$-value range & $\hat{A}_{12}$ range &&
        \ding{52}/\ding{109}/\ding{54} & $p$-value range & $\hat{A}_{12}$ range &&
        \ding{52}/\ding{109}/\ding{54} & $p$-value range & $\hat{A}_{12}$ range \\
        \hline

        \multirow{9}{*}{\makecell[c]{Extracted\\Context}}
        & \GeminiM & 0/8/0 & [0.063, 0.808] & [0.45, 0.61] && 3/3/2 & [<0.001, 0.754] & [0.25, 0.73] && 5/1/2 & [<0.001, 0.160] & [0.25, 0.83] \\
        & \ClaudeM & 1/7/0 & [0.036, 1.000] & [0.46, 0.62] && 7/0/1 & [<0.001, 0.043] & [0.36, 0.82] && 7/0/1 & [<0.001, 0.043] & [0.36, 0.95] \\
        & \GrokM & 0/8/0 & [0.131, 0.818] & [0.42, 0.58] && 0/2/6 & [<0.001, 0.893] & [0.15, 0.58] && 1/3/4 & [<0.001, 0.738] & [0.11, 0.67] \\
        & \GPTM & 1/7/0 & [0.019, 0.803] & [0.47, 0.64] && 8/0/0 & [<0.001, 0.043] & [0.64, 0.87] && 8/0/0 & [<0.001, 0.043] & [0.64, 0.96] \\
        & \DoubaoM & 1/7/0 & [0.004, 0.789] & [0.51, 0.66] && 3/3/2 & [<0.001, 0.424] & [0.21, 0.68] && 2/2/4 & [<0.001, 0.795] & [0.09, 0.70] \\
\end{tabular}
    \vspace{1mm}

    \setlength\tabcolsep{1.8mm}
    \begin{tabular}{lccc c ccc c ccc}
        \multicolumn{12}{l}{\textbf{Panel B: Prompting-setting comparisons within the same LLM}} \\
        \hline
        \multirow{2}{*}{\textbf{LLM}}
        & \multicolumn{3}{c}{\textbf{BugReveal@30}}
        && \multicolumn{3}{c}{\textbf{Token}}
        && \multicolumn{3}{c}{\textbf{Latency}} \\
        \cline{2-4}\cline{6-8}\cline{10-12}
        & Ext./XML & Ext./Vis. & XML/Vis.
        && Ext./XML & Ext./Vis. & XML/Vis.
        && Ext./XML & Ext./Vis. & XML/Vis. \\
        \hline

        \GeminiM & \ding{109} (0.50) & \ding{109} (0.51) & \ding{109} (0.51) && \ding{52} (0.77) & \ding{52} (0.66) & \ding{54} (0.27) && \ding{109} (0.55) & \ding{109} (0.62) & \ding{109} (0.57) \\
        \ClaudeM & \ding{109} (0.51) & \ding{109} (0.47) & \ding{109} (0.46) && \ding{52} (0.90) & \ding{52} (0.70) & \ding{54} (0.11) && \ding{109} (0.62) & \ding{109} (0.59) & \ding{109} (0.47) \\
        \GrokM & \ding{109} (0.51) & \ding{109} (0.51) & \ding{109} (0.50) && \ding{52} (0.71) & \ding{109} (0.63) & \ding{54} (0.35) && \ding{109} (0.47) & \ding{109} (0.50) & \ding{109} (0.54) \\
        \GPTM & \ding{109} (0.57) & \ding{109} (0.54) & \ding{109} (0.47) && \ding{52} (0.91) & \ding{52} (0.66) & \ding{54} (0.16) && \ding{109} (0.58) & \ding{109} (0.62) & \ding{109} (0.55) \\
        \DoubaoM & \ding{109} (0.57) & \ding{109} (0.54) & \ding{109} (0.47) && \ding{52} (0.80) & \ding{109} (0.61) & \ding{54} (0.22) && \ding{109} (0.47) & \ding{109} (0.45) & \ding{109} (0.48) \\
\end{tabular}
\end{table*}

\begin{table*}[!h]
    \centering
    \scriptsize
    \captionsetup{skip=2pt}
    \caption{Average effectiveness and token/time overhead per attempt across prompt settings for PPTR.}
    \label{TAB:resultpptr-RQ2-summary}
    \setlength\tabcolsep{2.35mm}
    
    \begin{tabular}{lrrrrrrr}
        \hline
        \textbf{Setting} & \textbf{@1} & \textbf{@2} & \textbf{@3} & \textbf{Input Token} & \textbf{Output Token} & \textbf{Total Token} & \textbf{Latency (s)} \\
        \hline
        Extracted Context & 66.8\% & 71.7\% & 73.3\% & 494.4 & 1,428.0 & 1,922.3 & 33.5 \\
        UI-hierarchy XML & 66.1\% & 73.3\% & 74.2\% & 6,538.2 & 1,246.5 & 7,784.8 & 31.4 \\
        Screenshot Vision & 60.5\% & 72.5\% & 73.8\% & 4,654.6 & 1,321.1 & 5,975.7 & 37.3 \\
        \hline
    \end{tabular}
\end{table*}

\begin{table*}[!h]
    \centering
    \scriptsize
    \captionsetup{skip=2pt}
    \caption{Average effectiveness and token/time overhead per attempt across prompt settings for BDR.}
    \label{TAB:resultbug-RQ2-summary}
    \begin{tabular}{lrrrrrrrr}
        \hline
        \textbf{Setting} & \textbf{@1} & \textbf{@5} & \textbf{@10} & \textbf{@30} & \textbf{Input Token} & \textbf{Output Token} & \textbf{Total Token} & \textbf{latency (s)} \\
        \hline
        Extracted Context & 33.3\% & 56.2\% & 61.3\% & 68.5\% & 9,059.5 & 22,781.1 & 31,840.6 & 500.8 \\
        UI-hierarchy XML & 34.5\% & 54.7\% & 59.2\% & 63.7\% & 21,174.5 & 23,915.7 & 45,090.1 & 508.1 \\
        Screenshot Vision & 32.7\% & 53.5\% & 59.8\% & 63.7\% & 25,187.0 & 24,008.3 & 49,195.3 & 509.3 \\
        \hline
    \end{tabular}
\end{table*}

\begin{figure}[!h]
	\centering
	\includegraphics[width=0.76\textwidth]{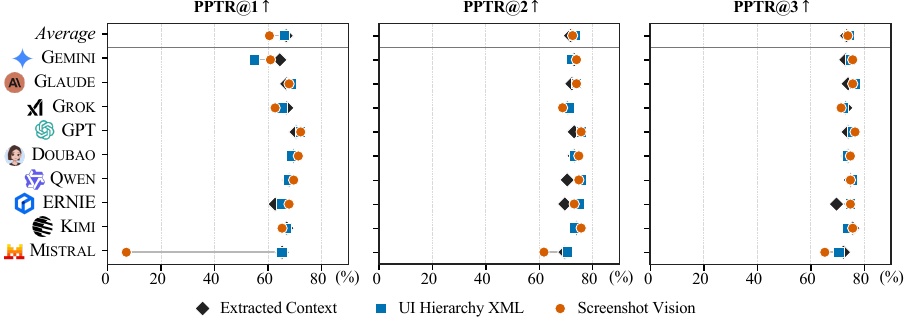}\captionsetup{skip=2pt}
	\caption{Prompt-setting impact on feedback-enhanced page pass-through effectiveness.}
	\label{fig:page_pass_feedback_prompt_setting_effectiveness}
\end{figure}

\begin{figure}[!h]
	\centering
	\includegraphics[width=\textwidth]{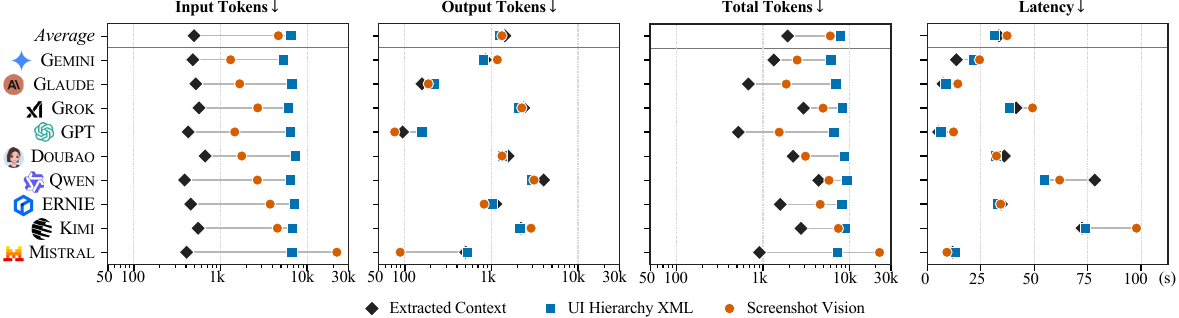}\captionsetup{skip=2pt}
	\caption{Prompt-setting token and time overhead per attempt for feedback-enhanced page pass-through.}
	\label{fig:page_pass_feedback_prompt_setting_overhead}
\end{figure}

\begin{figure}[!h]
	\centering
	\includegraphics[width=\textwidth]{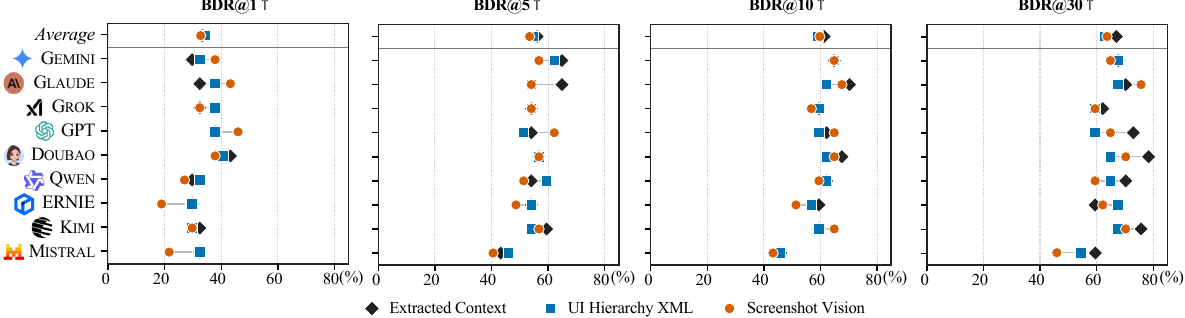}\captionsetup{skip=2pt}
	\caption{Prompt-setting impact on feedback-enhanced bug-revealing effectiveness.}
	\label{fig:bug_reveal_feedback_prompt_setting_effectiveness}
\end{figure}

\begin{figure}[!h]
	\centering
	\includegraphics[width=\textwidth]{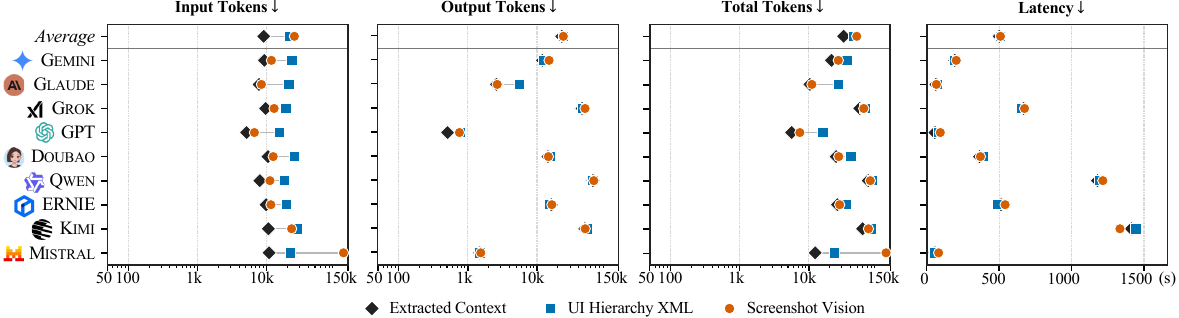}\captionsetup{skip=2pt}
	\caption{Prompt-setting token and time overhead per attempt for feedback-enhanced bug revealing.}
	\label{fig:bug_reveal_feedback_prompt_setting_overhead}
\end{figure}

\begin{figure}[!h]
	\centering
	\includegraphics[width=\textwidth]{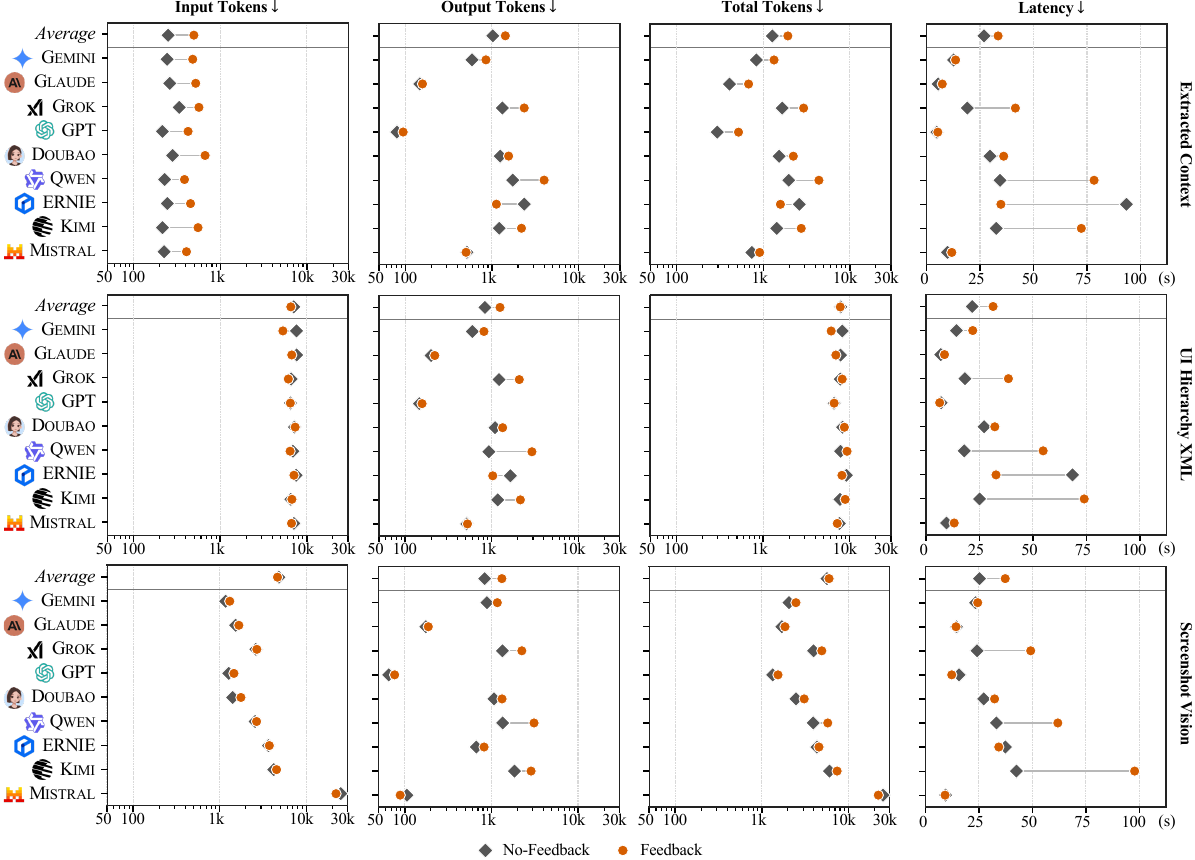}\captionsetup{skip=2pt}
	\caption{Token and time overhead per attempt for no-feedback and feedback-enhanced page pass-through.}
	\label{fig:page_pass_no_feedback_feedback_overhead}
\end{figure}

\begin{figure}[!h]
	\centering
	\includegraphics[width=\textwidth]{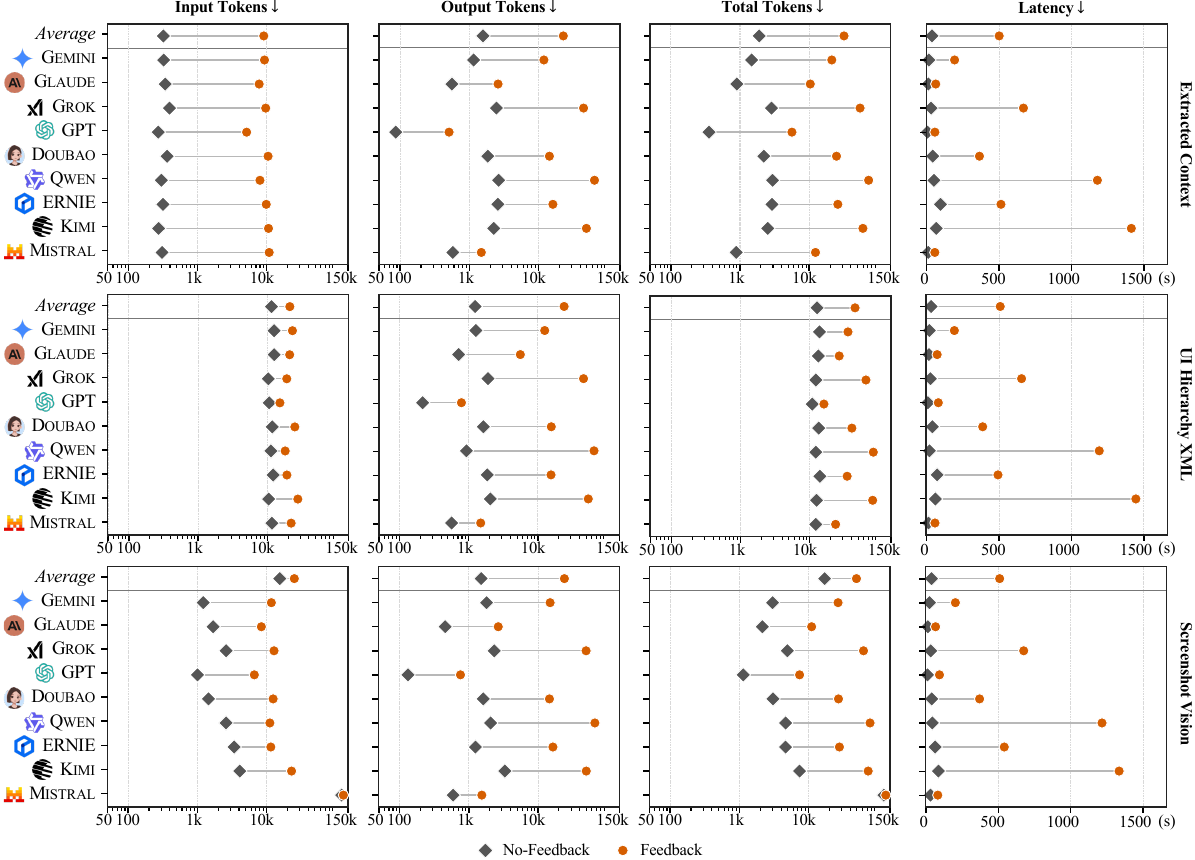}\captionsetup{skip=2pt}
	\caption{Token and time overhead per attempt for no-feedback and feedback-enhanced bug revealing.}
	\label{fig:bug_reveal_no_feedback_feedback_overhead}
\end{figure}

\end{document}